\pdfoutput=1
\PassOptionsToPackage{unicode}{hyperref}
\PassOptionsToPackage{hyphens}{url}
\PassOptionsToPackage{dvipsnames,svgnames,x11names}{xcolor}
\documentclass[
  12pt]{article}

\usepackage{amsmath,amssymb}

\usepackage[T1]{fontenc}
\usepackage[utf8]{inputenc}
\usepackage{textcomp} 
\usepackage{lmodern}

\IfFileExists{upquote.sty}{\usepackage{upquote}}{}
\IfFileExists{microtype.sty}{
  \usepackage[]{microtype}
  \UseMicrotypeSet[protrusion]{basicmath} 
}{}
\makeatletter
\@ifundefined{KOMAClassName}{
  \IfFileExists{parskip.sty}{%
    \usepackage{parskip}
  }{
    \setlength{\parindent}{0pt}
    \setlength{\parskip}{6pt plus 2pt minus 1pt}}
}{
  \KOMAoptions{parskip=half}}
\makeatother
\usepackage{xcolor}
\makeatletter
\ifx\paragraph\undefined\else
  \let\oldparagraph\paragraph
  \renewcommand{\paragraph}{
    \@ifstar
      \xxxParagraphStar
      \xxxParagraphNoStar
  }
  \newcommand{\xxxParagraphStar}[1]{\oldparagraph*{#1}\mbox{}}
  \newcommand{\xxxParagraphNoStar}[1]{\oldparagraph{#1}\mbox{}}
\fi
\ifx\subparagraph\undefined\else
  \let\oldsubparagraph\subparagraph
  \renewcommand{\subparagraph}{
    \@ifstar
      \xxxSubParagraphStar
      \xxxSubParagraphNoStar
  }
  \newcommand{\xxxSubParagraphStar}[1]{\oldsubparagraph*{#1}\mbox{}}
  \newcommand{\xxxSubParagraphNoStar}[1]{\oldsubparagraph{#1}\mbox{}}
\fi
\makeatother

\usepackage{longtable,booktabs,array}
\usepackage{calc} 
\usepackage{etoolbox}

\makeatletter
\patchcmd\longtable{\par}{\if@noskipsec\mbox{}\fi\par}{}{}
\makeatother
\IfFileExists{footnotehyper.sty}{\usepackage{footnotehyper}}{\usepackage{footnote}}
\makesavenoteenv{longtable}
\usepackage{graphicx}
\makeatletter
\def\maxwidth{\ifdim\Gin@nat@width>\linewidth\linewidth\else\Gin@nat@width\fi}
\def\maxheight{\ifdim\Gin@nat@height>\textheight\textheight\else\Gin@nat@height\fi}
\makeatother

\setkeys{Gin}{width=\maxwidth,height=\maxheight,keepaspectratio}

\makeatletter
\def\fps@figure{htbp}
\makeatother

\makeatletter
\@ifpackageloaded{caption}{}{\usepackage{caption}}
\AtBeginDocument{%
\ifdefined\contentsname
  \renewcommand*\contentsname{Table of contents}
\else
  \newcommand\contentsname{Table of contents}
\fi
\ifdefined\listfigurename
  \renewcommand*\listfigurename{List of Figures}
\else
  \newcommand\listfigurename{List of Figures}
\fi
\ifdefined\listtablename
  \renewcommand*\listtablename{List of Tables}
\else
  \newcommand\listtablename{List of Tables}
\fi
\ifdefined\figurename
  \renewcommand*\figurename{Figure}
\else
  \newcommand\figurename{Figure}
\fi
\ifdefined\tablename
  \renewcommand*\tablename{Table}
\else
  \newcommand\tablename{Table}
\fi
}
\@ifpackageloaded{float}{}{\usepackage{float}}
\floatstyle{ruled}
\@ifundefined{c@chapter}{\newfloat{codelisting}{h}{lop}}{\newfloat{codelisting}{h}{lop}[chapter]}
\floatname{codelisting}{Listing}

\makeatother
\makeatletter
\@ifpackageloaded{caption}{}{\usepackage{caption}}
\@ifpackageloaded{subcaption}{}{\usepackage{subcaption}}
\makeatother

\usepackage[]{natbib}
\usepackage{bibunits}
\defaultbibliographystyle{agsm}

\usepackage{bookmark}

\IfFileExists{xurl.sty}{\usepackage{xurl}}{} 
\hypersetup{
  pdftitle={Title},
  pdfauthor={Author 1; Author 2},
  pdfkeywords={3 to 6 keywords, that do not appear in the title},
  colorlinks=true,
  linkcolor={blue},
  filecolor={Maroon},
  citecolor={Blue},
  urlcolor={Blue},
  pdfcreator={LaTeX via pandoc}}

\newcommand{\anon}{1}

\usepackage{amssymb}
\usepackage{color, graphics, graphicx}
\usepackage{xcolor,colortbl}
\usepackage{amsmath}
\usepackage{bm}
\usepackage{fontenc}
\usepackage{subcaption}
\usepackage{lscape}
\usepackage{url}
\usepackage{bbm}
\usepackage{dsfont}
\usepackage{amsthm}
\usepackage{enumitem}
\usepackage{afterpage} 
\usepackage{longtable} 
\usepackage{booktabs} 
\usepackage{bbm}
\usepackage{flafter}
\usepackage{rotating}
\usepackage{tocbibind}
\usepackage[title, toc]{appendix}
\usepackage{mathtools}
\usepackage{authblk}
\usepackage{booktabs}
\usepackage{multirow}
\usepackage{array}
\usepackage{etoolbox}
\usepackage[ruled,vlined]{algorithm2e}

\definecolor{Gray}{gray}{0.85}

\newtheorem{assumptionA}{Assumption}

\newtheorem{assumptionB}{Assumption}

\newtheorem{condition}{Condition}

\newtheorem*{assumption*}{Assumption}
\newtheorem*{definition*}{Definition}
\newtheorem{theorem}{Theorem}
\newtheorem{model}{Model}

\newtheorem{corollary}{Corollary}

\newtheorem{definition}{Definition}

\newtheorem{lemma}{Lemma}

\newtheorem{proposition}{Proposition}

\DeclareMathOperator*{\argmin}{arg\,min}

\newcommand{\N}{\mathbb{N}}
\newcommand{\Z}{\mathbb{Z}}
\newcommand{\R}{\mathbb{R}}

\newcommand{\E}{\mathbb{E}}
\newcommand{\p}{\mathbb{P}}

\DeclarePairedDelimiter{\norm}{\lVert}{\rVert}
\DeclarePairedDelimiter{\abs}{\lvert}{\rvert}

\newcommand{\normp}[2][]{\norm*{#2}_{#1}}

\newcommand{\bhat}[1]{\hat{\bm{#1}}}  

\newcommand{\epsdelta}{$(\rate, \local)$}
\newcommand{\edstationary}{\epsdelta-stationary }
\newcommand{\llasso}{local lasso}

\newcommand{\fractn}{ \left( \frac{t}{n} \right) }
\newcommand{\Deltatn}{\left( \left\vert \frac{t}{n}-u \right\vert + \frac{1}{n} \right)}

\newcommand{\localComp}{\phi_\local(u)}
\newcommand{\vanishing}{\vartheta_n}
\newcommand{\lambdau}{\lambda(u)}

\newcommand{\rate}{\epsilon_n}
\newcommand{\rateErr}{\zeta_n} 
 
\newcommand{\local}{\delta}

\newcommand{\localSet}{S_\local(u)}
\newcommand{\s}{s_\local(u)}
\newcommand{\sbar}{\bar s_\local(u)}
\newcommand{\smax}{\check s_\local(u)}
\newcommand{\sj}{s_{j,\local}(u)}

\newcommand{\Fset}{\mathcal{F}_\lambda(u)}
\newcommand{\Fsetj}{\mathcal{F}_{j,\lambda_j}(u)}

\newcommand{\Sset}{\mathcal{S}_\delta(u)}
\newcommand{\Ssetbar}{\bar{\mathcal{S}}_\delta(u)}

\newcommand{\Asetj}{\mathcal A_{j,\local}(u)}

\newcommand{\x}[1][\relax]{X\tn\ifx\relax#1\relax\else^{(#1)}\fi}
\newcommand{\X}[1][\relax]{\bm{X}\tn\ifx\relax#1\relax\else^{(#1)}\fi}
\newcommand{\xu}[1][\relax]{ \tilde{X}\tn\ifx\relax#1\relax\else^{(#1)}\fi(u)}
\newcommand{\Xu}[1][\relax]{\tilde{\bm{X}}\tn\ifx\relax#1\relax\else^{(#1)}\fi(u)}
\newcommand{\epsu}{\tilde\varepsilon\tn(u)}

\newcommand{\Sig}{\bm{\Sigma} }
\newcommand{\Beta}{\bm{\beta}}
\newcommand{\gammaZero}[1][j]{\bm{\gamma}^{0}_{#1}}

\newcommand{\ndeps}[1][j]{v_{#1 t,n}}
\newcommand{\ndepsu}[1][j]{\tilde{v}_{#1 t,n}(u)}

\newcommand{\kernel}{K\left(\frac{t/n-u}{h}\right)}
\newcommand{\kernelmat}{ \bm{\mathcal{K}}(u) }

\newcommand{\sumt}{\sum_{t=1}^n}
\newcommand{\meant}{\frac{1}{nh} \sumt}

\newcommand{\tn}{_{t,n}}

\newcommand{\order}{m}
\newcommand{\lagorder}{d_n}
\newcommand{\nedsize}{\varphi}
\newcommand{\exodim}{q_n}
\newcommand{\exocoeff}{\eta}
\newcommand{\localEig}{\Lambda_\local(u)}
\newcommand{\effEig}{A_n(u)}

\newcommand{\bgamma}{\bm{\gamma}}
\newcommand{\mj}{^{-j}}
\newcommand{\mjtop}{^{-j\top}}
\newcommand{\pj}{^{(j)}}

\newcommand{\Betazero}[1][]{\Beta^{0}_{#1}}
\newcommand{\BetazeroU}[1][]{\Beta^{0}_{#1}(u)}

\newcommand{\BetahatU}[1][]{\hat{\Beta}_{#1}(u)}

\title{\bf Local Inference for \\
High-dimensional Time Series}

\author[1,2]{Julien Bodelet}
\author[3]{Davide La Vecchia}
\author[4]{Johannes Lederer}
\author[1,5]{Oliver Y. Chén} 

\affil[1]{Centre for Neural AI, Lausanne University Hospital, Lausanne, Switzerland}
\affil[2]{Department of Statistics, Stanford University, Stanford, CA, United States}
\affil[3]{Geneva School of Economics and Management, Geneva, Switzerland}
\affil[4]{Department of Mathematics, University of Hamburg, Hamburg, Germany}
\affil[5]{Faculty of Biology and Medicine, University of Lausanne, Lausanne, Switzerland}

\date{}

\begin{document}

\def\spacingset#1{\renewcommand{\baselinestretch}%
{#1}\small\normalsize} \spacingset{1}

\if1\anon
{
  \title{\bf Local Inference for \\
High-dimensional Time Series}
  \maketitle
} \fi

\if0\anon
{
  \bigskip
  \bigskip
  \bigskip
  \begin{center}
    {\LARGE\bf Local Inference for \\
High-dimensional Time Series}
\end{center}
  \medskip
} \fi

\bigskip

\begin{abstract}
Local stationarity provides a natural framework for time series whose dependence structure evolves smoothly over time. In high-dimensional settings, however, its classical definition can break down, as approximation errors accumulate across time-varying coefficients. 
To address this limitation, we introduce $(\epsilon_n,\delta)$-stationarity, a generalization of local stationarity that holds locally in time and allows approximation errors to grow with the dimension, and we combine it with local sparsity, which lets the set of relevant predictors change over time.
We show that, for high-dimensional time-varying autoregressions, the rate of stationary approximation is driven by local sparsity, so that local sparsity governs not only estimation but also the stationary approximation itself. 
We then develop a \textit{local lasso} with time-varying penalties that selects relevant predictors and their periods of activity, and we derive error bounds that depend on the local sparsity, quantify the cost of nonstationarity, and guide penalty and bandwidth selection. 
For inference, we propose a desparsified local lasso based on local nodewise regressions and establish a linear representation with an explicit smoothing bias.
Asymptotic normality then follows under two bandwidth conditions, one for coefficient variation and one for nonstationarity.
Simulations and applications to macroeconomic and neuroscience data illustrate the methods.
\end{abstract}

\noindent%
\vfill

\newpage
\spacingset{1.8} 

\begin{bibunit}

\section{Introduction}

Recent years have witnessed increasingly large time series datasets in many research areas. Alongside technological developments, both the number of observed series and the time horizons 
have been expanding. As time series observed over long horizons rarely keep the same dynamics,
stationarity, on which classical time series inference relies, is becoming progressively hard to justify, and so is the use of high-dimensional tools developed for studying stationary processes.

Only the observations close to a given time carry information about the behavior at that time,
so that the number of variables can be large relative to the information available locally, even when it is small relative to the sample size.
Consequently, long nonstationary time series are locally high-dimensional.
Given the widespread use of time series across research areas, this problem is of central importance to several disciplines.

For example, macroeconomic panels contain more than a hundred monthly series observed over decades, during which the relationships between indicators evolve with policy regimes and structural changes \citep{stock1999forecasting,StockWatson2007}. Multichannel electroencephalography (EEG) records dozens of signals over hours, and the dependence between channels varies across sleep stages \citep{achermann1998coherence} or cognitive states \citep{Ombao2005}. In both cases, the questions of interest are often inferential, such as identifying the predictors of inflation that matter in a given economic regime or the brain connections that are active during a given sleep stage.

Inference on a few parameters in the presence of many others is well developed for independent data \citep{belloni2014inference,zhang2017simultaneous,belloni2018uniformly} and stationary time series \citep{chernozhukov2021lasso,adamek2023lasso}, but has received far less attention for nonstationary time series. Answering such questions requires methods that jointly handle nonstationarity and high dimensionality while supporting valid inference.

The theory of locally stationary processes, introduced by \citet{Dahlhaus1996,Dahlhaus1997}, provides a powerful framework for modeling time series that behave approximately stationary over short periods but whose dynamics change smoothly over time. This framework has been developed for a wide range of models, including time-varying ARCH processes \citep{Dahlhaus2006}, nonparametric regression \citep{Vogt2012}, diffusion processes \citep{koo2012estimation,Deleamont2019}, and nonlinear models \citep{dahlhaus2019towards}.
Existing results, however, mostly concern models of fixed dimension, and do not cover inference on individual time-varying coefficients when the number of coefficients grows with the sample size.

This restriction matters because the quality of the classical stationary approximation can deteriorate with dimension, even in simple models. To see this, consider the Gaussian time-varying model
\begin{equation}\label{eq. example}
Y_{t,n} = \sum_{j=1}^{p_n} \beta_{j}^0\!\left(\frac{t}{n}\right) X_{t}^{(j)} + \varepsilon_{t},
\qquad t=1,\dots,n,
\end{equation}
where $p_n$ may grow with $n$.
A classical definition of local stationarity requires that, for each $u \in [0,1]$, there exists a stationary process $\tilde Y_t(u)$ such that $\E\bigl|Y_{t,n}-\tilde Y_t(u)\bigr| \leq C(1/n+|t/n - u|)$, with $C$ independent of $t$, $n$, and $u$.  The natural candidate for \eqref{eq. example} is $\tilde Y_t(u)=\sum_{j=1}^{p_n}\beta_j^0(u)X_t^{(j)}+\varepsilon_{t}$. Suppose that the covariates are stationary and Gaussian with mean zero and covariance matrix $\bm\Sigma_n$, whose smallest eigenvalue $\lambda_{\min}(\bm\Sigma_n)$ is bounded away from zero. The difference $Y_{t,n}-\tilde Y_t(u)=\sum_{j=1}^{p_n}\{\beta_j^0(t/n)-\beta_j^0(u)\}X_t^{(j)}$ is then a centered Gaussian variable, and
\begin{equation}\label{eq. non local stationarity example}
\E\bigl|Y_{t,n}-\tilde Y_t(u)\bigr|
\geq \sqrt{\frac{2}{\pi}\,\lambda_{\min}(\bm\Sigma_n)}\,
\Bigl\{\sum_{j=1}^{p_n}\Bigl(\beta_j^0\Bigl(\frac{t}{n}\Bigr)-\beta_j^0(u)\Bigr)^2\Bigr\}^{1/2}.
\end{equation}
Suppose further that, for some $\kappa, c>0$, at least $\kappa p_n$ coefficients vary at least linearly in time, that is, $\abs{\beta_j^0(u)-\beta_j^0(v)}\geq c\abs{u-v}$. The right-hand side of \eqref{eq. non local stationarity example} is then of order $\sqrt{p_n}\,\abs{t/n-u}$. When $p_n\to\infty$, the process therefore violates the bound $C(\abs{t/n-u}+1/n)$ imposed by classical local stationarity, with $C$ independent of $n$, even when every coefficient is Lipschitz with a common constant. This deterioration becomes an issue when lags of the response enter as covariates, as in time-varying autoregressions, because the design then inherits it.

A natural remedy is sparsity, a central assumption in high-dimensional statistics \citep{Tibshirani1996,buhlmann2011}. In the example above, only nonzero coefficients contribute to the lower bound in \eqref{eq. non local stationarity example}, so sparsity mitigates this deterioration by replacing the dimension with the number of active coefficients. In time-varying models, however, requiring the same few coefficients to remain active over the entire sample can be restrictive, as it fails to accommodate the time-varying nature of the process under consideration. For example, active brain connections may change across sleep stages. We therefore use local sparsity \citep{Wang2015,Kong2015}, which relaxes global sparsity by allowing the set of active predictors to change over time. Near a given time point, the approximation error then depends only on the locally active coefficients, which can be far fewer than those active over the entire period.  However, this reduction holds only within a neighborhood of the time point of interest, and the local sparsity level may still grow with the sample size. The lower bound in \eqref{eq. non local stationarity example} may therefore still diverge, and classical local stationarity fails even under local sparsity.

This paper addresses these issues in three steps. First, we connect high-dimensional statistics and the theory of locally stationary processes through a theoretical framework for locally high-dimensional time series. We introduce the notion of \epsdelta-stationarity, under which a process is approximated by a stationary process only within a $\local$-neighborhood of each time point, at a rate $\rate$ that may diverge with the sample size. The resulting class includes stationary and classical locally stationary processes as special cases. As a leading example, we show that high-dimensional time-varying autoregressive processes with exogenous covariates (tv-ARX) are \epsdelta-stationary under mild conditions, at a rate driven by the local sparsity levels and only mildly by the number of lags.
Second, building on this framework, we propose the \textit{\llasso}, a kernel-weighted lasso with a time-varying penalty that adapts to the local structure of the model.
Applied across time points, it performs variable selection and  indicates the periods during which each coefficient vanishes. Third, for inference, we construct a desparsified version of the \llasso\ using local nodewise regressions \citep{geer2014asymptotically} and estimate its asymptotic variance through a local estimator of the long-run variance.
To our knowledge, this is the first high-dimensional inference procedure for time-varying models in which the covariates themselves are allowed to be time series, as in time-varying autoregressions.

On the theoretical side, we bound the estimation and prediction errors of the \llasso\ in terms of the local sparsity level, that is, the number of coefficients active near the time point of interest, which can be much smaller than the global sparsity level when the active set changes over time. The bounds separate the error due to noise from the smoothing bias, which depends on the local sparsity level and the largest eigenvalue of the local covariance matrix of the active covariates. This dependence provides a theoretical rationale for allowing the penalty to vary over time. The bounds also show how the approximation rate enters the choice of both the penalty and the bandwidth, thereby quantifying the cost of nonstationarity.
Beyond \epsdelta-stationarity, our estimation and inference results impose no condition on the observed process, as moments and near-epoch dependence are assumed only for its stationary approximations.

For inference, the nonstationarity of the process raises two difficulties that are absent in the stationary setting. First, desparsification does not eliminate the smoothing bias. Second, the nodewise regression errors inherit the nonstationarity of the design. We show that these errors are \epsdelta-stationary and establish a linear representation of the desparsified estimator with an explicit bias term. Asymptotic normality then follows under distinct undersmoothing conditions for coefficient variation and for nonstationarity.

\subsection{Related Literature}
\label{sec. related literature}

Sparsity patterns that change over time have been modelled by bounding the number of active coefficients at each time point \citep{zhou2010time,kolar2010estimating,Chen2017}.
Local sparsity, under which coefficient functions vanish on subregions of their domain, was introduced by \citet{tu2012} and studied further by \citet{Wang2015}.
It has been used to identify the null subregions of coefficient functions in functional linear and generalized linear models \citep{Lin2015,sun2022adaptive} and in varying-coefficient models \citep{Kong2015}.
These works consider a fixed number of functions and use local sparsity only for domain selection, whereas in our framework it is a structural assumption that controls the approximation rate of the process and the rates of our estimators.

Kernel-weighted lasso estimators have been proposed for time-varying networks from independent observations \citep{zhou2010time,kolar2010estimating} and for time-varying vector autoregressions \citep{ding2017sparse}.
\citet{Kong2015} combine local linear smoothing with a group penalty in fixed dimension with independent observations, \citet{Lee2016} combine kernel smoothing with a group lasso penalty in high-dimensional varying-coefficient models under global sparsity, and \citet{yousuf2021boosting} use boosting for high-dimensional time-varying regressions.
None of these works considers inference, and \citet{ding2017sparse} and \citet{yousuf2021boosting} assume classical local stationarity, which can fail in high dimensions, as the example in \eqref{eq. non local stationarity example} shows.
Our analysis accounts for this deterioration, yields rates driven by the local sparsity level, and provides a rationale for a time-varying penalty.

The desparsified lasso has been extended to time series with time-invariant coefficients \citep{chernozhukov2021lasso,adamek2023lasso}, and inference based on kernel-weighted regularized estimators has been developed for independent observations in time-varying graphical models \citep{lu2018post} and sparse additive models \citep{lu2020kernel}.
Like \citet{adamek2023lasso}, we use near-epoch dependence, but we impose it on the stationary approximations only, not on the observed process.
\citet{Chen2017} propose pointwise tests with conservative familywise error control for high-dimensional time-varying linear models with a deterministic design, under independent errors or Gaussian errors with known serial covariance.
Our procedure allows nonstationary covariates, including lags of the response, and yields asymptotically normal estimators of individual coefficients under possibly non-Gaussian errors with unknown serial dependence.

\subsection{Organization of the Article}

The rest of the article is organized as follows.
Section~\ref{sec. local stationarity and sparsity} introduces \epsdelta-stationarity, formalizes $\local$-local sparsity, and shows that high-dimensional tv-ARX processes are \epsdelta-stationary under local sparsity.
Section~\ref{sec. penalized kernel estimation} presents the \llasso\ and its desparsified version.
Section~\ref{sec. theory} derives error bounds for the \llasso\ and establishes the asymptotic normality of the desparsified \llasso.
Section~\ref{sec. simulations} evaluates the finite-sample performance of the proposed methods, and Section~\ref{sec. real data} applies them to macroeconomic and EEG data.
Section~\ref{sec. discussion} concludes.
All proofs are given in the Supplementary Materials.

\subsection{Notation}

For a vector $\bm{a} \in \R^d$ and $q \ge 1$,
$\Vert \bm{a} \Vert_q:= (\sum_{j=1}^d |a_j|^q)^{1/q}$ denotes the $\ell_q$-norm and
$\Vert \bm{a}\Vert_\infty = \max_{1\leq j \leq d} \vert a_j \vert$ denotes the maximum norm.
For a matrix $\bm{A}$, we let $\Vert \bm{A} \Vert_{\text{max}}= \max_{i,j} \abs{A_{i,j}} $ be its maximum norm.
A superscript $\top$ denotes the transpose of a vector or matrix.
For two sequences $a_n$ and $b_n$, we write $a_n \asymp b_n$ if the ratio $a_n/b_n$ is bounded away from $0$ and infinity.
For two real numbers $a$ and $b$, we let $a \wedge b := \min(a,b)$, $a \vee b := \max(a, b)$, and $\lfloor a \rfloor$ denote the largest integer not exceeding $a$.
For a set $\mathcal A$, $|\mathcal A|$ denotes its cardinality and $\mathcal{P}(\mathcal A)$ its power set.
Throughout the article, $C$ denotes a generic positive constant, whose value may change from line to line.

\section{High-dimensional Time-varying Time Series}
\label{sec. local stationarity and sparsity}

Consider the following time-varying model 
\begin{equation}\label{eq. model}
Y\tn = \sum_{j=1}^{p_n}  \beta_{j}^0\fractn X_{t,n}^{(j)} + \varepsilon\tn \qquad \text{for } t=1, \dots, n,
\end{equation}
where $\x\pj$ and $Y\tn$ are triangular arrays, indexed by both the time $t$ and sample size $n$. 
The model is general and contains the time-varying autoregressive model as a special case, as we allow the covariates $\x\pj$ to contain lags of $Y_{t,n}$.
As is standard in the literature on locally stationary processes, the parameters $\beta_j:[0,1]\rightarrow \R$ do not depend on the time $t$ but on the rescaled time $t/n$ and are allowed to change smoothly over time.
However, unlike the literature on locally stationary processes, we allow the number $p_n$ of coefficients $\beta_j(\cdot)$, to be larger than the effective sample size and diverge as $n\rightarrow \infty$.

\subsection{Local Stationarity}

A locally stationary process is a triangular array $\{Z_{t,n}: t= 1, \dots, n\}_{n=1}^\infty$ that can be locally approximated, in a stochastic sense, by a family of stationary processes $\{\tilde Z_t(u):t \in \Z \}$, for $u \in [0,1]$ (see \citealp{dahlhaus2012} for a review).
Despite a growing interest in locally stationary processes, their properties in the high-dimensional setting remain largely unexplored, specifically when the dimension $p_n$ diverges to infinity as $n\rightarrow \infty$. 
We seek a notion of local stationarity that strikes a balance between applicability and analytical strength. 
It must be weak enough to be verified in commonly used high-dimensional models, yet strong enough to support meaningful asymptotic guarantees for the proposed methods.

\begin{definition}[$(\rate, \local)$-stationarity] \label{def. local stationarity}
Let $\rate$ and $d_n$ be nonnegative sequences and $\local \in (0,1]$.
The triangular array of processes $\{ \bm{Z}\tn: t=1,...,n\}_{n=1}^\infty$ is an 
$(\epsilon_n,\local)$-stationary process with respect to the norm $\|\cdot\|$ if, for each rescaled time point $u \in [0,1]$ and $n \in \N$, there exists an associated process 
$\{\tilde{\bm{Z}}\tn(u): t\in \mathbb{Z}\}$, such that, 
\begin{itemize}
    \item[(i)] the process $\{ \tilde{\bm Z}\tn(u) :t\in \mathbb{Z}\}$ is strictly stationary;
    \item [(ii)] for all $t$ with $|t/n -u|\leq \local $ it holds that
\begin{equation*}
    \left(\E\left[\Vert \bm{Z}\tn - \tilde{\bm Z}\tn(u) \Vert^\rho\right]\right)^{1/\rho} \leq C \rate \left( \left| \frac{t}{n}-u \right|+ \frac{\lagorder}{n}\right),
\end{equation*}
\end{itemize}
where $\rate \lagorder/n \rightarrow 0$, $\rho \ge 1$, and
$C<\infty$ independent of $t, n$, and $u$.
\end{definition}

The above definition generalizes the standard definition of local stationarity in three ways.
First, the local approximations $\tilde{\bm Z}\tn(u)$ are allowed to depend on $n$.
Second, for standard locally stationary processes, the inequality is usually supposed to hold uniformly for all $t$; in the above definition the inequality holds only in a $\local$-neighborhood of $u$.
Third, we introduce two sequences $\rate$ and 
$\lagorder$ that are allowed to diverge with $n$.
Furthermore, it is important to note that, although the stationary approximation $\tilde{\bm Z}\tn(u)$ is indexed by $n$, it is not a triangular array, but a family of time series indexed by $t\in \Z$.
Definition~\ref{def. local stationarity} encompasses a wide range of high-dimensional time-varying models. It includes stationarity ($\rate = 0$) and the classical local stationarity setting ($\epsilon_n=\epsilon$ constant, $\lagorder =1$, and $\delta =1$) as special cases.

\subsection{Local Sparsity}
\label{sec. local sparsity}

Sparsity is central to high-dimensional estimation.
The usual notion is global sparsity, which counts the coefficients that are nonzero at some point in time.
This notion is restrictive when the set of relevant covariates changes over time, since a coefficient that is active on a short period only then counts as active everywhere.
We therefore use a local notion, which we define as follows.

\begin{definition}[$\local$-Local Sparsity] \label{def. local sparsity}
Let $\delta >0$.
A continuous function $\beta:[0,1]\rightarrow \R$ is $\local$-locally sparse at 
$u$, if $\beta(v)=0$ for all $v \in [0,1]$ with $ |v- u| \leq \local$.
\end{definition}

Unlike the condition $\beta(u)=0$, this definition distinguishes a function that vanishes on an interval from a function that only crosses zero at $u$.
For each $\local>0$ and $u\in[0,1]$, we define the local active set and the local sparsity level
\begin{equation}\label{eq. active set}
\localSet:=\big\{j\in\{1,\dots,p\}:\ \beta^0_j \text{ is not } \local\text{-locally sparse at } u\big\},
\qquad
\s:=|\localSet| .
\end{equation}

The set $\localSet$ grows with $\local$, i.e. 
$S_\delta(u) \subseteq S_{\delta'}(u) \subseteq S_1(u) \text{ for } 0< \delta \leq \delta'$, so that $\s$ is non-decreasing in $\local$.
For $\local\ge1$, it contains every coefficient that is nonzero at some point in time and no longer depends on $u$.
The set $S_1:=S_1(u)$ is then the usual active set, and $s_1:=|S_1|$ is the global sparsity level.
Hence $\s\le s_1$ for all $\local$ and $u$, with equality when the coefficients are constant over time.

\subsection{High-dimensional Time-varying Autoregressive Processes}
\label{sec. tv arx}

We now specialize our general framework by considering a high-dimensional tv-ARX.
The tv-ARX is an important model as it encompasses both time-varying autoregressive and time-varying regression models and has proved useful in time series analysis (see, for example, \citealp{robinson1991,Dahlhaus1997,zhang2012}).
We consider a tv-ARX defined as follows:
\begin{equation}\label{eq. tv ARX}
Y\tn = \sum_{j=1}^{\lagorder} \alpha_{j}^0\fractn Y_{t-j,n} + \sum_{k=1}^{\exodim}  \exocoeff_{k}^0\fractn Z_{t,n}^{(k)} + \varepsilon\tn,
\end{equation}
where $p_n = \exodim + \lagorder$ and
$\bm Z_{t,n}:=\bigl(Z_{t,n}^{(1)},\dots,Z_{t,n}^{(\exodim)}\bigr)^\top$
is an exogenous process.
The full regressor vector is therefore
$\bm X_{t,n}
:=\bigl(Y_{t-1,n},\dots,Y_{t-\lagorder,n},
Z_{t,n}^{(1)},\dots,Z_{t,n}^{(\exodim)}\bigr)^\top
\in\mathbb R^{p_n}$.
Here we consider the high-dimensional model where both $\exodim$ and $\lagorder$ are allowed to diverge to infinity.
We denote the local active sets $S_{\local}^\alpha(u) :=\left\{j\in \{1,...,\lagorder\}: \alpha_j^0 \text{ is not $\local$-locally sparse at } u \right\}$ and 
$S_{\local}^\exocoeff(u) :=\left\{k\in \{1,...,\exodim\}: \exocoeff_k^0 \text{ is not $\local$-locally sparse at } u \right\}$.
To study the properties of the process, we assume that $\bm Z\tn$ and $\varepsilon\tn$ are defined for all $t\in\Z$, and we extend $Y\tn$ to $t\le0$ by setting $\alpha^0_j(u)=\alpha^0_j(0)$ and $\exocoeff^0_k(u)=\exocoeff^0_k(0)$ for $u\le0$, as is standard in the local stationarity literature \citep[e.g.,][]{Dahlhaus2006}.

\begin{condition}[Lipschitz condition]\label{cond. tvarx Lipschitz}
There exist constants $L_\alpha$, $L_\exocoeff$, and $C_{\exocoeff}<\infty$, such that
$ \vert \alpha^0_j(u) - \alpha^0_j(v)\vert \leq L_\alpha |u-v|$, $ \vert \exocoeff^0_k(u) - \exocoeff^0_k(v)\vert \leq L_\exocoeff |u-v|,$ $|\exocoeff_{k}^0(u)|\le C_{\exocoeff} $
for all $u,v\in[0,1]$, $j\le \lagorder$, $k\le \exodim$, $n\ge 1$.
\end{condition}

\begin{condition}[\epsdelta-stationarity]\label{cond. tvarx local stationarity}
Let  $0<\local<\local'$ and $\rho\geq 1$. For all $u \in [0,1]$, there exist jointly strictly stationary processes 
$\tilde{\bm Z}_{t,n}(u)$ and $\tilde\varepsilon_t(u)$ satisfying, for all $|t/n-u|\leq \local'$,
\begin{enumerate}[label=(\roman*)]
    \item $\displaystyle \E[ \Vert \bm  Z_{t,n}-\tilde{\bm  Z}\tn(u) \Vert_\infty^\rho]^{1/\rho} 
\leq C \rate^z \left( \left| \frac{t}{n}-u \right|+ \frac{\lagorder^z}{n}\right) $,
    \item $ \displaystyle \E[|\varepsilon_{t,n}-\tilde\varepsilon_t(u)|^\rho]^{1/\rho} \le C \rateErr\Bigl(\Bigl|\frac{t}{n}-u\Bigr|+\frac{1}{n}\Bigr)$,
\end{enumerate}
for some constant $C$, where $\rate^z\lagorder^z/n \rightarrow 0$ and $\zeta_n / n\rightarrow 0$.
\end{condition}

\begin{condition}[\emph{$\ell^1$-stability of the AR part}]
\label{cond. l1 stability}
There exists $\iota\in(0,1)$ such that
$\sum_{j=1}^{\lagorder}|\alpha_{j}^0(u)|\le 1-\iota
$ for all $ n\ge 1$ and $\ u\in[0,1]$.
\end{condition}

\begin{condition}[\emph{Uniform moment bounds}]
\label{cond. uniform bounds}
The moments of the innovations and of the exogenous covariates are bounded uniformly in $n$, that is,
$M_{\varepsilon}:=\sup_{t\in\Z}\big(\E|\varepsilon\tn|^\rho\big)^{1/\rho}\vee\sup_{u\in[0,1]}\big(\E|\tilde\varepsilon_t(u)|^\rho\big)^{1/\rho}$
and
$M_{z}:=\max_{k\le\exodim}\Big\{\sup_{t\in\Z}\big(\E|Z^{(k)}\tn|^\rho\big)^{1/\rho}\vee\sup_{u\in[0,1]}\big(\E|\tilde Z^{(k)}\tn(u)|^\rho\big)^{1/\rho}\Big\}$
are finite and do not depend on $n$.
\end{condition}

\begin{condition}[\emph{Geometric companion decay}]
\label{cond. geometric decay}
For all $u\in [0,1]$ and $n\geq 1$, there exists a sequence $\{\psi_{\ell,n}(u), \ell \geq 0\}$ satisfying
$\frac{1}{1-\sum_{j=1}^{\lagorder}\alpha_{j}^0(u)\,z^j}
=\sum_{\ell=0}^\infty\psi_{\ell,n}(u)\,z^\ell$
for $\abs{z}\leq 1$,
such that $|\psi_{\ell,n}(u)|\le C_\psi\,\kappa^\ell$, for all $\ell\ge 0$,
where $C_\psi<\infty$ and $\kappa\in(0,1)$ are uniform in $n$ and $u$.
\end{condition}

Condition \ref{cond. tvarx Lipschitz} is a standard Lipschitz condition with a bound on the exogenous coefficients.
Condition~\ref{cond. tvarx local stationarity} imposes
$(\zeta_n,\local')$-stationarity on the innovations and
$(\rate^z,\local')$-stationarity of the exogenous covariates with respect to the maximum norm.
A slightly larger radius $\local'$ is needed because $Y\tn$ depends on the past, and as this dependence decays geometrically, the margin $\local'-\local$ may vanish with $n$.
Condition \ref{cond. l1 stability} provides a uniform stability margin for the autoregressive component. 
Condition \ref{cond. uniform bounds} imposes uniform $L^\rho$-moment bounds on the innovations, the exogenous regressors, and their stationary approximations.
Condition \ref{cond. geometric decay} requires the impulse-response coefficients of the stationary approximation to decay geometrically, uniformly in $n$ and $u$,
and guarantees the existence of the stationary approximation
$\tilde Y\tn(u)$.

\begin{theorem}[$(\rate, \local)$-stationarity of the tv-ARX process]\label{th. local stationary}
Let $\{Y\tn\}$ satisfy Model \eqref{eq. tv ARX} and suppose Conditions
\ref{cond. tvarx Lipschitz}--\ref{cond. geometric decay} hold.
Let $\bar s_\alpha:=\sup_u |S_{\delta'}^\alpha(u)| \vee 1$ and
$\bar s_\exocoeff:=\sup_u |S^\exocoeff_{\delta'}(u)| \vee 1$.
Assume that $\lagorder^z \leq C \lagorder$, $\lagorder^{1+1/\rho} \left(\bar s_\alpha  \bar s_\exocoeff\, + \bar s_\exocoeff\,\epsilon_n^z  +\rateErr \right)/n \rightarrow 0$, and
$
n(\local'-\local) \ge 2\lagorder + (1-\kappa)^{-1}\,\log\Bigl(\frac{n}{(1-\kappa)\lagorder}\Bigr).
$
Let $\tilde Y\tn(u)$ be the stationary
solution of $ \displaystyle \tilde Y\tn(u)
=\sum_{j=1}^{\lagorder}\alpha_{j}^0(u)\,\tilde Y_{t-j,n}(u)
+\sum_{k=1}^{\exodim}\exocoeff_{k}^0(u)\,\tilde Z_{t,n}^{(k)}(u)
+\tilde\varepsilon_t(u)$
for each $u\in[0,1]$, and define
$\tilde{\bm X}\tn(u)
:=\bigl(\tilde Y_{t-1,n}(u),\dots,\tilde Y_{t-\lagorder,n}(u),
\tilde Z_{t,n}^{(1)}(u),\dots,\tilde Z_{t,n}^{(\exodim)}(u)
\bigr)^\top.$
Then for all $t$ with $|t/n-u|\le\delta$, we have
\[
\Bigl(\E\|\bm X_{t,n}-\tilde{\bm X}\tn(u)\|_\infty^\rho
\Bigr)^{1/\rho}
\;\le\;
C\,\epsilon_n
\Bigl(\Bigl|\frac{t}{n}-u\Bigr|+\frac{\lagorder}{n}\Bigr),
\]
where
$
\epsilon_n \;:=\; \lagorder^{1/\rho} \left( 
\bar s_\alpha  \bar s_\exocoeff\, 
+ \bar s_\exocoeff\,\epsilon_n^z
+\rateErr
\right)
$ and $C <\infty$ is independent of $n$, $t$ and $u$.
\end{theorem}

Theorem~\ref{th. local stationary} establishes \epsdelta-stationarity of the whole regressor vector $\bm X\tn$ with respect to the maximum norm, when $\lagorder$, $\lagorder^z$ and $\exodim$ diverge with $n$.
Applied with $\rho=2\bar\order$, which requires Conditions~\ref{cond. tvarx local stationarity} and~\ref{cond. uniform bounds} with $2\bar\order$ moments, Theorem~\ref{th. local stationary} verifies Assumption~\ref{as. local stationarity} of Section~\ref{sec. error bounds} for the tv-ARX model, and the number of lags then enters $\epsilon_n$ only through the factor $\lagorder^{1/(2\bar\order)}$.
In the literature, local stationarity is usually established for the outcome $Y\tn$ alone \citep{Dahlhaus1996,Dahlhaus1997}.
For fixed $\lagorder$ and $\exodim$, this is equivalent to local stationarity of $\bm X\tn$, as the factors depending on the dimensions are absorbed in the constant $C$, but this equivalence no longer holds when $\lagorder$ and $p_n$ grow with $n$.

\section{Penalized Local Estimation}
\label{sec. penalized kernel estimation}

\subsection{The Local Lasso}

Locally stationary processes are usually estimated by applying a stationary method, with kernel weights, to the observations in a segment $\{t:|t/n-u|\le h / 2\}$ on which the process is almost stationary.
The effective sample size is then of order $nh$, so local least squares is not uniquely defined when $p_n$ exceeds the number of observations in the window, and its $\ell_2$ error does not vanish unless $p_n/(nh) \to 0$.
We therefore define the \llasso\ estimator of $\Beta^0(u)\in\R^{p_n}$ in Model~\eqref{eq. model} as
\begin{equation}\label{eq. min problem}
\hat \Beta(u) = \argmin_{\Beta \in \R^{p_n}} \meant \kernel (Y\tn-\X^\top \Beta )^2 + 2\lambda(u) \Vert \Beta \Vert_1 ,
\end{equation}
where $\X:= (\x^{(1)}, \dots, \x^{(p_n)})^\top$, $K$ is a kernel, and $h$ is the bandwidth.
Here, unlike the fixed penalty usually used (e.g. in \citealp{Kong2015}), we let the penalty $\lambda: [0,1] \rightarrow \R$ vary with rescaled time.
A time-varying penalty is natural in our locally sparse framework, since the penalty should depend on the local sparsity level $s_\delta(u)$, which varies over time.

\subsection{The Desparsified Local Lasso}

The \llasso\ is not directly suited to inference, since its shrinkage bias is not negligible at the $\sqrt{nh}$ scale and its limiting distribution depends on the unknown sparsity pattern \citep{knight2000asymptotics}.
We therefore adapt the desparsified lasso \citep{zhang2014confidence,geer2014asymptotically,javanmard2014confidence}, extended to stationary time series by \citet{chernozhukov2021lasso} and \citet{adamek2023lasso}, to the locally stationary setting.

\textbf{The bias of the local lasso}. Let $\kernelmat$ denote the $n \times n$ diagonal matrix with diagonal elements $K((t/n - u)/h)$, $t = 1, \dots, n$, let $\bm X := (\bm X_{1,n}, \dots, \bm X_{n,n})^\top$ be the $n \times p_n$ matrix of covariates, and let $\bm Y := (Y_{1,n}, \dots, Y_{n,n})^\top$ and $\bm \varepsilon := (\varepsilon_{1,n}, \dots, \varepsilon_{n,n})^\top$. The KKT conditions of \eqref{eq. min problem} read
$ \bm X^\top \kernelmat (\bm Y - \bm X \hat\Beta(u)) / (nh) = \lambda(u) \hat{\bm\kappa}(u),$
where $\hat{\bm\kappa}(u)$ is a subgradient of the $\ell_1$-norm at $\hat\Beta(u)$. Substituting model \eqref{eq. model} for $\bm Y$, we obtain
\begin{equation}\label{eq. KKT conditions with Sigma}
\bhat \Sigma(u) (\hat\Beta(u) - \Beta^0(u)) + \lambda(u) \hat{\bm\kappa}(u) = \bm X^\top \kernelmat \bm \varepsilon / (nh) + \bm \omega(u),
\end{equation}
where $\bhat \Sigma(u) := \bm X^\top \kernelmat \bm X / (nh)$ is the local Gram matrix and
\begin{equation}
\bm\omega(u) := \meant \kernel \X \left(\X^\top \Beta^0(t/n) - \X^\top \Beta^0(u)\right)
\end{equation}
is the smoothing bias induced by nonstationarity. It vanishes when $\Beta^0$ is constant on the support of the kernel.

\textbf{An approximate inverse by local nodewise regression.}
When $p_n$ exceeds the number of observations in the window, $\bhat\Sigma(u)$ is not invertible.
The idea behind the de-biased estimator is to use an approximate inverse.
For this aim, we introduce a local version of the approach of \citet{meinshausen2006high} and \citet{geer2014asymptotically}. 
Specifically, for $j \in \{1,\dots,p_n\}$, we consider the \llasso\ 
$$
\hat \bgamma_j(u) := \argmin_{\bgamma \in \R^{p_n-1}} 
\meant \kernel
\left( \x\pj - \X\mjtop \bgamma \right)^2 + 2\lambda_j(u) \Vert \bm \bgamma \Vert_1,
$$
where $\X\mj$ is $\X$ with the $j$th {element} removed and $\bhat\gamma_j 
= \{\hat\gamma_{j,k}(u); k=1,\dots, p_n, k\neq j\}$.
Subsequently, we define the approximate inverse of $\bhat \Sigma(u)$ by
the matrix $\bhat \Theta(u)$ with rows
$
\bhat \Theta_j := \left(-\hat\gamma_{j,1}(u), \dots, -\hat\gamma_{j,j-1}(u), 1, -\hat\gamma_{j,j+1}(u), \cdots, -\hat\gamma_{j,p_n}(u) \right) / \hat\tau_j^2(u)
$
where 
$$
\hat \tau_j^2(u):= 
\meant \kernel
\left( \x\pj - \X\mjtop \bhat\gamma_j \right)^2
+ \lambda_j(u) \Vert \hat \bgamma_j(u) \Vert_1.
$$
In practice, only the rows of $\bhat\Theta(u)$ corresponding to the coefficients of interest need to be computed.
By the KKT conditions of the nodewise regressions, $\bhat\Theta(u)$ is an approximate inverse of $\hat{\bm\Sigma}(u)$, in the sense that
$\|\hat{\bm\Sigma}(u)\bhat\Theta_j(u)^\top-\bm e_j\|_\infty\le\lambda_j(u)/\hat\tau_j^2(u)$.

\textbf{The desparsified estimator.}
{We define the desparsified \llasso\ estimator}
\begin{equation}\label{eq. desparsified lasso}
\bhat b(u) {:=} \hat\Beta(u) + \bhat\Theta(u) \bm X^\top \kernelmat(\bm Y - \bm X \hat\Beta(u) ) / (nh),
\end{equation}
{a} one-step update of the \llasso\ in which the approximate inverse $\bhat\Theta(u)$ replaces the inverse of the local {Gram matrix} $\bhat\Sigma(u)$.
{Multiplying \eqref{eq. KKT conditions with Sigma} by $\bhat\Theta(u)$ and using the KKT conditions, we obtain the decomposition}
$$
{\bhat b(u)-\Beta^0(u)} = \bhat\Theta(u)\frac{\bm X^\top \kernelmat \bm \varepsilon}{nh} + \bhat\Theta(u) \bm \omega(u)
- \frac{\bm \Delta_n(u)}{\sqrt{nh}},
$$
where $\bm \Delta_n(u):= \sqrt{nh}(\bhat\Theta(u)\bhat \Sigma(u)-\bm I)(\hat\Beta(u)- \Beta^0(u))$ {is} the error due to the approximation of the inverse \citep{geer2014asymptotically}.
{The term $\bhat\Theta(u)\bm\omega(u)$ is the smoothing bias, which has no counterpart in the stationary setting; it arises because the coefficients are estimated from observations whose parameters $\Beta^0(t/n)$ differ from the target $\Beta^0(u)$.}
In Section~\ref{sec. asymptotic analysis}, we show that $\bm\Delta_n(u)$ is negligible when the local lasso and the nodewise regressions are sufficiently sparse{,} and that the bias term $\sqrt{nh}\,\hat{\bm\Theta}(u)\bm\omega(u)$ is negligible if $h$ is small enough.

\section{Theory}
\label{sec. theory}

We establish the rates of convergence for the estimation and prediction errors of the \llasso\ in Section \ref{sec. error bounds} and study the asymptotic normality of the desparsified estimator in Section \ref{sec. asymptotic analysis}.

\subsection{Error Bounds for the Local Lasso}
\label{sec. error bounds}

We first derive an oracle inequality for the estimation and prediction errors of the \llasso\ at a fixed rescaled time $u$, for a general bandwidth $h$ and penalty $\lambda(u)$.
We then provide an optimized choice of the bandwidth and the penalty and derive the resulting rates of convergence.
We consider the following assumptions.

\begin{assumptionA}[Lipschitz condition]\label{as. Lipschitz}
There is an absolute constant $L$, such that
$\max_j \vert \beta^0_j(u) - \beta^0_j(v)\vert \leq L |u-v|,$
for all $u,v \in [0,1]$.
\end{assumptionA}

\begin{assumptionA}[Kernel]\label{as. kernel}
The kernel $K$ is nonnegative, symmetric about zero, and supported on $[-1/2, 1/2]$, that is, $K(v) = 0$ for all $|v| > 1/2$, with $\int K(v)\,dv = 1$. Moreover, $K$ is Lipschitz: $|K(v) - K(v')| \leq C|v - v'|$ for all $v, v' \in \R$.
\end{assumptionA}

\begin{assumptionA}[Local stationarity]\label{as. local stationarity}
There exist constants $C > 0$ and $\rho \geq 2$, and sequences $\rate, \rateErr \geq 0$ with $\rate\,\lagorder/n \to 0$ and $\rateErr/n \to 0$, such that, for each $u \in [0,1]$ and $n \in \N$, there exist jointly strictly stationary processes $\{\Xu : t \in \Z\}$ and $\{\epsu : t \in \Z\}$ satisfying, for all $t$ with $|t/n - u| \leq \local$,
\begin{itemize}
\item[(i)] $\displaystyle \E\left[\Vert \X - \Xu \Vert_\infty^\rho\right]^{1/\rho} \leq C\rate \left(\left| \frac{t}{n} - u \right| + \frac{\lagorder}{n}\right)$.
\item[(ii)] $\displaystyle \E\left[\vert \varepsilon\tn - \epsu \vert^\rho\right]^{1/\rho} \leq C\rateErr \Deltatn$.
\end{itemize}
\end{assumptionA}

\begin{assumptionA}[Moments and dependence]\label{as. NED}
Let $\bm W\tn(u) := (\Xu^\top, \epsu)^\top$.
There exist constants $C > 0$, $\bar\order > \order > 2$ and $\nedsize \geq 1 \vee (\bar\order/\order - 1)/(\bar\order - 2)$ such that, for each $u \in [0,1]$ and $n \in \N$:
\begin{itemize}
    \item[(i)] $\E[\bm W\tn(u)] = \bm 0$, $\E[\Xu\,\epsu] = \bm 0$, and $\max_{1 \leq j \leq p_n+1} \E\big| W\tn\pj(u)\big|^{2\bar\order} \leq C$.
    \item[(ii)] There exists a process $\{\bm U\tn(u) : t \in \Z\}$ that is $\alpha$-mixing of size $-\nedsize/(1/\order - 1/\bar\order)$, with $\sigma$-field $\mathcal{F}^U\tn(u) := \sigma\{\bm U\tn(u), \bm U_{t-1,n}(u), \dots\}$ such that $\bm W\tn(u)$ is $\mathcal{F}^U\tn(u)$-measurable. The process $\{ W\tn\pj(u)\}$ is $L_{2\order}$-near-epoch-dependent (NED) of size $-\nedsize$ on $\{\bm U\tn(u)\}$ with positive bounded NED constants, uniformly over $j = 1, \dots, p_n+1$.
\end{itemize}
The mixing rate in (ii) and the NED constants are uniform in $u$ and $n$.
\end{assumptionA}

\begin{assumptionA}[Local compatibility]\label{as. compatibility condition}
Let $\bm\Sigma(u) := \E\left[\Xu \Xu^\top\right]$ and $S_\local := \localSet$. There exists a constant $\localComp \geq 1/C$ such that, for all $\bm b \in \R^{p_n}$ with $\Vert \bm b_{S_\local^{c}}\Vert_1 \leq 3\Vert \bm b_{S_\local}\Vert_1$,
$$
\Vert \bm b_{S_\local}\Vert_1^2 \leq \frac{\s}{\localComp^2}\; \bm b^\top \bm\Sigma(u)\, \bm b .
$$
\end{assumptionA}

\begin{assumptionA}[Rate conditions]\label{as. bound on s}
There exists a sequence $\vanishing \in [(nh)^{-1/2},1]$ with $\vanishing \to 0$ such that
$$
\s \leq C \vanishing^{\frac{\nedsize + \order - 1}{\nedsize\order + \order - 1}}
\left(\frac{1}{p_n^{\frac{2}{\nedsize} + \frac{2}{\order-1}}}
\sqrt{\frac{nh}{\log(p_n \vee \sqrt{nh})}}\right)^{\frac{1}{\frac{1}{\nedsize} + \frac{\order}{\order-1}}}.
$$
\end{assumptionA}

Note that Assumption \ref{as. Lipschitz} and $\delta$-local sparsity together imply that, for $ |t/n- u| \leq \local$,
$\Vert \BetazeroU- \Betazero(t/n) \Vert_2 \leq L \sqrt{\s}|u-t/n|$.
For the rest of the paper the bandwidth is assumed to converge to zero at a polynomial rate; that is, $ h \asymp n^{-\alpha}$ for some $ 0<\alpha < 1/2$.

Assumptions \ref{as. Lipschitz} and \ref{as. kernel} are standard in the literature on locally stationary processes.
Note that we do not assume that the coefficients are differentiable.
Assumption~\ref{as. local stationarity} states that the covariates and the errors are \edstationary with respect to the maximum norm. The term $\lagorder/n$ in (i) accounts for possible lagged regressors: the $j$th lag of the response is observed at rescaled time $(t-j)/n$, so that its distance to $u$ is at most $|t/n - u| + \lagorder/n$. Assumption~\ref{as. NED} is Assumption~1 of \citet{adamek2023lasso}, imposed on the stationary approximations rather than on the observed process. It allows for general forms of weak dependence and requires the variables to be centered. Assumption~\ref{as. compatibility condition} is a local version of the compatibility condition.
Assumption~\ref{as. bound on s} ensures that $\hat \Sig(u)$ concentrates around $\bm\Sigma(u)$ at a rate faster than $1/\s$. It has the form of the sparsity condition in \citet{adamek2023lasso}. 
As $\nedsize$ and $\order$ grow, it approaches the familiar requirement $\s \leq C\sqrt{nh/\log p_n}$.

\begin{theorem}[Local oracle inequality]\label{th. error bound}
Let $\Sig_{S_\local}(u)$ be the submatrix of $\Sig(u)$ with rows and columns in $\localSet$, and let $\localEig$ denote its largest eigenvalue.
Suppose Assumptions \ref{as. Lipschitz}--\ref{as. bound on s} hold, 
$u\in(0,1)$, $h \leq 2(\local\wedge u\wedge(1-u))$, $\lagorder \leq Cnh$, $\s\ge1$, 
$p_n^{1/(2\bar\order)}\,\s\,h\,\rate \leq C\vanishing$, and
\[
\lambda(u) \geq C\left(\vanishing^{-1/\order}\frac{p_n^{1/\order}}{\sqrt{nh}}
\vee \vanishing^{-1} h \left(\rate + \rateErr\,p_n^{1/(2\bar\order)}\right)\right).
\]
Then, for $n$ large enough, with probability at least $1 - C\vanishing$,
\begin{align}
\meant \kernel \left(\X^\top(\hat\Beta(u) - \Beta^0(t/n))\right)^2
&\leq C\left(\lambda(u)^2\,\s + \localEig\,\s\,h^2\right), \label{eq. th3 error bound prediction}\\
\Vert\hat\Beta(u) - \Beta^0(u)\Vert_1
&\leq C\left(\lambda(u)\,\s + \frac{\localEig\,\s\,h^2}{\lambda(u)}\right). \label{eq. th3 error bound beta}
\end{align}
\end{theorem}

The bounds are local in time. They depend on the local sparsity level 
$\s$ and on the degree of nonstationarity $\rate$ and $\rateErr$. 
The first term in \eqref{eq. th3 error bound prediction} and \eqref{eq. th3 error bound beta} is the estimation error of the lasso, in which the effective sample size $nh$ enters through $\lambda(u)$. The second is the approximation error incurred by treating the coefficients as constant over the kernel window. It follows from Assumption~\ref{as. Lipschitz} and local sparsity, since for $h \leq 2\local$ only the coefficients in $\localSet$ vary over the window. As $\s$ is non-decreasing in $\local$, one would like to take $\local$ as small as possible, but controlling the approximation error requires $\local \geq h/2$. 
By the moment bound in Assumption \ref{as. NED}, $\localEig \leq C\s$, so that this term is at most of order $\s^2h^2$; it is of order $\s h^2$ when the active covariates are weakly correlated.

The rates $\rate$ and $\rateErr$ of the stationary approximation enter the theorem in two places. First, the condition $p_n^{1/(2\bar m )}\,\s\,h\,\rate \leq C\vanishing$ ensures that, within the kernel window, the Gram matrix of the observed process is close to that of its stationary approximation at the scale $1/\s$ required by the compatibility condition. It allows $\rate$ to diverge, as happens in high-dimensional models where the approximation error grows with the number of active coefficients (see Theorem~\ref{th. local stationary} for the tv-ARX model), provided the bandwidth shrinks fast enough. Second, the penalty must dominate not only the score of the stationary approximation, of order $p_n^{1/\order}/\sqrt{nh}$, but also the contribution of non-stationarity to the score, of order $h(\rate + \rateErr\,p_n^{1/(2\bar m )})$. A stronger departure from stationarity therefore calls for a larger penalty and inflates the estimation error, unless $\sqrt{nh^3}\,(\rate + \rateErr\,p_n^{1/(2\bar m )}) \leq C\vanishing^{1-1/\order}p_n^{1/\order}$, in which case the first term of the lower bound dominates.

\begin{corollary}\label{cor. bandwidth}
Let $\effEig := \localEig \vee \vanishing^{-2}\left(\rate + \rateErr\,p_n^{1/(2\bar\order)}\right)^{2}$, and define
$h^{*} := \left(\frac{\vanishing^{-2/\order}\,p_n^{2/\order}}{n\,\effEig}\right)^{1/3}$ and $\lambda^{*}(u) := C\sqrt{\effEig}\;h^{*}$.
Suppose Assumptions \ref{as. Lipschitz}--\ref{as. compatibility condition} hold, and Assumption~\ref{as. bound on s} holds with $h = h^{*}$. 
Assume further that $u\in(0,1)$, $h^{*} \leq 2(\local\wedge u\wedge(1-u))$, $\lagorder \leq Cnh^{*}$, $\s\ge1$, and $\rate \leq C\,\vanishing^{1+2/\order}\,n / (p_n^{2/\order + 3/(2\bar\order)}\,\s^{3} ) $.
Then, for $n$ large enough, the \llasso\ with bandwidth $h^{*}$ and penalty $\lambda^{*}(u)$ satisfies, with probability at least $1 - C\vanishing$,
\begin{align*}
\meant \kernel \left(\X^\top(\hat\Beta(u) - \Beta^0(t/n))\right)^2
&\leq C\,\s\,\effEig^{1/3}\left(\frac{\vanishing^{-2/\order}\,p_n^{2/\order}}{n}\right)^{2/3},\\
\Vert\hat\Beta(u) - \Beta^0(u)\Vert_1
&\leq C\,\s\,\effEig^{1/6}\left(\frac{\vanishing^{-2/\order}\,p_n^{2/\order}}{n}\right)^{1/3}.
\end{align*}
\end{corollary}

The bandwidth $h^{*}$ balances the estimation and approximation terms of Theorem \ref{th. error bound}, and no other choice of $h$ and $\lambda(u)$ satisfying the conditions of that theorem improves the bounds \eqref{eq. th3 error bound prediction} and \eqref{eq. th3 error bound beta} by more than a constant factor. The factor $\effEig$ collects the two features of the local model that enter these bounds. 
When $\effEig$ is bounded, both bounds are linear in $\s$ and, for fixed $p_n$ and $\s$, the $\ell_1$ bound is of order $n^{-1/3}$, the minimax rate for Lipschitz coefficient functions.
$\lambda^{*}(u)$ depends on $u$ through $\effEig$, which motivates a local choice of the tuning parameters.

\subsection{Asymptotic Theory for the Desparsified Local Lasso}
\label{sec. asymptotic analysis}

We now derive the asymptotic distribution of the desparsified \llasso\ at a fixed rescaled time $u\in(0,1)$, for a finite set $\mathcal J\subset\{1,\dots,p_n\}$ of coefficients of interest.
Our starting point is the decomposition of Section~\ref{sec. penalized kernel estimation}, in which $\bhat b(u)-\Beta^0(u)$ is the sum of a stochastic term, a smoothing bias and a remainder $\bm\Delta_n(u)$.
To make the remainder negligible, we need error bounds for both the \llasso\ and the nodewise regressions, and Theorem~\ref{th. error bound} provides the first.
The second is not straightforward because the errors of the nodewise regressions inherit the non-stationarity of the covariates.
In Proposition~\ref{prop. projection errors}, we show that they are nevertheless \epsdelta-stationary, so that Theorem~\ref{th. error bound} applies to the nodewise regressions as well.
With both bounds, the analysis proceeds in two steps.
Theorem~\ref{th. linear representation} gives an asymptotically linear representation of $\bhat b(u)$ with an explicit smoothing bias.
Corollary~\ref{cor. asymptotic normality} shows that the estimator is asymptotically normal when the bandwidth is small enough for both the smoothing bias and the effect of non-stationarity on the stochastic term to be negligible.

Let $\bm\Theta(u) := \bm\Sigma(u)^{-1}$, with rows $\bm\Theta_j(u)$. For each $j$, define the local projection
\[
\gammaZero(u) := \argmin_{\bm\gamma \in \R^{p_n-1}} \E\big(\xu[j] - \Xu[-j]^\top \bm\gamma\big)^2.
\]
We define the local nodewise errors by
$ \ndepsu := \xu[j] - \Xu[-j]^\top \gammaZero(u),$
and let $\tau_j^2(u) := \E\big[\ndepsu^2\big]$, so that $\Theta_{j,j}(u) = 1/\tau_j^2(u)$ and $\Theta_{j,k}(u) = -\gamma^0_{j,k}(u)/\tau_j^2(u)$ for $k \neq j$. 
The corresponding errors in the observed process are $\ndeps := \x[j] - \X\mjtop \gammaZero(t/n)$. 
We also define the local active set for the nodewise regressions:
$S_{j,\local}(u) :=\left\{k\in \{1,...,p_n\}: \gamma_{j,k}^0 \text{ is not $\local$-locally sparse at } u \right\}$, 
where we let $s_{j,\local}(u):= |S_{j,\local}(u)|$.
Finally, we write 
$\sbar := 1\vee\max_{j \in \mathcal{J}} \sj$ and $\smax := \s \vee \sbar$.

\begin{assumptionB}[Local precision matrix]\label{as. precision matrix}
There exist constants $C, L_\Theta > 0$ such that, for all $u, u' \in [0,1]$ and $n \in \N$, (i) the smallest eigenvalue of $\bm\Sigma(u)$ is at least $1/C$;
(ii) $\Vert \bm\Theta_j(u) - \bm\Theta_j(u') \Vert_2 \leq L_\Theta\,|u - u'|$ for all $j \in \mathcal{J}$.
\end{assumptionB}

\begin{assumptionB}[Projection errors]\label{as. projection errors}
For each $j \in \mathcal{J}$, $\E|\ndepsu|^{2\bar\order} \leq C$, and $\{\ndepsu\}$ is $L_{2\order}$-NED of size $-\nedsize$ on $\{\bm U\tn(u)\}$ with bounded NED constants, uniformly in $u$ and $n$.
\end{assumptionB}

Assumption~\ref{as. precision matrix}(i) is standard for the desparsified lasso \citep{geer2014asymptotically,adamek2023lasso} and implies the compatibility condition of Assumption~\ref{as. compatibility condition} for the \llasso\ and the nodewise regressions.
Assumption~\ref{as. precision matrix}(ii) is the counterpart of Assumption~\ref{as. Lipschitz} for the nodewise regressions, and holds when $\bm\Sigma(u)$ is Lipschitz in spectral norm.
Assumption~\ref{as. projection errors} imposes on the projection errors the moment and dependence conditions that Assumption~\ref{as. NED} imposes on $\epsu$.

\begin{proposition}[Local stationarity of the projection errors]\label{prop. projection errors}
Suppose Assumptions \ref{as. local stationarity}, \ref{as. NED} and \ref{as. precision matrix} hold, and $\rho \leq 2\bar \order$. Then, for all $j \in \mathcal{J}$ and all $t$ with $|t/n - u| \leq \local$,
\[
\E\left[\vert \ndeps - \ndepsu \vert^\rho\right]^{1/\rho}
\leq C\,\rate^v \left(\left|\frac{t}{n} - u\right| + \frac{\lagorder}{n}\right),
\qquad
\rate^v := \sqrt{\sbar}\,(\rate \vee L_\Theta).
\]
\end{proposition}

The observed nodewise errors differ from their stationary counterparts for two reasons, the non-stationarity of the covariates and the variation of the projection coefficients over the window.
Proposition~\ref{prop. projection errors} controls both, so that Theorem~\ref{th. error bound} applies to the nodewise regressions with $\rate^v$ in place of $\rateErr$.
The rate $\rate^v$ grows with the sparsity of the projections even when the covariates are locally stationary in the classical sense, that is, $\rate\asymp1$.

\begin{theorem}[Linear representation]\label{th. linear representation}
Suppose Assumptions \ref{as. Lipschitz}--\ref{as. NED} and \ref{as. precision matrix}--\ref{as. projection errors} hold, and Assumption \ref{as. bound on s} holds with $\smax$ in place of $\s$. Let $u \in (0,1)$, $h \leq 2\local$, $\lagorder \leq Cnh$ and $\lambda_j(u) \asymp \lambdau$ for all $j \in \mathcal{J}$, and assume that
(a) $p_n^{1/(2\bar\order)}\,\smax\,h\,\rate \to 0$;
(b) $p_n^{1/\order}/ \left( \lambdau \sqrt{nh} \right) \to 0$ and $ h\,\big(\rate + (\rateErr \vee \rate^v)\,p_n^{1/(2\bar\order)}\big)/ \lambdau \to 0$;
(c) $\sqrt{nh}\,\big(\lambdau^2 + \bar\Lambda_\local(u)\,h^2\big)\,\smax \to 0$,
where $\bar\Lambda_\local(u):= \localEig \vee \max_{j\in\mathcal J}\Lambda_{j,\local}(u),$ and
$\Lambda_{j,\local}(u)$ is the largest eigenvalue of the submatrix of $\bm\Sigma(u)$ with rows and columns in $S_{j,\local}(u)$.
Then, for $n$ large enough,
\[
\sqrt{nh}\,\big(\hat b_j(u) - \beta_j^0(u)\big)
= \frac{1}{\sqrt{nh}} \sumt \kernel\, \bm\Theta_j(u)\,\X\, \varepsilon\tn
+\sqrt{nh}\,\bhat\Theta_j(u)\bm\omega(u)
+ o_\p(1),
\]
for all $j \in \mathcal{J}$. Furthermore, we have $\sqrt{nh}\,\abs{\bhat\Theta_j(u)\bm\omega(u)}
\leq C \sqrt{nh^3\,\localEig\,\s}$, with probability tending to one.
\end{theorem}

The first term is a kernel-weighted sum of scores, in which the estimated row $\bhat\Theta_j(u)$ is replaced by its population counterpart.
The second term is the smoothing bias, which does not vanish in general and is kept explicit.
Conditions (a) and (b) are the conditions of Theorem~\ref{th. error bound} for the main and the nodewise regressions, and condition (c) makes the remainder $\bm\Delta_n(u)$ of the desparsification negligible.

To obtain asymptotic normality from Theorem~\ref{th. linear representation}, it remains to apply a central limit theorem to the leading term and to choose the bandwidth so that the smoothing bias vanishes. As the scores are serially dependent, the variance of the leading term is a local long-run variance, which we require to be nondegenerate.

\begin{assumptionB}[Local long-run variance]\label{as. long-run variance}
Let $\tilde{\bm w}\tn(u) := \bm\Theta_{\mathcal{J}}(u)\,\Xu\,\epsu$, where $\bm\Theta_{\mathcal{J}}(u)$ collects the rows $\bm\Theta_j(u)$, $j \in \mathcal{J}$. The smallest eigenvalue of
\[
\bm V_n(u) := \E\left[\frac{1}{nh}\Big(\sumt \kernel\, \tilde{\bm w}\tn(u)\Big)\Big(\sumt \kernel\, \tilde{\bm w}\tn(u)\Big)^{\!\top}\right]
\]
is at least $1/C$.
\end{assumptionB}

\begin{corollary}[Asymptotic normality]\label{cor. asymptotic normality}
Suppose the conditions of Theorem~\ref{th. linear representation} and Assumption~\ref{as. long-run variance} hold, and that
(d) $\sqrt{nh^3}\,\big(\rateErr + \sqrt{\sbar}\,\rate\big) \to 0$, and
(e) $nh^3\,\localEig\,\s \to 0$.
Then
$\displaystyle \sqrt{nh}\; \bm V_n(u)^{-1/2} \big(\bhat b_{\mathcal{J}}(u) - \Beta^0_{\mathcal{J}}(u)\big) \overset{d}{\longrightarrow} \mathcal{N}(\bm 0, \bm I_{|\mathcal{J}|}).
$
In particular, for each $j \in \mathcal{J}$,
$\sup_{z\in\R} \big|\p\big(\sqrt{nh}\,(\hat b_j(u)-\beta_j^0(u))/\sqrt{V_{n,jj}(u)}\le z\big)-\Phi(z)\big| \to 0$.
\end{corollary}

Condition (d) allows the scores to be replaced by those of the stationary approximation, to which a central limit theorem for NED processes applies \citep{adamek2023lasso}, and condition (e) makes the smoothing bias negligible.
In practice, $\bm V_n(u)$ is unknown, and we estimate it by a kernel-weighted Bartlett estimator $\bhat V(u)$ built from the products of the nodewise and \llasso\ residuals \citep{neweyWest1986}.
More details can be found in the supplementary materials.

\section{Numerical Experiments}
\label{sec. simulations}

We compare the \llasso\ with existing methods and evaluate the coverage of the confidence intervals of the desparsified \llasso.

We consider two data-generating processes. The first is a time-varying regression with
exogenous covariates whose coefficients are cosine bumps. The number of bumps $K$ controls
how long each coefficient is active, which moves the model from global to local
sparsity while the signal and the local sparsity stay fixed. The second model is a time-varying ARX process, as used in our applications. It adds lagged responses with globally and locally sparse coefficients.

To generate locally sparse coefficients, we use the compactly supported functions $\phi_k(u):=\phi(Ku-k+1/2)$, $k=1,\dots,K$, where $\phi(x):=\cos(\pi x/2)\mathbbm{1}(|x|\le1)$.
Each $\phi_k$ is nonzero on an interval of length $2/K$, and the supports of $\phi_k$ and $\phi_{k+1}$ are shifted by $1/K$.
Hence, at most two coefficients are nonzero at each time point while the global sparsity level increases with $K$.

\begin{model}\label{model sparse setting}
We generate $Y\tn=\sum_{k=1}^{K}\phi_k(t/n)Z\tn^{(k)}+\varepsilon\tn$ with $p_n$ covariates, $n=400$, $p_n\in\{50,200,1000\}$ and $K\in\{2,10\}$.
\end{model}

\begin{model}\label{model time series}
We generate
$$
Y\tn=\alpha_1(t/n)Y_{t-1,n}+\alpha_2(t/n)Y_{t-2,n}+\eta_1(t/n)Z\tn^{(1)}+\sum_{k=1}^{10}\phi_k(t/n)Z\tn^{(2k+1)}+\varepsilon\tn,
$$
with $\alpha_1(u)=0.7\sin(3\pi u)/\{2+\sin(4\pi u)\}$, $\alpha_2(u)=-0.5\,\phi(6u-4)$ and $\eta_1(u)=\sin(2\pi u)$.
The regressors are $Y_{t-1,n},\dots,Y_{t-4,n}$, $\bm Z\tn$ and $\bm Z_{t-1,n}$, where $\bm Z\tn$ has $(p_n-4)/2$ components, and all other coefficients are zero.
We consider $n\in\{400,800,1600\}$ and $p_n\in\{50,200,1000\}$.
\end{model}

\begin{table}[h!]
\centering\small
\caption{Estimation accuracy with $n=400$ of the oracle, the local lasso (LL) and the group lasso (GL).
The $\ell_2$ and $\ell_1$ errors are $n^{-1}\sum_t\|\hat\Beta(t/n)-\Beta^0(t/n)\|_2^2$ and $n^{-1}\sum_t\|\hat\Beta(t/n)-\Beta^0(t/n)\|_1$, and FN and FP are the average numbers of false negatives and false positives per time point.
Entries are medians over 200 replications, with median absolute deviations in parentheses.}
\label{tab. estimation}
\begin{tabular}{ll ccc ccc cc cc}
\toprule
 & & \multicolumn{3}{c}{$\ell_2$} & \multicolumn{3}{c}{$\ell_1$} & \multicolumn{2}{c}{FN} & \multicolumn{2}{c}{FP}\\
\cmidrule(lr){3-5}\cmidrule(lr){6-8}\cmidrule(lr){9-10}\cmidrule(lr){11-12}
Model & $p_n$ & Oracle & LL & GL & Oracle & LL & GL & LL & GL & LL & GL\\
\midrule
1, $K=2$   & 50   & 0.01 & 0.04 (0.01) & 0.03 (0.01) & 0.09 & 0.25 & 0.30 & 0.05 & 0.00 & 2.4 & 10.5\\
           & 200  & 0.01 & 0.05 (0.01) & 0.04 (0.01) & 0.09 & 0.28 & 0.34 & 0.06 & 0.00 & 2.6 & 14.5\\
           & 1000 & 0.01 & 0.06 (0.01) & 0.04 (0.01) & 0.09 & 0.31 & 0.34 & 0.07 & 0.00 & 2.9 & 17.5\\
\midrule
1, $K=10$  & 50   & 0.04 & 0.12 (0.02) & 0.27 (0.03) & 0.23 & 0.51 & 1.45 & 0.01 & 0.00 & 3.2 & 32.1\\
           & 200  & 0.05 & 0.15 (0.03) & 0.32 (0.03) & 0.24 & 0.55 & 1.60 & 0.01 & 0.00 & 3.3 & 51.1\\
           & 1000 & 0.05 & 0.19 (0.03) & 0.38 (0.03) & 0.26 & 0.61 & 1.73 & 0.02 & 0.00 & 3.7 & 67.6\\
\midrule
2          & 50   & 0.07 & 0.26 (0.04) & 0.35 (0.04) & 0.43 & 1.08 & 2.02 & 0.29 & 0.00 & 6.0 & 35.8\\
           & 200  & 0.08 & 0.38 (0.05) & 0.43 (0.04) & 0.46 & 1.25 & 2.30 & 0.48 & 0.00 & 6.5 & 64.8\\
           & 1000 & 0.10 & 0.54 (0.08) & 0.55 (0.06) & 0.51 & 1.45 & 2.46 & 0.75 & 0.00 & 6.6 & 83.8\\
\bottomrule
\end{tabular}
\end{table}

In both models, the covariates follow $\bm Z\tn=a\bm Z_{t-1,n}+\sqrt{1-a^2}\,\bm g\tn$ with $a=0.25$, where the $\bm g\tn\sim\mathcal N(\bm 0,\bm R(t/n))$ are independent, $R_{j,k}(u)=r(u)^{|j-k|}$ and $r(u)=0.3+0.4u$, so that their correlation varies over time. The errors $\varepsilon\tn$ are i.i.d.\ $\mathcal N(0,0.25)$.

All estimators use the Epanechnikov kernel.
The \llasso\ and the group lasso use the bandwidth $h^*=(p_n^{1/5}/n)^{1/3}$, which follows the rate of Corollary~\ref{cor. bandwidth} with $\order=10$, and the desparsified \llasso\ uses $h=h^*n^{-1/15}$.
The penalty $\lambda(u)$ minimizes the local BIC
\begin{equation}\label{eq. BIC criterion}
\text{BIC}(u) = \log\left( \frac{\text{RSS}(u)}{\tilde{n}(u)}\right) + \text{df}(u) \frac{\log \tilde{n}(u) }{\tilde{n}(u)},
\end{equation}
where $\text{RSS}(u)$ is the kernel-weighted residual sum of squares, $\tilde{n}(u):=(\sum_t K_t)^2/\sum_t K_t^2$ with $K_t:=K((t/n-u)/h)$ is the effective sample size of the window, and $\text{df}(u)$ is the number of nonzero coefficients of $\hat\Beta(u)$.
We compute $\lambda(u)$ on 50 equispaced grid points and smooth $\log\lambda(u)$ over $u$.
The nodewise penalties $\lambda_j(u)$ are selected by the same criterion, and the long-run variance estimator is given in the Supplementary Materials.
Each setting is replicated 200 times.

\begin{table}[t]
\centering\small
\setlength{\tabcolsep}{4pt}
\caption{Coverage of the 95\% confidence intervals of the desparsified local lasso in Model~\ref{model time series} over 200 replications.
}
\label{tab. inference}
\begin{tabular}{rr ccc ccc ccc}
\toprule
 & & \multicolumn{3}{c}{$u = 0.25$} & \multicolumn{3}{c}{$u = 0.5$} & \multicolumn{3}{c}{$u = 0.75$}\\
\cmidrule(lr){3-5}\cmidrule(lr){6-8}\cmidrule(lr){9-11}
$n$ & $p$ & $\alpha_1$ & $\alpha_2$ & $\eta_1$ & $\alpha_1$ & $\alpha_2$ & $\eta_1$ & $\alpha_1$ & $\alpha_2$ & $\eta_1$\\
\midrule
400  & 50   & 0.83 & 0.85 & 0.81 & 0.81 & 0.95 & 0.94 & 0.80 & 0.73 & 0.78\\
     & 200  & 0.79 & 0.92 & 0.84 & 0.79 & 0.97 & 0.96 & 0.84 & 0.75 & 0.81\\
     & 1000 & 0.79 & 0.95 & 0.84 & 0.77 & 0.96 & 0.98 & 0.87 & 0.66 & 0.78\\
\addlinespace
800  & 50   & 0.91 & 0.93 & 0.92 & 0.86 & 0.95 & 0.97 & 0.88 & 0.91 & 0.84\\
     & 200  & 0.95 & 0.91 & 0.95 & 0.86 & 0.97 & 0.97 & 0.91 & 0.84 & 0.91\\
     & 1000 & 0.90 & 0.98 & 0.94 & 0.83 & 0.96 & 0.97 & 0.90 & 0.81 & 0.91\\
\addlinespace
1600 & 50   & 0.95 & 0.95 & 0.93 & 0.90 & 0.95 & 0.95 & 0.93 & 0.93 & 0.90\\
     & 200  & 0.92 & 0.94 & 0.95 & 0.90 & 0.96 & 0.95 & 0.90 & 0.90 & 0.93\\
     & 1000 & 0.92 & 0.94 & 0.98 & 0.88 & 0.97 & 0.97 & 0.91 & 0.92 & 0.95\\
\bottomrule
\end{tabular}
\end{table}

\begin{figure}[h]
    \centering
    \includegraphics[trim = 0 0 0 1cm, clip = TRUE]{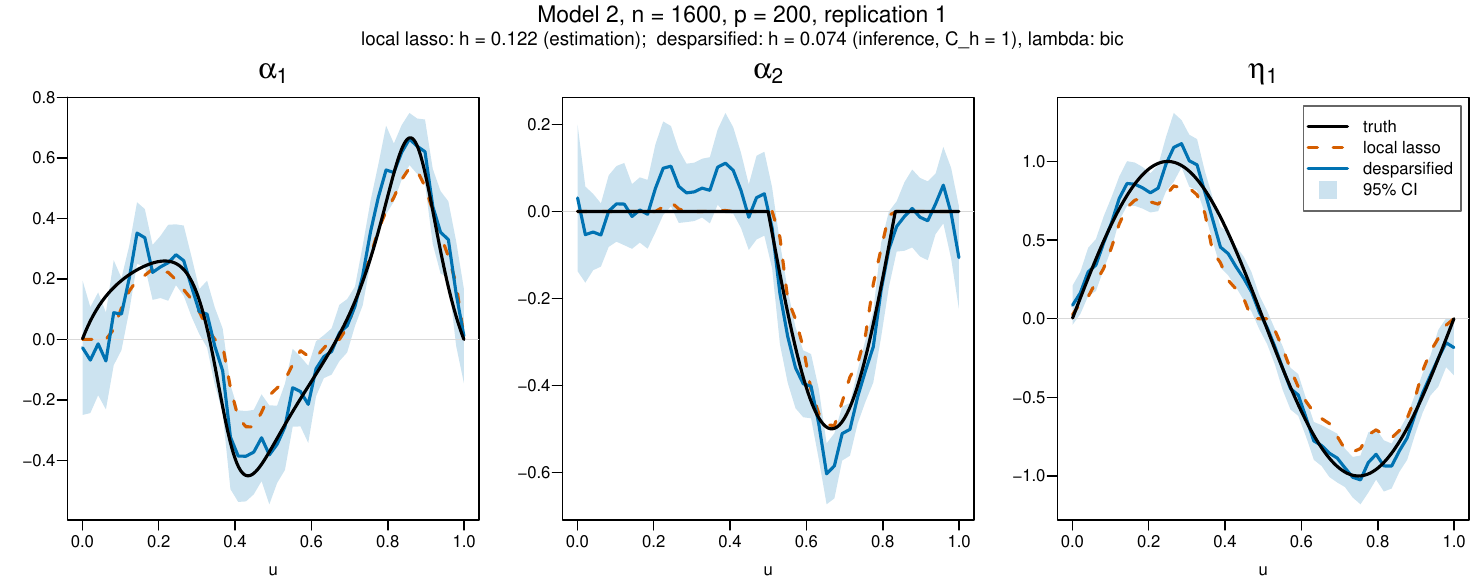}
    \caption{Desparsified \llasso\ estimates for Model~\ref{model time series} with $p_n=200$ and $n=1600$.}
    \label{Fig. inference model}
\end{figure}

We compare the \llasso\ with the local-constant version of the kernel group lasso of \citet{Lee2016}.
Its penalty is chosen by five-fold cross-validation.
We also report an oracle, the kernel-weighted least-squares estimator on the true active set at each time point.

The results are summarized in Table~\ref{tab. estimation}, Table~\ref{tab. inference}, and Figure \ref{Fig. inference model}.
In Model~\ref{model sparse setting} with $K=2$, the two estimators have similar errors, and the \llasso\ selects four to six times fewer false positives.
For $K=10$, the $\ell_1$ error of the \llasso\ decreases from 2.8--3.6 to about 2.3 times that of the oracle, whereas that of the group lasso increases from 3.4--4.0 to 6.2--6.6 times.
With $K=10$, the \llasso\ halves the $\ell_2$ error of the group lasso and divides its $\ell_1$ error by almost three.
The group lasso gives 32 to 68 false positives on average against fewer than four for the \llasso.
In Model~\ref{model time series}, the \llasso\ has a smaller $\ell_1$ error and fewer false positives, at the cost of  slightly more false negatives.

Table~\ref{tab. inference} reports the coverage of the 95\% confidence intervals in Model~\ref{model time series} at $u=0.25$, 0.5 and 0.75.
Coverage approaches the nominal level as $n$ grows, and depends little on $p_n$.
At $n=400$, undercoverage is confined to nonzero coefficients, which is consistent with residual shrinkage or smoothing bias, both of which vanish for zero coefficients.

\section{Real Data Analysis}
\label{sec. real data}

\subsection{The Time-varying Phillips Curve}

It is important for policymakers to understand how unemployment and inflation are related, particularly because many central banks are charged with supporting both high levels of employment and price stability. 
The negative relation documented by \citet{phillips1958} has proved unstable over time \citep{CogleySargent2005}, and inflation has become harder to forecast \citep{StockWatson2007}.
Whether unemployment remains informative about inflation has therefore been the subject of considerable debate among economists.
Moreover, inflation is related to many macroeconomic indicators beyond unemployment \citep{stock1999forecasting,medeiros2021forecasting}, so assessing the role of unemployment calls for conditioning on a large set of controls.
Combined with time variation, this yields a locally high-dimensional problem in which the number of regressors exceeds the effective local sample size.

\begin{figure}[h!]
    \centering
        \begin{subfigure}[t]{0.45\textwidth}
        \centering
        \includegraphics[trim = 0 0 0 4.8cm, clip = TRUE]{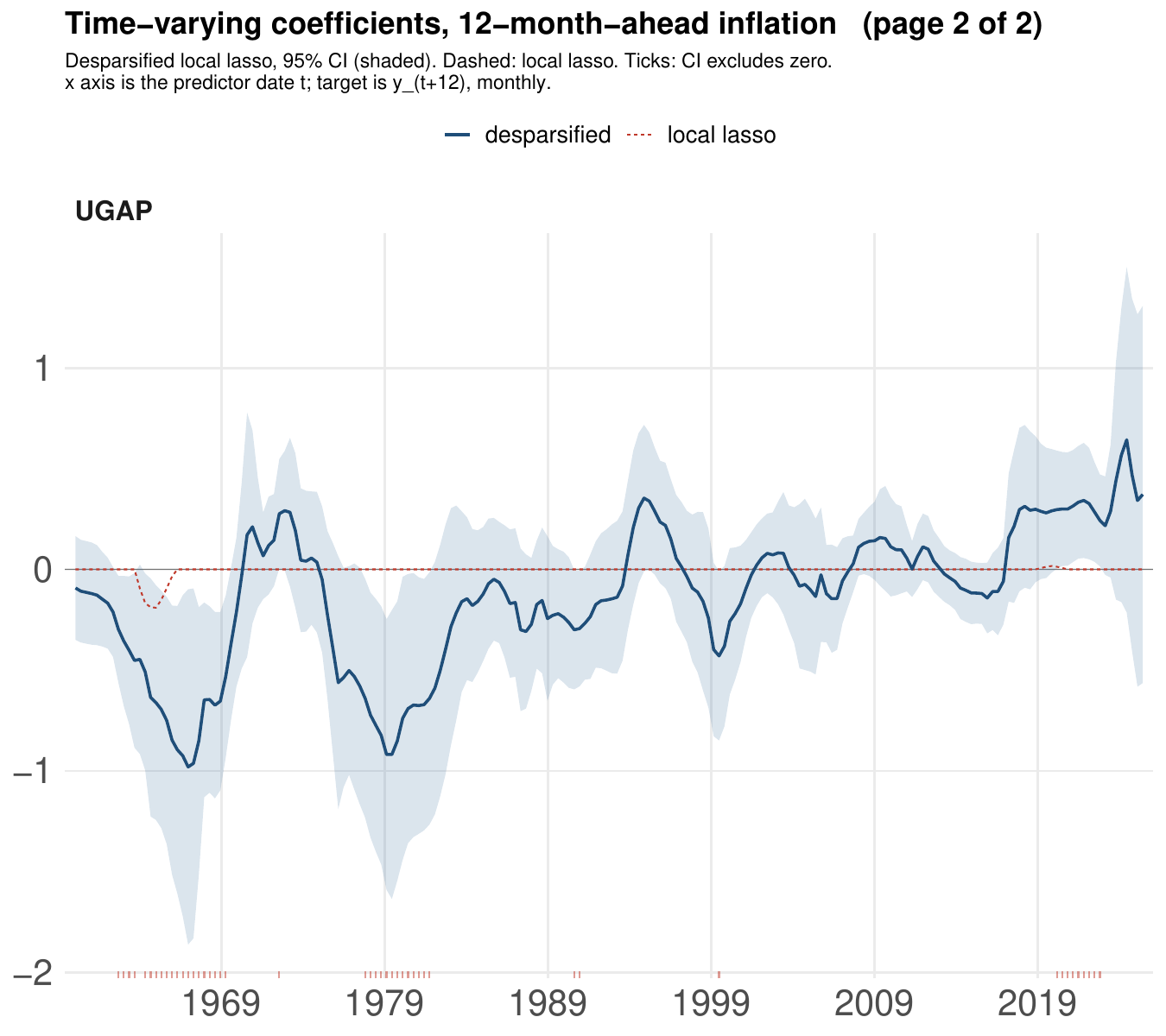}
        \caption{Unemployment gap}
        \label{fig:phillips desparsified coeff ugap}
    \end{subfigure}
    \hfill
    \begin{subfigure}[t]{0.45\textwidth}
        \centering
        \includegraphics[trim = 0 0 0 4.8cm, clip = TRUE]{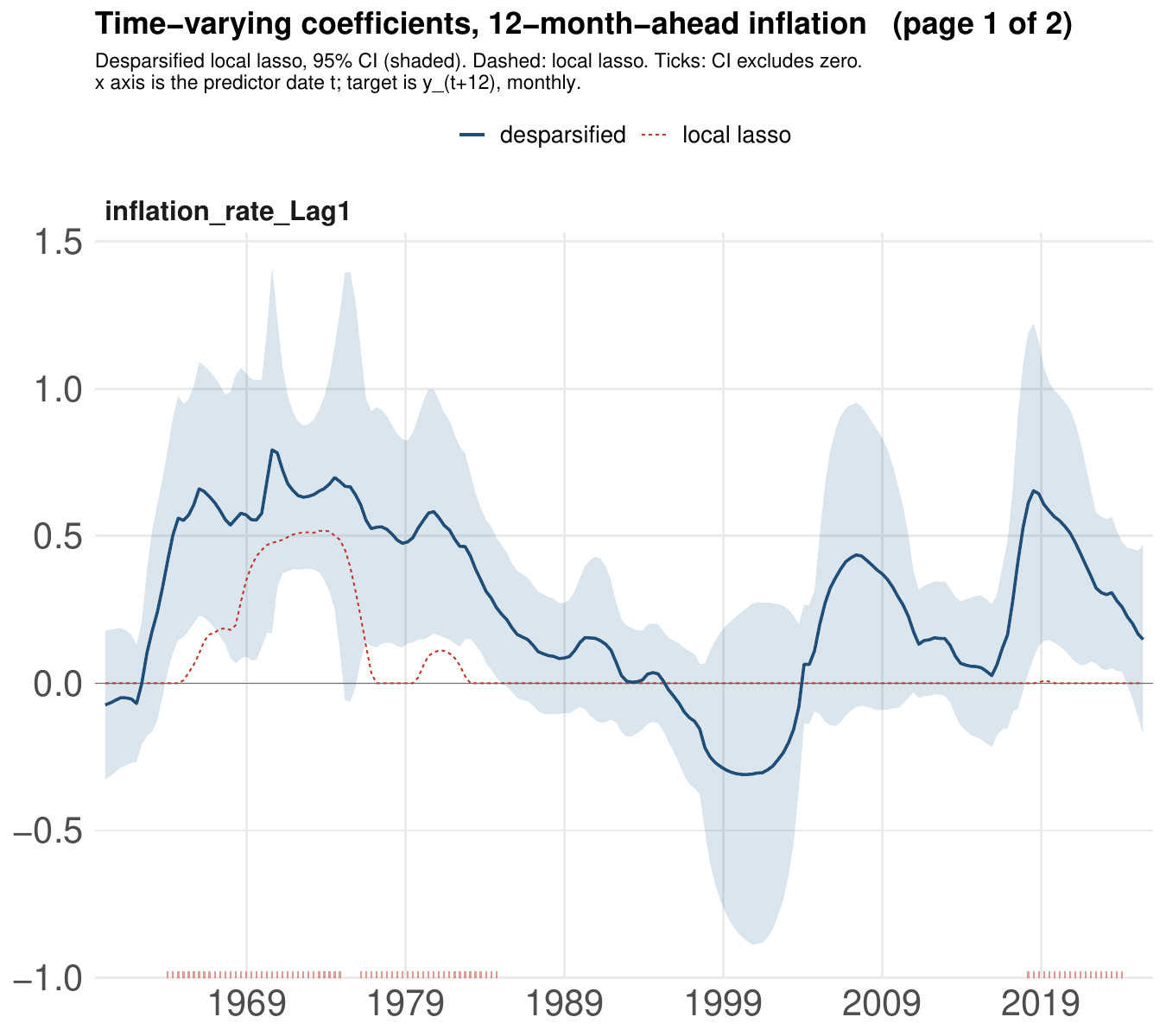}
        \caption{Inflation first lag}
        \label{fig:phillips desparsified coeff inflation lag}
    \end{subfigure}
    \caption{Desparsified local lasso (solid) with pointwise 95\% confidence intervals (shaded), and local lasso (dashed). Ticks mark the dates at which the interval excludes zero.}
    \label{fig: phillips desparsified coeff}
\end{figure}

We revisit the Phillips curve using the September 2025 vintage of the FRED-MD database \citep{mccracken2016fred}, with the aim of assessing whether unemployment remains conditionally associated with inflation.
We estimate the tv-ARX model~\eqref{eq. tv ARX} with an intercept and $d_n=12$ inflation lags.
The exogenous regressors are the unemployment gap, which is the difference between unemployment rate and the Noncyclical Rate of Unemployment, at lag $12$ only and $118$ FRED-MD series, with no missing values, at lags $12$ and $13$. 
In total the model contains $p_n = 249$ regressors, and the sample size is $n = 785$.

Figure~\ref{fig: phillips desparsified coeff} reports the estimates and the 95\% confidence intervals of the coefficient of the unemployment gap and the lagged inflation.
The coefficient of the first inflation lag declines sharply after the 1970s, in line with the changes in U.S.\ inflation dynamics documented by \citet{StockWatson2007}.
The unemployment-gap coefficient is negative in the late 1960s and around 1980, as the conventional trade-off predicts.
It weakens from the mid-1980s onward, in line with a flatter Phillips curve and better-anchored inflation expectations \citep{blanchard2015inflation}.
Finally, it turns positive around the COVID-19 episode, when the pandemic rise in unemployment was followed, twelve months later, by the inflation surge of 2021--22 \citep{bernanke2025caused}.

\subsection{Application to EEG Data}

Epileptic seizures are a natural example of nonstationarity.
In focal epilepsy, abnormal electrical activity starts in a localized seizure-onset zone (SOZ) and then evolves in amplitude, rhythm and spatial extent as the seizure progresses.
Stereo-electroencephalography (sEEG) records this evolution at high frequency through depth electrodes implanted in many brain regions, which yields a high-dimensional time series whose dependence structure changes over the course of a seizure.

\begin{figure}[h!]
    \centering
    \includegraphics[width=0.8\textwidth, trim = 0cm 0cm 0cm 0cm, clip = TRUE]{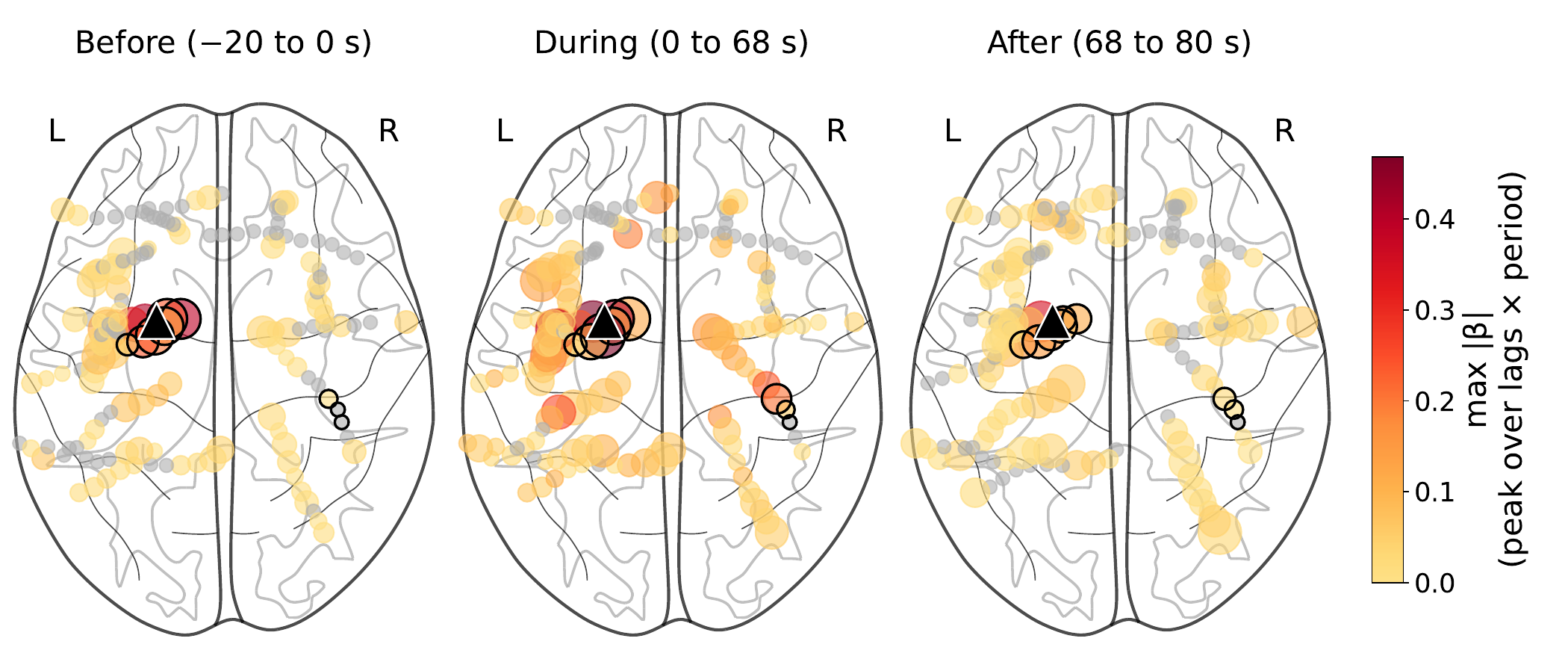}
\caption{Local lasso estimates for LA3 (black triangle), before, during and after the seizure. Each marker is a recording contact, black outlines mark the SOZ, and grey markers are never selected. Color gives the largest absolute coefficient of the contact's lags over the period, and size the share of the period during which at least one of them is nonzero.
}
\label{fig. eeg brain plot}
\end{figure}

We analyze publicly available sEEG recordings from the Hospital of the University of Pennsylvania (HUP), in which each electrode contact is clinically annotated as lying inside or outside of the SOZ, providing an external clinical reference for interpreting the estimated dependence structure \citep{bernabei2023quantitative}. 
We focus on one seizure from subject HUP162 and use LA3, one of the identified SOZ contacts, as the response channel.
We analyze a peri-ictal window extending from 20 s before to 80 s after clinical seizure onset. 
The recordings are sampled at 128 Hz, and each channel enters through 10 lags, corresponding to approximately 78 ms of recent history. With 169 recorded channels, this yields $p_n=1690$ candidate predictors.

\newcommand{\width}{0.32}
\begin{figure}[h!]
    \centering
    \begin{subfigure}[t]{\width\textwidth}
        \centering
        \includegraphics[trim = 1cm 1cm 1cm 1.9cm, clip = TRUE]{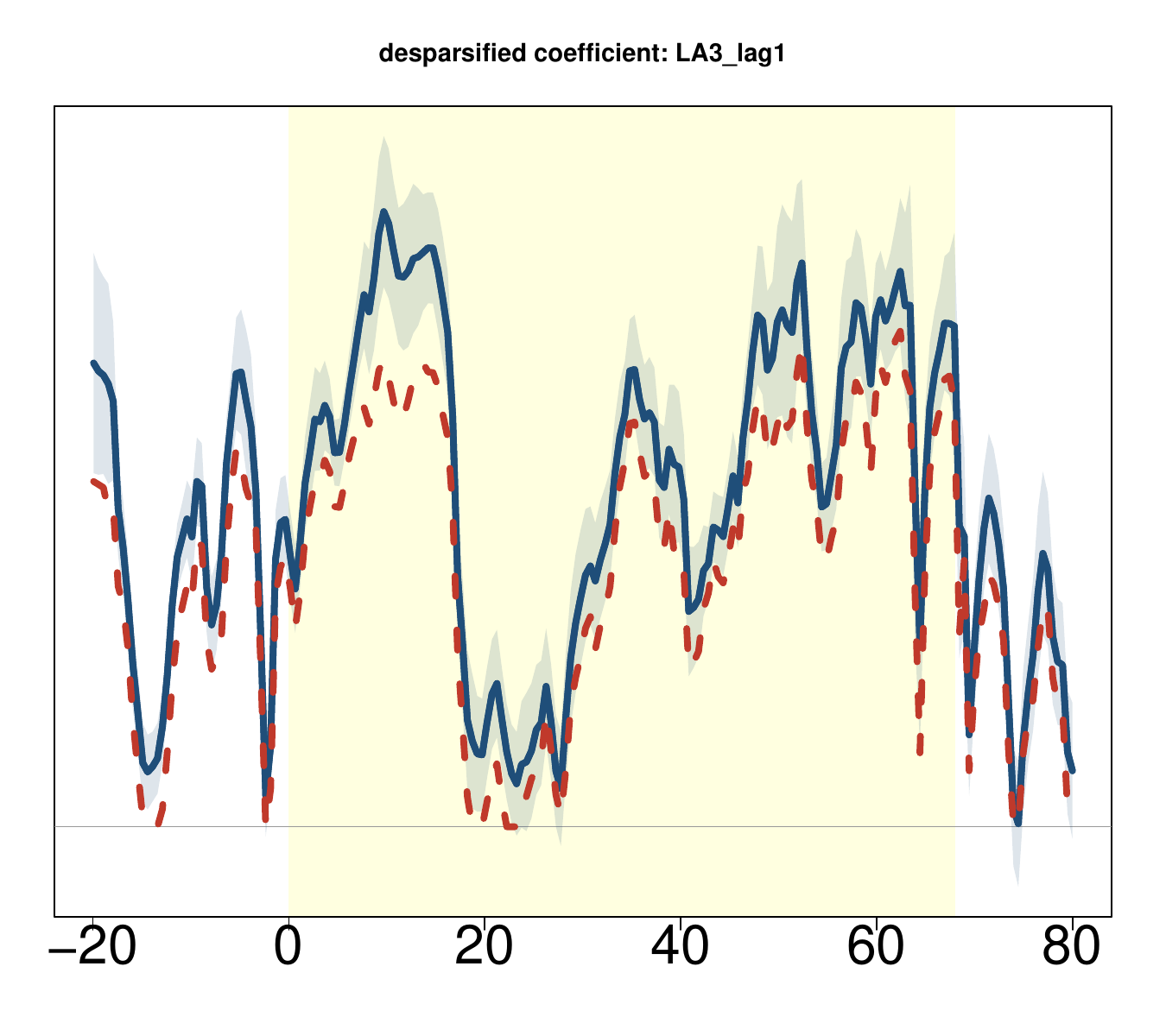}
        \caption{LA3 (non sparse)}
    \end{subfigure}
    \hfill
    \begin{subfigure}[t]{\width\textwidth}
        \centering
        \includegraphics[trim = 1cm 1cm 1cm 1.9cm, clip = TRUE]{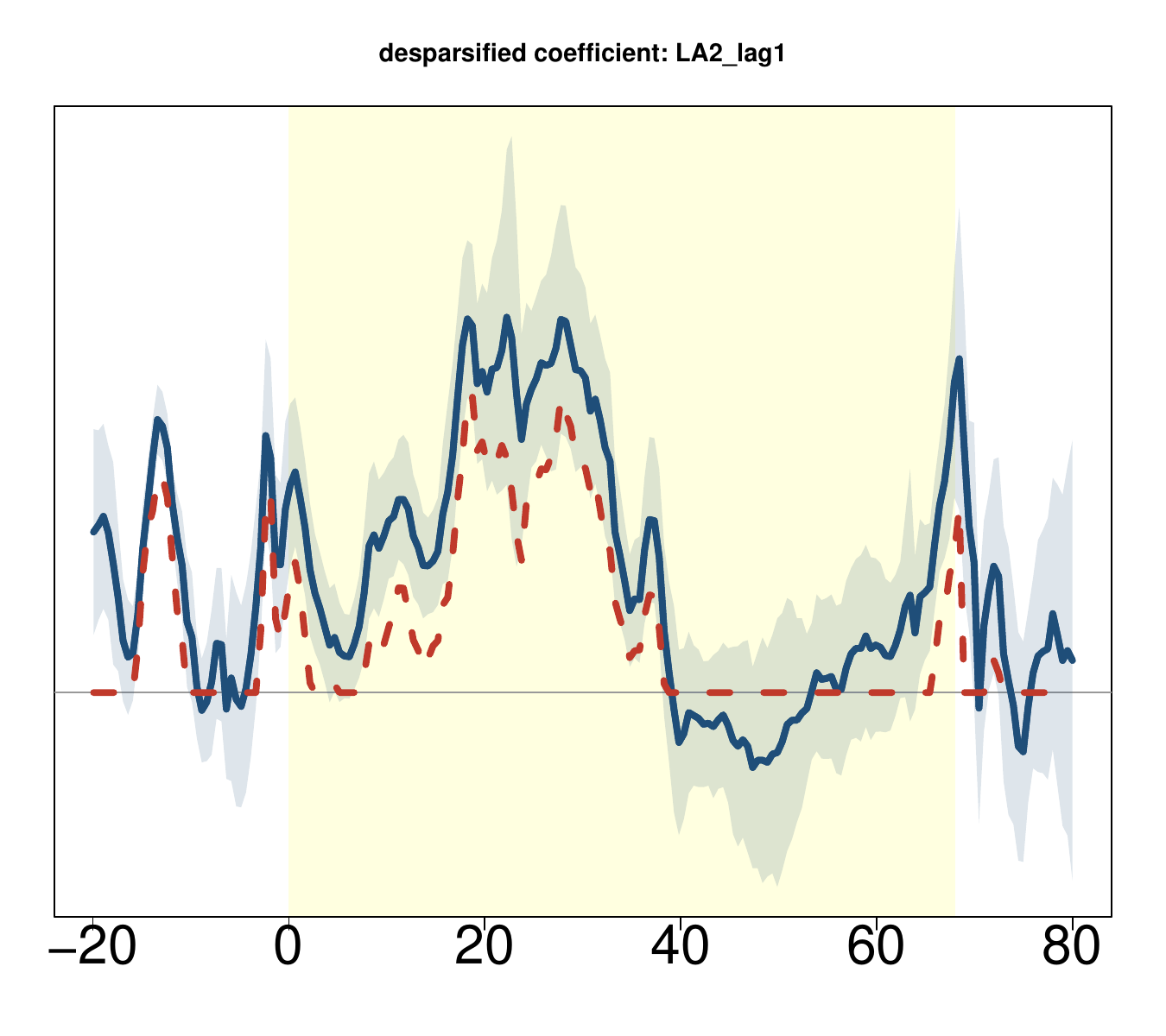}
        \caption{LA2 (locally sparse)}
    \end{subfigure}
        \hfill
    \begin{subfigure}[t]{\width\textwidth}
        \centering
        \includegraphics[trim = 1cm 1cm 1cm 1.9cm, clip = TRUE]{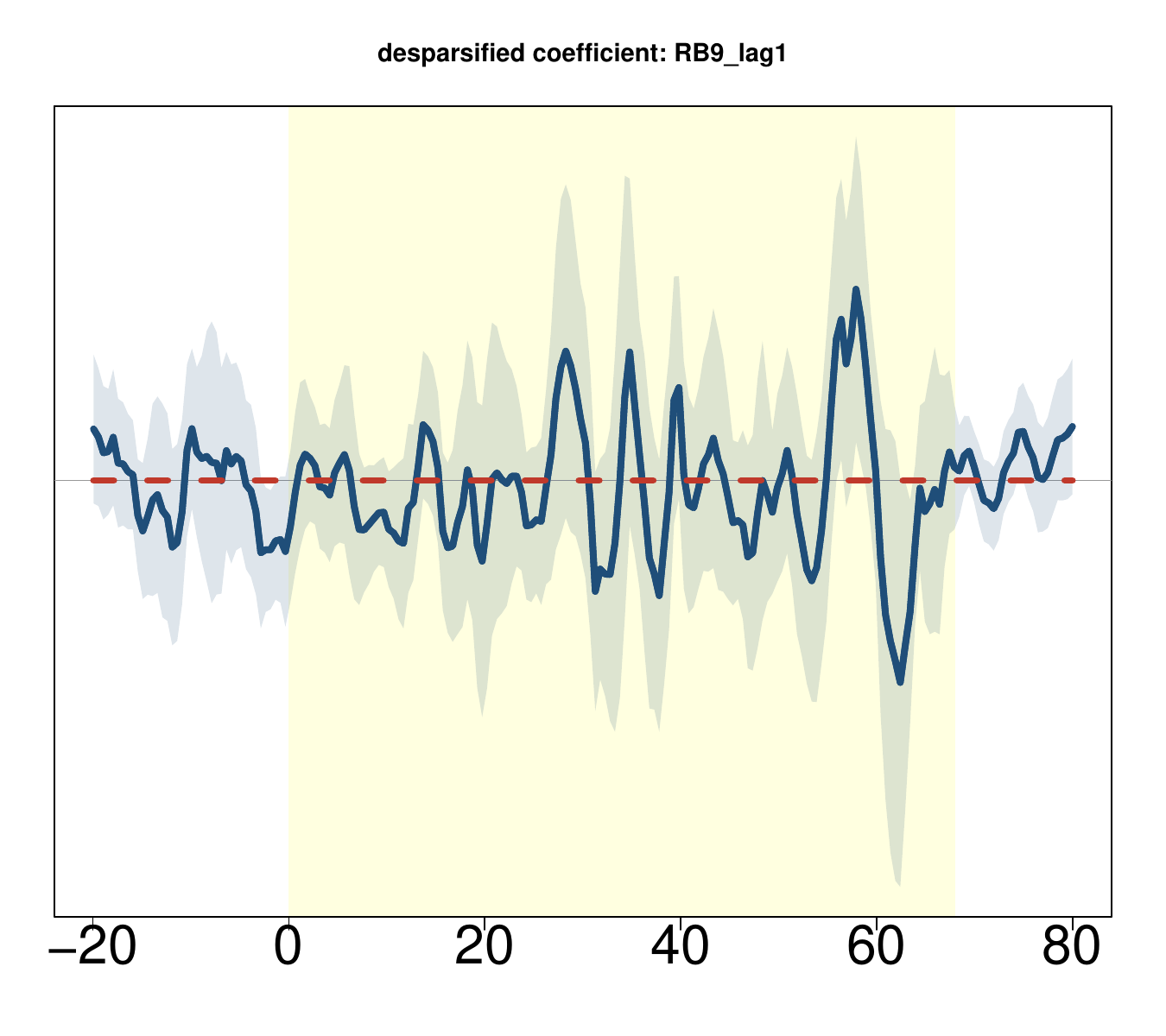}
        \caption{RB9 (globally sparse)}
    \end{subfigure}
    \caption{
    Desparsified estimates (solid) with 95\% confidence intervals (shaded), and local lasso estimates (dotted), for three lagged contacts. The yellow band marks the seizure. 
    }
    \label{fig: eeg desparsified coeff}
\end{figure}

The dependence of LA3 on the other contacts strengthens during the seizure, is concentrated around the SOZ but extends to other contacts, and fades after the seizure ends (Figure~\ref{fig. eeg brain plot}).
Figure~\ref{fig: eeg desparsified coeff} illustrates three forms of local sparsity.
The autoregressive coefficient of LA3 is active over the whole period and varies strongly in magnitude.
The coefficient of the neighboring contact LA2 is active during the first half of the seizure and near its end, and close to zero in between. 
The coefficient of RB9 is never selected by the \llasso, and the intervals of its desparsified estimate contain zero at most time points.

\section{Discussion}
\label{sec. discussion}

In this paper, building on \epsdelta-stationarity and local sparsity, we developed estimation and inference methods for high-dimensional time series whose dependence structure changes over time. We conclude with several natural extensions.
Our confidence intervals are pointwise in time.
Simultaneous confidence bands would require the distribution of the maximum of the estimator over time, which is usually obtained by Gaussian approximation \citep{zhou2010simultaneous,karmakar2022simultaneous}, and these results remain to be extended to \epsdelta-stationary processes.
The \llasso\ also indicates the periods during which each coefficient is zero.
Showing that these periods are consistently estimated is delicate near the times at which a coefficient becomes active, where it is close to zero by continuity.
Future work could also consider local polynomial versions of the \llasso, which would estimate the derivatives of the coefficients, at the price of differentiability conditions on the coefficients and on the stationary approximation \citep{dahlhaus2019towards}.
Finally, the \llasso\ could be applied equation by equation to time-varying vector autoregressions and graphical models.
The main step would then be to establish \epsdelta-stationarity for these processes, as Theorem~\ref{th. local stationary} does for the tv-ARX model.

\section*{Supplementary Materials}

The supplementary materials contain the proofs of the theoretical results and the technical lemmas supporting them, the implementation details of the estimators, and further details on the simulations and the data analyses.

\if1\anon{

\section*{Acknowledgements} J.B. and O.Y.C.~acknowledge support from the Swiss National Science Foundation (grants 3200-0-239967 and CR00I5-235987). This work was also supported as part of the Swiss AI Initiative by a grant from the Swiss National Supercomputing Centre (CSCS) under project ID 130 on Alps. J.L.~acknowledges support from the German Research Foundation (DFG) under project numbers 573018111 and 520388526 (TRR391 projects A02 and A05). D.L.V.~acknowledges support from the Swiss National Science Foundation, grant number CR00-5L-23981.
}
\fi

  \putbib[Lasso]
\end{bibunit}

{ 
\begin{center}
\bf \Large
Supplementary Material for \\
Local Inference for \\
High-dimensional Time Series
\end{center}
}

\counterwithin*{lemma}{section}         
\renewcommand{\thelemma}{\thesection\arabic{lemma}}      
\renewcommand{\theHlemma}{\thesection.\arabic{lemma}}

\begin{bibunit}

\begin{appendices}

\numberwithin{equation}{section}
\renewcommand{\theequation}{\thesection\arabic{equation}}

Appendix~\ref{app. proof of th 1} proves Theorem~\ref{th. local stationary}.
Appendix~\ref{app. technical lemmas} introduces additional notation and the technical lemmas used for the proofs of Section~\ref{sec. theory}.
Appendix~\ref{app. proof error bounds} contains the proofs of Theorem~\ref{th. error bound} and Corollary~\ref{cor. bandwidth}.
Appendix~\ref{app. proofs asymptotics} provides the auxiliary lemmas and the proofs of Proposition~\ref{prop. projection errors}, Theorem~\ref{th. linear representation} and Corollary~\ref{cor. asymptotic normality}.
Appendix~\ref{app. proofs of lemmas} contains the proofs of the lemmas of Appendices~\ref{app. technical lemmas} and~\ref{app. proofs asymptotics}.
Appendix~\ref{appendix. simulations Competing methods} describes the implementation details.

\section{Proof of Theorem \ref{th. local stationary}}
\label{app. proof of th 1}

Write $\Delta_{t,n}(u):=|t/n-u|+\lagorder/n$ and denote
$\Vert W \Vert_{L^\rho}:= (\E|W|^\rho)^{1/\rho}$ the $L_\rho$-norm for some random variable $W$.

By Conditions~\ref{cond. tvarx Lipschitz}--\ref{cond. geometric decay}, $\tilde Y\tn(u)$ is well defined in $L^\rho$ as a moving average of the jointly stationary process $(\tilde{\bm Z}\tn(u),\tilde\varepsilon_t(u))$, and is therefore strictly stationary, as is $\tilde{\bm X}\tn(u)$.
The proof then follows in six steps.

\medskip
\noindent\textbf{Step 1: Uniform $L^\rho$ bounds.}
By Minkowski's inequality, we have
$$
\normp[L^\rho]{Y\tn}
\le \sum_{j=1}^{\lagorder} \abs{\alpha_j^0(t/n)} \normp[L^\rho]{Y_{t-j,n} }
+ \sum_{k=1}^{\exodim}\abs{\exocoeff_k^0(t/n)} \normp[L^\rho]{Z\tn^{(k)}}
+\normp[L^\rho]{\varepsilon\tn}.
$$
Using Conditions \ref{cond. tvarx Lipschitz}, \ref{cond. l1 stability}, \ref{cond. uniform bounds} and taking the supremum on both sides yields:
\begin{align*}
\sup_t \|Y_{t,n}\|_{L^\rho}
\le
(1-\iota)\,M_{y,n}
+C_\exocoeff \bar s_\exocoeff\,M_{z}+M_{\varepsilon},
\end{align*}
where we let $M_{y,n}:=\sup_{t\leq n}\|Y_{t,n}\|_{L^\rho} \vee
\sup_{u\in [0,1] }\| \tilde{Y}\tn(u)\|_{L^\rho}
$.
Similarly, we have
$$
\sup_{u\in [0,1] } \| \tilde{Y}\tn(u)\|_{L^\rho}
\le
(1-\iota)\,M_{y,n}
+C_\exocoeff\,\bar s_\exocoeff\,M_{z}+M_{\varepsilon}.
$$
By induction, one can easily show that $M_{y,n} < \infty$ for every $n$.
We thus conclude that
\begin{align}\label{eq:moment-y}
M_{y,n} &\le  (1-\iota)\,M_{y,n} +C_\exocoeff \bar s_\exocoeff\,M_{z}+M_{\varepsilon}
\nonumber \\
&\le\iota^{-1}\bigl(C_{\exocoeff}\bar s_\exocoeff\,M_{z}
+M_{\varepsilon}\bigr).
\end{align}

\medskip
\noindent\textbf{Step 2: Scalar difference equation.}

Define $e_{t,n}(u):=Y_{t,n}-\tilde Y\tn(u)$.
By subtracting the stationary approximation, we obtain:
\begin{equation}\label{eq. proofth1 difference recursion}
e_{t,n}(u)
=
\sum_{j=1}^{\lagorder}\alpha_{j}^0(u)\,e_{t-j,n}(u)
+R_{t,n}(u),
\end{equation}
where
\begin{align*}
R_{t,n}(u)
&:=
\underbrace{
\sum_{j=1}^{\lagorder}
\bigl(\alpha_{j}^0(t/n)-\alpha_{j}^0(u)\bigr)\,Y_{t-j,n}
}_{\text{(I)}}
+ \underbrace{
\sum_{k=1}^{\exodim}
\bigl(\exocoeff_{k}^0(t/n)-\exocoeff_{k}^0(u)\bigr)\,Z_{t,n}^{(k)}
}_{\text{(II)}}  \\
&+
\underbrace{
\sum_{k=1}^{\exodim}\exocoeff_{k}^0(u)
\bigl(Z_{t,n}^{(k)}-\tilde Z_{t,n}^{(k)}(u)\bigr)
}_{\text{(III)}}
+
\underbrace{
\bigl(\varepsilon_{t,n}-\tilde\varepsilon\tn(u)\bigr)
}_{\text{(IV)}}.
\end{align*}

\medskip
\noindent\textbf{Step 3: Bounding $\normp[L^\rho]{R\tn(u)}$.}

We first bound $\normp[L^\rho]{R\tn(u)}$ for the case where $\abs{t/n -u }\le \delta'$:

\emph{Term~(I).} By Condition \ref{cond. tvarx Lipschitz} and~\eqref{eq:moment-y}:
$
\| \text{(I)} \|_{L^\rho}
\le \bar s_\alpha\,L_\alpha\,M_{y,n} |t/n-u|.
$

\emph{Term~(II).}
By Condition \ref{cond. tvarx Lipschitz},
$\|\text{(II)}\|_{L^\rho}
\le\bar s_\exocoeff\,L_\exocoeff\,M_{z}\,|t/n-u|$.

\emph{Term~(III).}
By Conditions \ref{cond. tvarx Lipschitz} and \ref{cond. tvarx local stationarity},
$\|\text{(III)}\|_{L^\rho}
\le C C_\exocoeff \bar s_\exocoeff\,\epsilon_n^z (|t/n-u|+\lagorder^z/n)$.

\emph{Term~(IV).}
By Condition \ref{cond. tvarx local stationarity},
$\|\text{(IV)}\|_{L^\rho}
\le C \rateErr\, (|t/n-u|+1/n) $.

\medskip\noindent For Term (I), we used the local sparsity of coefficients, which implies that, if $j\notin S^\alpha_{\delta'}(u)$,
then $\alpha_j(t/n) = \alpha_j(u) = 0$ for all $\abs{t/n-u}\le \delta'$.
We do the same for Term (II).
By combining all the above, we get, whenever $ \abs{t/n-u}\le\delta'$ and
$\lagorder^z \leq C \lagorder$:
\begin{align}\label{eq: proofth1 R bound 1}
\|R_{t,n}(u)\|_{L^\rho}
&\le C \left( \bar s_\alpha\,L_\alpha\,M_{y,n}+
\bar s_\exocoeff\, L_\exocoeff\,M_{z}
+C_\exocoeff \bar s_\exocoeff\,\epsilon_n^z
+\rateErr\right)
\Delta_{t,n}(u) \nonumber \\
&\le C \pi_n \Delta_{t,n}(u)
\end{align}
where, by \eqref{eq:moment-y}, we let
$\pi_n :=
\bar s_\alpha  \bar s_\exocoeff\,
+ \bar s_\exocoeff\,\epsilon_n^z
+\rateErr$.

We also use the following crude bounds, valid for all $t\le n$:

\emph{Term~(I).} By Condition \ref{cond. l1 stability} and~\eqref{eq:moment-y}:
$
\| \text{(I)} \|_{L^\rho}
\le 2(1-\iota) M_{y,n}.
$

\emph{Term~(II).}
By Conditions \ref{cond. tvarx Lipschitz} and \ref{cond. uniform bounds},
$\|\text{(II)}\|_{L^\rho}
\le 2 \bar s_\exocoeff  C_\exocoeff  M_{z} $.

\emph{Term~(III).}
By Conditions \ref{cond. tvarx Lipschitz} and \ref{cond. uniform bounds},
$\|\text{(III)}\|_{L^\rho}
\le 2 C_\exocoeff \bar s_\exocoeff M_{z}$.

\emph{Term~(IV).}
By Condition \ref{cond. uniform bounds},
$\|\text{(IV)}\|_{L^\rho}
\le 2 M_\varepsilon $.

By \eqref{eq:moment-y}, we obtain the following bound:
\begin{align}\label{eq: proofth1 R bound 2}
\|R_{t,n}(u)\|_{L^\rho}
&\le
2 (1-\iota) M_{y,n} +
4 C_\exocoeff \bar s_\exocoeff M_{z} + 2 M_\varepsilon \nonumber \\
&\le
C \left( \bar s_\exocoeff M_{z} + M_\varepsilon \right).
\end{align}

As $\bar s_\alpha,\bar s_\exocoeff\ge1$, we have $\bar s_\exocoeff M_{z}+M_\varepsilon\le(M_z+M_\varepsilon)\,\pi_n$.

\medskip
\noindent\textbf{Step 4: $MA(\infty)$ representation.}

As the recursion \eqref{eq. proofth1 difference recursion} holds for all $t\leq n$, $e\tn(u)$ is bounded in $L^\rho$ by Step~1, and $1-\sum_{j}\alpha_j^0(u)z^j$ has no root in $\{|z|\le1\}$ by Condition~\ref{cond. geometric decay}, $e\tn(u)$ is the unique bounded solution of this recursion and admits the representation
$$
e_{s,n}(u)=\sum_{\ell=0}^\infty
\psi_{\ell,n}(u) R_{s-\ell,n}(u).
$$
To prove the theorem, we need to bound $\|e_{t-j,n}(u)\|_{L^\rho}$ for $j=1,\dots, \lagorder$ and $\abs{t/n-u}\le \delta$.
Minkowski's inequality gives
\[
\|e_{t-j,n}(u)\|_{L^\rho}
\le
\sum_{\ell=0}^\infty
|\psi_{\ell,n}(u)|\, \normp[L^\rho]{R_{t-j-\ell,n}(u)}.
\]
We then split the sum in two parts:
$$
\|e_{t-j,n}(u)\|_{L^\rho}
\le
\sum_{\ell=0}^{L_{t,n}}
|\psi_{\ell,n}(u)|\, \normp[L^\rho]{R_{t-j-\ell,n}(u)}
+
\sum_{\ell=L_{t,n}+1}^\infty
|\psi_{\ell,n}(u)|\, \normp[L^\rho]{R_{t-j-\ell,n}(u)}
$$
where we define $L_{t,n}:= \max\{\ell \in \Z: \abs{(t-\lagorder-\ell)/n-u} \le \delta' \}$.
We have thus $L_{t,n} = \lfloor t-\lagorder-nu+n\delta'\rfloor$.
For $0\le\ell\le L_{t,n}$ and $1\le j\le\lagorder$, we have $u-\local'\le(t-j-\ell)/n\le t/n\le u+\local'$, so that \eqref{eq: proofth1 R bound 1} applies to $R_{t-j-\ell,n}(u)$, whereas \eqref{eq: proofth1 R bound 2} applies for $\ell>L_{t,n}$.
As $|t/n-u|\le\local$, we have $L_{t,n}+1\ge n(\local'-\local)-\lagorder$.

We then apply the facts that
$|\psi_{\ell,n}(u)| \le C_\psi \kappa^\ell$ and $\Delta_{t-j-\ell,n}(u)\le\Delta_{t,n}(u)+(j+\ell)/n$:
\begin{equation}\label{eq. proofth1 y diff sum split}
\|e_{t-j,n}(u)\|_{L^\rho}
\le
C\underbrace{\sum_{\ell=0}^{L_{t,n}}
C_\psi \kappa^\ell \, \pi_n \left(
\Delta_{t,n}(u) + \frac{\ell +j}{n} \right)
}_{A}
+ \;
C  \underbrace{\sum_{\ell=L_{t,n}+1}^\infty
C_\psi \kappa^\ell \, \left( \bar s_\exocoeff M_{z} + M_\varepsilon \right) }_{B},
\end{equation}
where $\pi_n$ is obtained in \eqref{eq: proofth1 R bound 1}.
We have:
$$ A= \pi_nC_\psi \left( \frac{1-\kappa^{L_{t,n}+1}}{1-\kappa} \left(\Delta_{t,n}(u) +\frac{j}{n} \right) +
\kappa\frac{1-(L_{t,n}+1)\kappa^{L_{t,n}}+L_{t,n} \kappa^{L_{t,n}+1} }{(1-\kappa)^2} \frac{1}{n} \right). $$
For the second term, we have simply:
\begin{align*}
B \le C_\psi \left( \bar s_\exocoeff M_{z} + M_\varepsilon \right) \frac{\kappa^{L_{t,n}+1}}{1-\kappa} .
\end{align*}

\medskip
\noindent\textbf{Step 5: Bounding $A$ and $B$.}

As $\bar s_\exocoeff M_{z}+M_\varepsilon\le(M_z+M_\varepsilon)\pi_n$, we have $B\le C\pi_n\kappa^{L_{t,n}+1}/(1-\kappa)$.
As $L_{t,n}+1\ge n(\local'-\local)-\lagorder$ and $\log(1/\kappa)\ge1-\kappa$, we have $\kappa^{L_{t,n}+1}\le(1-\kappa)\lagorder/n$ as soon as
$$
n(\local'-\local)\;\ge\;\lagorder+\frac{1}{1-\kappa}\log\Bigl(\frac{n}{(1-\kappa)\lagorder}\Bigr),
$$
which is implied by the condition of Theorem~\ref{th. local stationary}.
Under this condition, $B\le C \pi_n\lagorder/n\le C\pi_n\Delta_{t,n}(u)$.

For $A$, we use $j/n\le\lagorder/n\le\Delta_{t,n}(u)$, $1/n\le\Delta_{t,n}(u)$ and $1-(L\tn+1)\kappa^{L\tn}+L\tn\kappa^{L\tn+1}\le1$:
$$
A\le C_\psi\pi_n\Delta_{t,n}(u)\Bigl(\frac{2}{1-\kappa}+\frac{\kappa}{(1-\kappa)^2}\Bigr)\le\frac{2C_\psi\pi_n}{(1-\kappa)^2}\,\Delta_{t,n}(u).
$$
Together with the bound on $B$ and \eqref{eq. proofth1 y diff sum split}, this gives, uniformly in $1\le j\le\lagorder$,
$$
\|e_{t-j,n}(u)\|_{L^\rho}\le C\pi_n\Delta_{t,n}(u),
$$
where $C$ depends only on $C_\psi$, $\kappa$, $\iota$, $L_\alpha$, $L_\exocoeff$, $C_\exocoeff$, $M_z$ and $M_\varepsilon$.

\medskip
\noindent\textbf{Step 6: Bounding the covariates.}

Taking $L^\rho$~norms and applying Minkowski's inequality,
\[
\Bigl(\E\|\bm X_{t,n}-\tilde{\bm X}\tn(u)\|_\infty^\rho
\Bigr)^{1/\rho}
\le
\underbrace{
\Bigl(\E\max_{1\le j\le \lagorder}|e_{t-j,n}(u)|^\rho\Bigr)^{1/\rho}
}_{\displaystyle=:\,T_Y}
\;+\;
\underbrace{
\Bigl(\E\|\bm Z_{t,n}-\tilde{\bm Z}_{t,n}(u)
\|_\infty^\rho\Bigr)^{1/\rho}
}_{\displaystyle=:\,T_Z}.
\]

\emph{Bounding $T_Y$.}
As $\max_j|e_{t-j,n}(u)|^\rho\le\sum_j|e_{t-j,n}(u)|^\rho$, Step~5 gives
$$
T_Y\le\lagorder^{1/\rho}\max_{1\le j\le\lagorder}\|e_{t-j,n}(u)\|_{L^\rho}\le C\,\lagorder^{1/\rho}\,\pi_n\,\Delta_{t,n}(u).
$$

\emph{Bounding $T_Z$.}
By Condition~\ref{cond. tvarx local stationarity} and $\lagorder^z\le C\lagorder$,
$T_Z\le C\epsilon_n^z\,\Delta_{t,n}(u)$.

\medskip\noindent
We thus conclude that, for all $t$ with $|t/n-u|\le\local$,
\[
\Bigl(\E\|\bm X_{t,n}-\tilde{\bm X}\tn(u)\|_\infty^\rho\Bigr)^{1/\rho}
\le C\bigl(\lagorder^{1/\rho}\pi_n+\epsilon_n^z\bigr)\Delta_{t,n}(u)
\le 2C\,\epsilon_n\,\Delta_{t,n}(u),
\]
as $\epsilon_n^z\le\bar s_\exocoeff\epsilon_n^z\le\pi_n\le\lagorder^{1/\rho}\pi_n=\epsilon_n$, where $C$ does not depend on $n$, $t$ and $u$.
Together with the strict stationarity of $\tilde{\bm X}\tn(u)$ and the rate condition $\epsilon_n\lagorder/n\to0$, this shows that $\bm X\tn$ is \epsdelta-stationary.

\section{Notation and Technical Lemmas for Section \ref{sec. theory}}
\label{app. technical lemmas}

This appendix introduces notation and the technical lemmas used in the proofs of Sections~\ref{sec. error bounds} and~\ref{sec. asymptotic analysis}. The proofs of these lemmas are in Appendix~\ref{app. proofs of lemmas}.

\subsection{Notation}

Throughout the appendices, we write $p:=p_n$, and $C$ denotes a positive constant that does not depend on $n$, $t$ and $u$, and whose value may change from line to line.
For $\rho >0$, we let $\Vert W \Vert_{L^\rho}:= (\E|W|^\rho)^{1/\rho}$.
For a vector $\bm b$, $\|\bm b\|_\infty:=\max_j|b_j|$. For a matrix $\bm A$, $\|\bm A\|_{\max}:=\max_{j,k}|A_{j,k}|$, and $\|\bm A\|$ denotes the spectral norm.
For a set $S \subseteq \{1,\dots,p\}$ and a vector $\Beta \in \R^p$,
we define $\Beta_S$ by
$\beta_{S,j}:= \beta_j \mathbbm{1}\{j\in S\}$ for all $j=1,\dots,p.$
Similarly, for a set-valued function $S:[0,1]\rightarrow 2^{\{1,\dots,p\}}$ and a vector-valued function $\Beta:[0,1]\rightarrow \R^p$,
we define $\Beta_S(u)$ by
$$
\beta_{S,j}(u):= \beta_j(u) \mathbbm{1}\{j\in S(u)\}, \quad j=1,\dots,p.
$$

In all the proofs, we denote the approximation bias by
\begin{equation}\label{eq. approximation bias definition}
B_n^2:= \meant \kernel \left( \X^\top (\Beta^0(t/n) - \Beta^0(u)) \right)^2.
\end{equation}

Since $K$ is Lipschitz, supported on $[-1/2,1/2]$ with $\int K=1$, a Riemann-sum argument gives, whenever $h\le 2\big(u\wedge(1-u)\big)$ and $nh\ge1$,
\begin{equation}\label{eq. riemann}
\Big|\meant\kernel-1\Big|\le\frac{C}{nh}.
\end{equation}

\subsection{Technical Lemmas}

\begin{lemma}[Triplex inequality]\label{lem. tech. triplex inequality}
Suppose Assumption~\ref{as. kernel} holds and $nh\ge1$.
Let $\{\chi_t\}$ be adapted to an increasing sequence of $\sigma$-fields $\{\mathcal F_t\}$, $L_r$-bounded for some $r>1$, and an $L_{\tilde r}$-mixingale of size $-\nedsize$ with respect to $\{\mathcal F_t\}$ for some $\tilde r\ge1$, with mixingale constants bounded by $C$ \citep{davidson2021stochastic}.
Let $\{w_t\}$ be deterministic with $|w_t|\le1$.
Then, for all $\xi>0$, all integers $q\ge1$ and all $\alpha_n>0$,
\[
\p\Big(\Big|\sumt \kernel w_t(\chi_t-\E\chi_t)\Big|>nh\,\xi\Big)
\le 2q\exp\Big(-\frac{nh\,\xi^2}{C q^2\alpha_n^2}\Big)+C\frac{q^{-\nedsize}}{\xi}+C\frac{\alpha_n^{1-r}}{\xi}.
\]
\end{lemma}

\begin{lemma}[Products of stationary approximations]\label{lem. tech. eps X mixingale}
Let $\tilde{\bm W}\tn(u)\in\R^{q}$ satisfy Assumption~\ref{as. NED}, with $p_n+1$ replaced by $q$.
Then, for all $j,k\le q$, the process $\chi\tn:=\tilde W\tn\pj(u)\,\tilde W\tn^{(k)}(u)$ satisfies:
\begin{enumerate}[label=(\roman*)]
\item $\E|\chi\tn|^{\bar\order}\le C$;
\item $\{\chi\tn\}$ is $L_{\order}$-NED of size $-\nedsize$ on $\{\bm U\tn(u)\}$;
\item $\{\chi\tn-\E\chi\tn\}$ is an $L_{\order}$-mixingale of size $-\nedsize$ with respect to $\{\mathcal F^U\tn(u)\}$. Since $\nedsize\ge1$, its mixingale sequence is summable.
\end{enumerate}
All constants are bounded uniformly in $j$, $k$, $t$, $u$ and $n$.
\end{lemma}

\begin{lemma}[Transfer to the stationary approximation]\label{lem. tech. transfer}
Suppose Assumption~\ref{as. kernel} holds.
Let $\bm A_{t,n}\in\R^{p_A}$ and $\bm B_{t,n}\in\R^{p_B}$ have stationary approximations $\tilde{\bm A}_{t,n}(u)$ and $\tilde{\bm B}_{t,n}(u)$ with lag orders $\lagorder^a$ and $\lagorder^b$ such that, for all $t$ with $|t/n-u|\le\local$,
\begin{enumerate}[label=(\roman*)]
\item $\E[\|\bm A_{t,n}-\tilde{\bm A}_{t,n}(u)\|_\infty^2]^{1/2}\le C\rate^a(|t/n-u|+\lagorder^a/n)$,
\item $\E[\|\bm B_{t,n}-\tilde{\bm B}_{t,n}(u)\|_\infty^2]^{1/2}\le C\rate^b(|t/n-u|+\lagorder^b/n)$,
\item $\max_{j}\E|\tilde A_{t,n}^{(j)}(u)|^{2\bar\order}\le C$ and $\max_{k}\E|\tilde B_{t,n}^{(k)}(u)|^{2\bar\order}\le C$.
\end{enumerate}
Then, for $h\le 2\local$, $nh\ge1$ and $\lagorder^a\vee\lagorder^b\le Cnh$,
\[
\E\Big[\max_{j,k}\Big|\meant\kernel\big(A_{t,n}^{(j)}B_{t,n}^{(k)}-\tilde A_{t,n}^{(j)}(u)\tilde B_{t,n}^{(k)}(u)\big)\Big|\Big]
\le C h\big(\rate^a p_B^{1/(2\bar\order)}+\rate^b p_A^{1/(2\bar\order)}+h\,\rate^a\rate^b\big).
\]
\end{lemma}

\begin{lemma}[Empirical process]\label{lem. tech. score}
Suppose Assumption~\ref{as. kernel} holds.
For some nonnegative sequences $\rate$ and $\zeta_n$, let $\bm A_{t,n}\in\R^{p_A}$ and $e_{t,n}\in\R$ be $(\rate,\local)$- and $(\zeta_n,\local)$-stationary with lag orders $\lagorder^a$ and $\lagorder^e$, respectively.
That is, for some $\rho\ge2$ and all $t$ with $|t/n-u|\le\local$,
\begin{enumerate}[label=(\roman*)]
\item $\E[\|\bm A_{t,n}-\tilde{\bm A}_{t,n}(u)\|_\infty^\rho]^{1/\rho}\le C\rate\,(|t/n-u|+\lagorder^a/n)$,
\item $\E[|e_{t,n}-\tilde e_{t,n}(u)|^\rho]^{1/\rho}\le C\zeta_n\,(|t/n-u|+\lagorder^e/n)$.
\end{enumerate}
Suppose further that $\lagorder^a\vee\lagorder^e\le Cnh$, that $\E[\tilde{\bm A}_{t,n}(u)\,\tilde e_{t,n}(u)]=\bm 0$,
and that Assumption~\ref{as. NED} holds with $\tilde{\bm W}_{t,n}(u) =(\tilde{\bm A}_{t,n}(u)^\top,\tilde e_{t,n}(u))^\top$.
Then, for $h\le 2\local$, $nh\ge1$ and all $\xi>0$,
\[
\p\Big(\max_{1\le k\le p_A}\Big|\meant\kernel A^{(k)}_{t,n}\,e_{t,n}\Big|>\xi\Big)
\le \frac{C h\big(\rate+\zeta_n\, p_A^{1/(2\bar m)}+h\,\rate\,\zeta_n\big)}{\xi}
+\frac{C\,p_A}{\xi^{\order}(nh)^{\order/2}} .
\]
\end{lemma}

\begin{lemma}[Covariance event]\label{lem. tech. covariance}
Suppose Assumption~\ref{as. kernel} holds.
For some nonnegative sequence $\rate$, let $\bm A_{t,n}\in\R^{q}$ be $(\rate,\local)$-stationary with respect to $\|\cdot\|_\infty$, with lag order at most $Cnh$ and stationary approximation $\tilde{\bm A}_{t,n}(u)$ satisfying Assumption~\ref{as. NED}, with $\tilde{\bm W}_{t,n}(u)$ replaced by $\tilde{\bm A}_{t,n}(u)$.
Let $\{w^{(j,k)}_t\}$ be deterministic weights with $|w^{(j,k)}_t|\le 1$, define $\bm\Sigma^A(u):=\E[\tilde{\bm A}_{t,n}(u)\tilde{\bm A}_{t,n}(u)^\top]$, and let $\vanishing\in[(nh)^{-1/2},1]$.
Then, for $h\le 2\local$, $n$ large enough, and all
\[
\xi \ge \xi_n:=\frac{C}{\vanishing^{\frac{\nedsize+\order-1}{\nedsize\order+\order-1}}}
\left(\frac{q^{\frac{2}{\nedsize}+\frac{2}{\order-1}}\sqrt{\log(q\vee\sqrt{nh})}}{\sqrt{nh}}\right)^{\frac{1}{\frac{1}{\nedsize}+\frac{\order}{\order-1}}},
\]
we have
\[
\p\Big(\max_{j,k}\Big|\meant\kernel w^{(j,k)}_t\big(A^{(j)}_{t,n}A^{(k)}_{t,n}-\Sigma^A_{j,k}(u)\big)\Big|>\xi\Big)
\le C\left(\frac{h\big(\rate\,q^{1/(2\bar m)}+h\,\rate^2\big)}{\xi}+\vanishing\right).
\]
\end{lemma}

\section{Proofs for Section \ref{sec. error bounds}}
\label{app. proof error bounds}

\subsection{Events}

Recall that $\bm\Sigma(u)=\E[\Xu\Xu^\top]$, which does not depend on $t$ since $\Xu$ is stationary.
Let $S:=\localSet$, and let $a_{t,j}:=2\big(\beta^0_j(t/n)-\beta^0_j(u)\big)/(Lh)$ if $j\in S$ and $|t/n-u|\le h/2$, and $a_{t,j}:=0$ otherwise.
These weights are deterministic and, by Assumption~\ref{as. Lipschitz}, satisfy $|a_{t,j}|\le1$.
For $\kappa>0$, we define the events
\begin{align*}
\Fset&:=\Big\{\max_{1\le j\le p}\Big|\meant\kernel\varepsilon\tn\x[j]\Big|\le\frac{\lambda(u)}{4}\Big\},\\
\Sset&:=\Big\{\s\,\big\|\hat{\bm\Sigma}(u)-\bm\Sigma(u)\big\|_{\max}\le\kappa\Big\},\\
\mathcal A_\local(u)&:=\Big\{\s\max_{j,k\in S}\Big|\meant\kernel a_{t,j}a_{t,k}\big(\x[j]\x[k]-\Sigma_{j,k}(u)\big)\Big|\le\kappa\Big\}.
\end{align*}
The event $\Fset$ controls the empirical process, $\Sset$ transfers the compatibility condition from $\bm\Sigma(u)$ to $\hat{\bm\Sigma}(u)$, and $\mathcal A_\local(u)$ controls the smoothing bias.
In the proof of Theorem~\ref{th. error bound}, $\kappa:=\localComp^2/32$.

\begin{lemma}[Empirical process event]\label{lem. error. event F}
Suppose Assumptions~\ref{as. kernel}, \ref{as. local stationarity} and~\ref{as. NED} hold, $h\le2\local$, $nh\ge1$ and $\lagorder\le Cnh$. Then, for all $\lambda(u)>0$,
\[
\p\big(\Fset^c\big)\le\frac{Ch\big(\rate+\rateErr\,p^{1/(2\bar\order)}+h\,\rate\,\rateErr\big)}{\lambda(u)}+\frac{C\,p}{\lambda(u)^{\order}(nh)^{\order/2}} .
\]
\end{lemma}

\begin{lemma}[Covariance event]\label{lem. error. event S}
Suppose Assumptions~\ref{as. kernel}, \ref{as. local stationarity} and~\ref{as. NED} hold, $h\le2\big(\local\wedge u\wedge(1-u)\big)$ and $\lagorder\le Cnh$.
Let $\kappa>0$, and suppose that Assumption~\ref{as. bound on s} holds with a sufficiently small constant, depending on $\kappa$, and that $nh\ge C_0\,\s/\kappa$ for a constant $C_0$ depending only on $K$ and $\max_j\Sigma_{j,j}(u)$.
Then, for $n$ large enough as in Lemma~\ref{lem. tech. covariance},
\[
\p\Big(\s\,\|\hat{\bm\Sigma}(u)-\bm\Sigma(u)\|_{\max}>\kappa\Big)
\le C\Big(\frac{\s\,h\big(\rate\,p^{1/(2\bar\order)}+h\,\rate^2\big)}{\kappa}+\vanishing\Big).
\]
\end{lemma}

\begin{lemma}[Bias event]\label{lem. error. event A}
Suppose Assumptions~\ref{as. Lipschitz}, \ref{as. kernel}, \ref{as. local stationarity} and~\ref{as. NED} hold, $h\le2\local$ and $\lagorder\le Cnh$.
Let $\kappa>0$, and suppose that Assumption~\ref{as. bound on s} holds with a sufficiently small constant, depending on $\kappa$.
Then, for $n$ large enough as in Lemma~\ref{lem. tech. covariance},
\[
\p\big(\mathcal A_\local(u)^c\big)\le C\Big(\frac{\s\,h\big(\rate\,\s^{1/(2\bar\order)}+h\,\rate^2\big)}{\kappa}+\vanishing\Big).
\]
\end{lemma}

\subsection{Proof of Theorem \ref{th. error bound}}
\label{proof. theorem error bound}

Throughout the proof, write $S:=\localSet$, $\bm\Delta:=\BetahatU-\BetazeroU$, $\kappa:=\localComp^2/32$,
\[
\hat{\bm\Sigma}(u):=\meant\kernel\X\X^\top,
\qquad
P_n:=\meant\kernel\big(\X^\top(\BetahatU-\Beta^0(t/n))\big)^2,
\]
and let $B_n^2$ be the approximation bias \eqref{eq. approximation bias definition}.
Step~1 shows that the bounds of Theorem~\ref{th. error bound} hold on $\Fset\cap\Sset\cap\mathcal A_\local(u)$, and Step~2 shows that this event has probability at least $1-C\vanishing$ for $n$ large enough.
Since $\localComp\ge1/C$, powers of $1/\localComp$ are absorbed into the constants $C$.

\medskip
\noindent\textbf{Step 1: deterministic part.}
By the definition of $\local$-local sparsity, $\beta^0_j(u)=0$ for $j\notin S$, so that $\|\BetazeroU\|_1=\|\Beta^0_S(u)\|_1$ and $\bm\Delta_{S^c}=\hat\Beta_{S^c}(u)$.

\noindent\textit{Basic inequality.}
Since $\BetahatU$ minimizes \eqref{eq. min problem},
\begin{align*}
&\meant\kernel(Y\tn-\X^\top\BetahatU)^2+2\lambda(u)\|\BetahatU\|_1\\
&\qquad\le\meant\kernel(Y\tn-\X^\top\BetazeroU)^2+2\lambda(u)\|\BetazeroU\|_1 .
\end{align*}
Substituting $Y\tn=\X^\top\Beta^0(t/n)+\varepsilon\tn$ and expanding both squares, the terms $\meant\kernel\varepsilon\tn^2$ cancel, and
\[
P_n+2\lambda(u)\|\BetahatU\|_1\le\frac{2}{nh}\sumt\kernel\varepsilon\tn\X^\top\bm\Delta+2\lambda(u)\|\BetazeroU\|_1+B_n^2 .
\]

\noindent\textit{Cone condition.}
By H\"older's inequality, on $\Fset$,
\[
\Big|\frac{2}{nh}\sumt\kernel\varepsilon\tn\X^\top\bm\Delta\Big|
\le 2\max_{j}\Big|\meant\kernel\varepsilon\tn\x[j]\Big|\,\|\bm\Delta\|_1
\le\frac{\lambda(u)}{2}\|\bm\Delta\|_1 .
\]
Using $\|\BetahatU\|_1\ge\|\Beta^0_S(u)\|_1-\|\bm\Delta_S\|_1+\|\bm\Delta_{S^c}\|_1$ and $\|\bm\Delta\|_1=\|\bm\Delta_S\|_1+\|\bm\Delta_{S^c}\|_1$, and multiplying by two, we obtain
\begin{equation}\label{eq. cone}
2P_n+3\lambda(u)\|\bm\Delta_{S^c}\|_1\le5\lambda(u)\|\bm\Delta_S\|_1+2B_n^2 .
\end{equation}

\noindent\textit{Case 1: $\lambda(u)\|\bm\Delta_S\|_1\le2B_n^2$.}
Then \eqref{eq. cone} gives $2P_n+3\lambda(u)\|\bm\Delta_{S^c}\|_1\le12B_n^2$, hence
$P_n+\lambda(u)\|\bm\Delta\|_1\le 6B_n^2+4B_n^2+2B_n^2=12B_n^2$.

\noindent\textit{Case 2: $\lambda(u)\|\bm\Delta_S\|_1>2B_n^2$.}
Then \eqref{eq. cone} gives $\|\bm\Delta_{S^c}\|_1\le2\|\bm\Delta_S\|_1$, so $\bm\Delta$ lies in the cone of Assumption~\ref{as. compatibility condition}.
We first transfer the compatibility condition to $\hat{\bm\Sigma}(u)$.
For $\bm\beta$ with $\|\bm\beta_{S^c}\|_1\le3\|\bm\beta_S\|_1$, we have $\|\bm\beta\|_1\le4\|\bm\beta_S\|_1$, so that on $\Sset$
\[
\big|\bm\beta^\top(\hat{\bm\Sigma}(u)-\bm\Sigma(u))\bm\beta\big|
\le\|\hat{\bm\Sigma}(u)-\bm\Sigma(u)\|_{\max}\|\bm\beta\|_1^2
\le\frac{16\kappa}{\s}\|\bm\beta_S\|_1^2=\frac{\localComp^2}{2\s}\|\bm\beta_S\|_1^2 .
\]
Assumption~\ref{as. compatibility condition} then gives $\|\bm\beta_S\|_1^2\le\frac{\s}{\localComp^2}\bm\beta^\top\hat{\bm\Sigma}(u)\bm\beta+\frac12\|\bm\beta_S\|_1^2$, that is,
\[
\|\bm\beta_S\|_1^2\le\frac{2\s}{\localComp^2}\,\bm\beta^\top\hat{\bm\Sigma}(u)\bm\beta
\]
\citep[see also][Corollary~6.8]{buhlmann2011}.
Moreover, since $(a+b)^2\le2a^2+2b^2$,
\[
\bm\Delta^\top\hat{\bm\Sigma}(u)\bm\Delta=\meant\kernel\big(\X^\top\bm\Delta\big)^2\le2P_n+2B_n^2 .
\]
Adding $3\lambda(u)\|\bm\Delta_S\|_1$ to both sides of \eqref{eq. cone} and using $2xy\le x^2+y^2$,
\begin{align*}
2P_n+3\lambda(u)\|\bm\Delta\|_1
&\le8\lambda(u)\|\bm\Delta_S\|_1+2B_n^2
\le\frac{16\lambda(u)\sqrt{\s}}{\localComp}\sqrt{P_n+B_n^2}+2B_n^2\\
&\le\frac{64\lambda(u)^2\s}{\localComp^2}+P_n+3B_n^2 ,
\end{align*}
so that $P_n+3\lambda(u)\|\bm\Delta\|_1\le64\lambda(u)^2\s/\localComp^2+3B_n^2$.

Combining the two cases, on $\Fset\cap\Sset$,
\begin{equation}\label{eq. oracle deterministic}
P_n+\lambda(u)\|\BetahatU-\BetazeroU\|_1\le\frac{64\lambda(u)^2\s}{\localComp^2}+12B_n^2 .
\end{equation}

\noindent\textit{Approximation error.}
Let $\bm r_t:=\Beta^0(t/n)-\BetazeroU$.
Since $\kernel=0$ for $|t/n-u|>h/2$ and $h/2\le\local$, only indices with $|t/n-u|\le\local$ contribute.
For these, $\bm r_t$ is supported on $S$ by local sparsity, and $r_{t,j}=(Lh/2)\,a_{t,j}$.
Hence $B_n^2=I_n+II_n$, where
\begin{align*}
I_n  &:=\meant\kernel\,\bm r_t^\top\bm\Sigma(u)\,\bm r_t, \\
II_n &:=\frac{L^2h^2}{4}\sum_{j,k\in S}\meant\kernel\,a_{t,j}a_{t,k}\big(\x[j]\x[k]-\Sigma_{j,k}(u)\big).
\end{align*}
Since $\bm r_t^\top\bm\Sigma(u)\bm r_t\le\localEig\|\bm r_t\|_2^2\le\localEig\,\s L^2h^2/4$ and $\meant\kernel\le C$, we have $I_n\le C\localEig\,\s\,h^2$.
On $\mathcal A_\local(u)$, $|II_n|\le(L^2h^2/4)\,\s^2\,(\kappa/\s)=\kappa L^2\s h^2/4$.
Taking $\bm\beta=\bm 1_S$ in Assumption~\ref{as. compatibility condition} gives $\s^2\le\s\cdot\localEig\,\s/\localComp^2$, that is, $\localComp^2\le\localEig$, so that $\kappa\le\localEig$.
Hence $B_n^2\le C\localEig\,\s\,h^2$ on $\mathcal A_\local(u)$, and \eqref{eq. oracle deterministic} yields, on $\Fset\cap\Sset\cap\mathcal A_\local(u)$,
\begin{equation}\label{eq. oracle final}
P_n+\lambda(u)\|\BetahatU-\BetazeroU\|_1\le C\Big(\lambda(u)^2\s+\localEig\,\s\,h^2\Big).
\end{equation}
The bound on $P_n$ gives \eqref{eq. th3 error bound prediction}, and dividing \eqref{eq. oracle final} by $\lambda(u)$ gives \eqref{eq. th3 error bound beta}.

\medskip
\noindent\textbf{Step 2: probabilistic part.}
Replacing $\vanishing$ by $M\vanishing$ for a large constant $M$ weakens the conditions of the theorem, multiplies the probability bound by $M$, and keeps $M\vanishing\le1$ for $n$ large enough.
Since $\kappa\ge1/(32C)$, we may therefore assume that Assumption~\ref{as. bound on s} holds with the small constant required by Lemmas~\ref{lem. error. event S} and~\ref{lem. error. event A}.
Moreover, the exponent of $nh$ in Assumption~\ref{as. bound on s} is smaller than $1/2$, and $nh\to\infty$ since $(nh)^{-1/2}\le\vanishing\to0$. Hence $\s/(nh)\to0$, and $nh\ge C_0\s/\kappa$ for $n$ large enough.

\noindent\textit{Consequences of the conditions.}
By the lower bound on $\lambda(u)$,
\begin{equation}\label{eq. th2 lambda consequences}
\frac{h\big(\rate+\rateErr\,p^{1/(2\bar\order)}\big)}{\lambda(u)}\le C\vanishing
\qquad\text{and}\qquad
\frac{p}{\lambda(u)^{\order}(nh)^{\order/2}}\le C\vanishing .
\end{equation}
By the condition on $\rate$ and $\s\ge1$,
\begin{equation}\label{eq. th2 h rate}
h\rate\le\s\,h\,\rate\,p^{1/(2\bar\order)}\le C\vanishing .
\end{equation}

\noindent\textit{Empirical process event.}
By Lemma~\ref{lem. error. event F}, \eqref{eq. th2 lambda consequences} and \eqref{eq. th2 h rate},
\[
\p\big(\Fset^c\big)\le C\Big(\frac{h\big(\rate+\rateErr\,p^{1/(2\bar\order)}\big)}{\lambda(u)}+h\rate\cdot\frac{h\rateErr}{\lambda(u)}+\frac{p}{\lambda(u)^{\order}(nh)^{\order/2}}\Big)\le C\vanishing .
\]

\noindent\textit{Covariance and bias events.}
By Lemmas~\ref{lem. error. event S} and~\ref{lem. error. event A}, $\s\le p$, the condition on $\rate$ and \eqref{eq. th2 h rate},
\[
\p\big(\Sset^c\big)+\p\big(\mathcal A_\local(u)^c\big)\le C\Big(\s\,h\,\rate\,p^{1/(2\bar\order)}+(\s h\rate)(h\rate)+\vanishing\Big)\le C\vanishing .
\]

\noindent\textit{Conclusion.}
The union bound gives $\p\big(\Fset\cap\Sset\cap\mathcal A_\local(u)\big)\ge1-C\vanishing$, and Step~1 completes the proof.

\subsection{Proof of Corollary~\ref{cor. bandwidth}}

Throughout, $\local$ is fixed, so that $\s$ and $\localEig$ do not depend on $h$.
We write $s:=\s$, $\Lambda:=\localEig$, $\Lambda^*:=\effEig$, $\lambda:=\lambda(u)$, and
\[
a_n:=\vanishing^{-1/\order}\frac{p^{1/\order}}{\sqrt n},
\qquad
b_n:=\vanishing^{-1}\big(\rate+\rateErr\,p^{1/(2\bar\order)}\big),
\]
so that $\Lambda^*=\Lambda\vee b_n^2$ and $h^*=(a_n^2/\Lambda^*)^{1/3}$.
Let $C_0$ be the constant in the lower bound on $\lambda$ in Theorem~\ref{th. error bound}, so that this bound reads
\begin{equation}\label{eq. proof cor lambda}
\lambda\ge C_0\max\big\{a_nh^{-1/2},\ b_nh\big\}.
\end{equation}
Let
\[
F_1(h,\lambda):=\lambda s+\frac{\Lambda s h^2}{\lambda},
\qquad
F_2(h,\lambda):=\lambda^2 s+\Lambda s h^2
\]
denote the right-hand sides of \eqref{eq. th3 error bound beta} and \eqref{eq. th3 error bound prediction}, up to the constant $C$.

\medskip\noindent\textbf{Step 1: an identity.}
By definition of $h^*$, $a_n^2=\Lambda^*h^{*3}$, hence
\begin{equation}\label{eq. proof cor identity}
a_n\,h^{*-1/2}=\sqrt{\Lambda^*}\,h^* .
\end{equation}

\medskip\noindent\textbf{Step 2: the conditions of Theorem~\ref{th. error bound} hold at $(h^*,\lambda^*)$.}
By \eqref{eq. proof cor identity}, the first term in \eqref{eq. proof cor lambda} at $h=h^*$ equals $C_0\sqrt{\Lambda^*}\,h^*$.
Since $b_n^2\le\Lambda^*$, the second term is at most $C_0\sqrt{\Lambda^*}\,h^*$.
Hence $\lambda^*:=C_0\sqrt{\Lambda^*}\,h^*$ satisfies \eqref{eq. proof cor lambda}.

Next, since $\Lambda^*\ge b_n^2\ge\vanishing^{-2}\rate^2$, we have
\[
h^{*3}=\frac{a_n^2}{\Lambda^*}\le\frac{\vanishing^{2-2/\order}\,p^{2/\order}}{n\,\rate^2}.
\]
By the assumed bound on $\rate$, it follows that
\[
\big(p^{1/(2\bar\order)}\,s\,h^*\rate\big)^3
\le\frac{\vanishing^{2-2/\order}\,p^{2/\order+3/(2\bar\order)}\,s^3\,\rate}{n}
\le C\vanishing^3 ,
\]
that is, $p^{1/(2\bar\order)}\,s\,h^*\rate\le C\vanishing$.
The remaining conditions of Theorem~\ref{th. error bound} hold at $h=h^*$ by assumption.
Hence \eqref{eq. th3 error bound prediction} and \eqref{eq. th3 error bound beta} hold for $(h,\lambda)=(h^*,\lambda^*)$ with probability at least $1-C\vanishing$, for $n$ large enough.

\medskip\noindent\textbf{Step 3: the rates.}
Using $\lambda^*=C_0\sqrt{\Lambda^*}h^*$, $\Lambda\le\Lambda^*$ and $h^*=(a_n^2/\Lambda^*)^{1/3}$,
\begin{align*}
F_2(h^*,\lambda^*)&=C_0^2\Lambda^*s\,h^{*2}+\Lambda s\,h^{*2}
\le(C_0^2+1)\,s\,\Lambda^*h^{*2}
=(C_0^2+1)\,s\,\Lambda^{*1/3}a_n^{4/3},\\
F_1(h^*,\lambda^*)&=C_0\sqrt{\Lambda^*}\,s\,h^*+\frac{\Lambda s\,h^{*2}}{C_0\sqrt{\Lambda^*}\,h^*}
\le(C_0+C_0^{-1})\,s\sqrt{\Lambda^*}\,h^*
=(C_0+C_0^{-1})\,s\,\Lambda^{*1/6}a_n^{2/3}.
\end{align*}
Since $a_n^{2/3}=(\vanishing^{-2/\order}p^{2/\order}/n)^{1/3}$, these are the rates stated in the corollary.

\medskip\noindent\textbf{Step 4: optimality.}
Let $(h,\lambda)$ satisfy \eqref{eq. proof cor lambda}.
By the inequality of arithmetic and geometric means, $F_1(h,\lambda)\ge2s\sqrt\Lambda\,h$, and $F_1(h,\lambda)\ge\lambda s\ge C_0 s\max\{a_nh^{-1/2},b_nh\}$.
Since $\sqrt{\Lambda^*}=\sqrt\Lambda\vee b_n$,
\[
F_1(h,\lambda)\ge c\,s\max\big\{a_nh^{-1/2},\sqrt{\Lambda^*}\,h\big\}\ge c\,s\,\Lambda^{*1/6}a_n^{2/3},
\]
where the last bound is the minimum over $h>0$, attained at $h=h^*$.
Similarly, $F_2(h,\lambda)\ge s\max\{\lambda^2,\Lambda h^2\}\ge c\,s\max\{a_n^2h^{-1},\Lambda^*h^2\}\ge c\,s\,\Lambda^{*1/3}a_n^{4/3}$.
Hence, up to constants, $(h^*,\lambda^*)$ minimizes both bounds of Theorem~\ref{th. error bound} over all $(h,\lambda)$ satisfying its conditions.

\section{Proofs for Section \ref{sec. asymptotic analysis}}
\label{app. proofs asymptotics}

This appendix proves the results of Section~\ref{sec. asymptotic analysis}. We first introduce notation and events, then state auxiliary lemmas, whose proofs are in Appendix~\ref{app. proofs of lemmas}, and finally prove Proposition~\ref{prop. projection errors}, Theorem~\ref{th. linear representation} and Corollary~\ref{cor. asymptotic normality}.

\subsection{Notation and events}
Throughout, $j\in\mathcal J$ and $|\mathcal J|\le C$. Recall that $\sbar\ge1$, so that $\smax\ge1$.
By Lyapunov's inequality, Assumption~\ref{as. local stationarity} still holds with $\rho$ replaced by $\rho\wedge2\bar\order\ge2$. Without loss of generality, we therefore assume that $\rho\le2\bar\order$, so that Proposition~\ref{prop. projection errors} applies.
Let $\hat{\bm\gamma}_j(u)$ denote the local nodewise lasso with penalty $\lambda_j(u)$, and let
\[
\hat v_{j,t}:=\x[j]-\X\mjtop\hat{\bm\gamma}_j(u),
\qquad
\hat\tau_j^2(u):=\meant\kernel\hat v_{j,t}^2+\lambda_j(u)\|\hat{\bm\gamma}_j(u)\|_1 ,
\]
so that $\bhat\Theta_j(u)\X=\hat v_{j,t}/\hat\tau_j^2(u)$. We also let
\[
\check v_{j,t}:=\x[j]-\X\mjtop\gammaZero(u)=\tau_j^2(u)\,\bm\Theta_j(u)\X ,
\]
which differs from $\ndeps$ in that the projection coefficients are evaluated at $u$.
Let $\bm\omega(u):=\meant\kernel\X\X^\top(\Beta^0(t/n)-\BetazeroU)$ and $\bm\Delta_n(u):=\sqrt{nh}\,(\bhat\Theta(u)\hat{\bm\Sigma}(u)-\bm I)(\BetahatU-\BetazeroU)$. Then
\begin{equation}\label{eq. S5 decomposition}
\sqrt{nh}\,(\hat{\bm b}(u)-\BetazeroU)=\frac{1}{\sqrt{nh}}\sumt\kernel\bhat\Theta(u)\X\varepsilon\tn+\sqrt{nh}\,\bhat\Theta(u)\bm\omega(u)-\bm\Delta_n(u).
\end{equation}
Let $\bar\kappa:=1/(32C)$, with $C$ as in Assumption~\ref{as. precision matrix}(i).
In addition to $\Fset$ and $\mathcal A_\local(u)$, defined in Appendix~\ref{app. proof error bounds}, we define, for $\nu>0$,
\begin{align*}
\Fsetj&:=\Big\{\max_{k\neq j}\Big|\meant\kernel\ndeps\x[k]\Big|\le\frac{\lambda_j(u)}{4}\Big\}, \\
\Ssetbar&:=\Big\{\smax\|\hat{\bm\Sigma}(u)-\bm\Sigma(u)\|_{\max}\le\bar\kappa\Big\},\\
\mathcal V_\nu(u)&:=\Big\{\max_{j\in\mathcal J}\Big|\meant\kernel\ndeps^2-\tau_j^2(u)\Big|\le\nu\Big\}, \\
\Asetj &:=\Big\{\sj \max_{k,k'\in S_{j,\local}(u) }\Big|\meant\kernel g_{t,jk}g_{t,jk'}\big(\x[k]\x[k']-\Sigma_{k,k'}(u)\big)\Big|\le \bar \kappa\Big\},
\end{align*}
where $g_{t,jk}:=2\big(\gamma^0_{j,k}(t/n)-\gamma^0_{j,k}(u)\big)/(C_\gamma L_\Theta h)$ if $k\in S_{j,\local}(u)$ and $|t/n-u|\le h/2$, and $g_{t,jk}:=0$ otherwise.
Here $C_\gamma L_\Theta$ is the Lipschitz constant of Lemma~\ref{lem. asymp. nodewise model}(v).

Let $\mathcal E_n$ be the intersection of $\Fset$, $\mathcal A_\local(u)$, $\Ssetbar$, $\mathcal V_{\nu_n}(u)$, and of $\Fsetj$ and $\mathcal A_{j,\local}(u)$ for $j\in\mathcal J$; the sequence $\nu_n$ is given in the proof of Theorem~\ref{th. linear representation}.
By Assumption~\ref{as. precision matrix}(i), Assumption~\ref{as. compatibility condition} holds with $\localComp^2=1/C$, so that $\kappa=\localComp^2/32=\bar\kappa$; since $\s\le\smax$, we have $\Ssetbar\subseteq\Sset$.

\subsection{Auxiliary lemmas}

\begin{lemma}[Local projections]\label{lem. asymp. nodewise model}
Suppose Assumption~\ref{as. NED}(i) holds. Then, for all $j\in\mathcal J$, $u\in[0,1]$ and $n$:
\begin{itemize}
\item[(i)] $\E\ndepsu=0$ and $\E\big[\ndepsu\,\xu[k]\big]=0$ for all $k\neq j$.
\end{itemize}
If, in addition, Assumptions~\ref{as. NED}(ii) and~\ref{as. projection errors} hold, then:
\begin{itemize}
\item[(ii)] $\max_{k}\E\big|\ndepsu\,\xu[k]\big|^{\bar\order}\le C$ and $\E\big|\ndepsu\,\epsu\big|^{\bar\order}\le C$.
\item[(iii)] For $k\neq j$, $\{\ndepsu\,\xu[k]\}$ is $L_{\order}$-NED of size $-\nedsize$ on $\{\bm U\tn(u)\}$ and an $L_{\order}$-mixingale of size $-\nedsize$ with respect to $\mathcal F^U\tn(u)$, with constants bounded uniformly in $j$, $k$, $u$ and $n$.
\item[(iv)] $\{\ndepsu\,\epsu\}$ is $L_{\order}$-NED of size $-\nedsize$ on $\{\bm U\tn(u)\}$ and an $L_{\order}$-mixingale of size $-\nedsize$ with respect to $\mathcal F^U\tn(u)$, with constants bounded uniformly in $j$, $u$ and $n$.
\end{itemize}
If, in addition to Assumption~\ref{as. NED}(i), Assumption~\ref{as. precision matrix} holds, then:
\begin{itemize}
\item[(v)] there is a constant $C_\gamma$, depending only on the constants in Assumptions~\ref{as. NED}(i) and~\ref{as. precision matrix}(i), such that $\|\gammaZero(u)-\gammaZero(u')\|_2\le C_\gamma L_\Theta\,|u-u'|$ for all $u,u'\in[0,1]$;
\item[(vi)] $1/C\le\tau_j^2(u)\le C$ and $\|\gammaZero(u)\|_2\le C$; hence $\|\gammaZero(u)\|_1\le C\sqrt{\sj}$ and $\|\bm\Theta_j(u)\|_1\le C(1+\sqrt{\sbar})$.
\end{itemize}
\end{lemma}

\begin{lemma}[Local nodewise lasso]\label{lem. asymp. nodewise lasso}
Suppose Assumptions~\ref{as. kernel}, \ref{as. NED} and~\ref{as. precision matrix} hold, $h\le2\local$, and $j\in\mathcal J$.
\begin{enumerate}[label=(\roman*)]
\item Whenever $\hat\tau_j^2(u)>0$,
\[
\|\bm e_j^\top-\bhat\Theta_j(u)\hat{\bm\Sigma}(u)\|_\infty\le\frac{\lambda_j(u)}{\hat\tau_j^2(u)}
\qquad\text{and}\qquad
\meant\kernel\big(\bhat\Theta_j(u)\X\big)^2\le\frac{1}{\hat\tau_j^2(u)} .
\]
\item On $\Fsetj\cap\Ssetbar\cap\mathcal A_{j,\local}(u)$,
\begin{align*}
\meant\kernel &\big(\X\mjtop(\hat{\bm\gamma}_j(u)-\gammaZero(t/n))\big)^2+\lambda_j(u)\|\hat{\bm\gamma}_j(u)-\gammaZero(u)\|_1 \\
& \le C\big(\lambda_j(u)^2\sj+\Lambda_{j,\local}(u)\,\sj\, h^2\big).
\end{align*}
\item If, in addition, $\lambda_j(u)^2\sj+\Lambda_{j,\local}(u)\,\sj\,h^2+\nu\le c$ for a sufficiently small constant $c>0$, then on $\Fsetj\cap\Ssetbar\cap\mathcal A_{j,\local}(u)\cap\mathcal V_\nu$,
\[
|\hat\tau_j^2(u)-\tau_j^2(u)|\le C\big(\lambda_j(u)\sqrt{\sj}+\sqrt{\sj\,\Lambda_{j,\local}(u)}\,h+\nu\big),
\]
and in particular $\hat\tau_j^2(u)\ge\tau_j^2(u)/2\ge1/(2C)$, so that the bounds in (i) are at most $2C\lambda_j(u)$ and $2C$.
\end{enumerate}
\end{lemma}

\begin{lemma}[Empirical process events]\label{lem. asymp. event F}
Suppose Assumptions~\ref{as. kernel}, \ref{as. local stationarity}, \ref{as. NED}, \ref{as. precision matrix} and~\ref{as. projection errors} hold, $h\le2\local$, $nh\ge1$ and $\lagorder\le Cnh$.
Then for all $j\in\mathcal J$ and $\lambda_j(u)>0$,
\[
\p\big(\Fsetj^c\big)\le\frac{Ch\big(\rate+\rate^v p^{1/(2\bar\order)}+h\,\rate\,\rate^v\big)}{\lambda_j(u)}+\frac{C\,p}{\lambda_j(u)^{\order}(nh)^{\order/2}} .
\]
\end{lemma}

\begin{lemma}[Covariance for the nodewise regressions]\label{lem. asymp. event S}
Suppose Assumptions~\ref{as. kernel}, \ref{as. local stationarity}, \ref{as. NED} and~\ref{as. precision matrix} hold, $h\le2\big(\local\wedge u\wedge(1-u)\big)$, $\lagorder\le Cnh$, and Assumption~\ref{as. bound on s} holds with $\smax$ in place of $\s$ and a sufficiently small constant. Then, for $n$ large enough,
\[
\p\big(\Ssetbar^c\big)\le C\Big(\smax\,h\big(\rate\,p^{1/(2\bar\order)}+h\,\rate^2\big)+\vanishing\Big).
\]
\end{lemma}

\begin{lemma}[Bias events for the nodewise regressions]\label{lem. asymp. event A}
Suppose Assumptions~\ref{as. kernel}, \ref{as. local stationarity}, \ref{as. NED} and~\ref{as. precision matrix} hold, $h\le2\local$, $\lagorder\le Cnh$, and Assumption~\ref{as. bound on s} holds with $\smax$ in place of $\s$ and a sufficiently small constant. Then, for $n$ large enough, for all $j\in\mathcal J$,
\[
\p\big(\mathcal A_{j,\local}(u)^c\big)\le C\Big(\sj\,h\big(\rate\,\sj^{1/(2\bar\order)}+h\,\rate^2\big)+\vanishing\Big).
\]
\end{lemma}

\begin{lemma}[Nodewise variance event]\label{lem. asymp. event V}
Suppose Assumptions~\ref{as. kernel}, \ref{as. local stationarity}, \ref{as. NED}, \ref{as. precision matrix} and~\ref{as. projection errors} hold, $h\le2\big(\local\wedge u\wedge(1-u)\big)$ and $\lagorder\le Cnh$. Then there is a constant $C_0$ such that, for $n$ large enough and all $\nu\ge C_0/(nh)$,
\[
\p\big(\mathcal V_\nu(u)^c\big)\le C\Big(\frac{h\,\rate^v(1+h\,\rate^v)}{\nu}+\frac{1}{\nu^{\order}(nh)^{\order/2}}\Big).
\]
\end{lemma}

\begin{lemma}[Central limit theorem]\label{lem. asymp. clt}
Suppose Assumptions~\ref{as. kernel}, \ref{as. NED}, \ref{as. precision matrix}, \ref{as. projection errors} and~\ref{as. long-run variance} hold, and $nh\to\infty$. Then
\[
\bm V_n(u)^{-1/2}\frac{1}{\sqrt{nh}}\sumt\kernel\tilde{\bm w}\tn(u)\overset{d}{\longrightarrow}\mathcal N(\bm0,\bm I_{|\mathcal J|}).
\]
\end{lemma}

\subsection{Proof of Proposition~\ref{prop. projection errors}}

Fix $j\in\mathcal J$ and $t$ with $|t/n-u|\le\local$, and write $\Delta_{t,n}:=|t/n-u|+\lagorder/n$ and $\bm d_t:=\gammaZero(t/n)-\gammaZero(u)$.
The process $\ndepsu=\xu[j]-\Xu[-j]^\top\gammaZero(u)$ is strictly stationary, since it is a fixed linear function of the strictly stationary process $\Xu$.
By the definitions of $\ndeps$ and $\ndepsu$,
\[
\ndeps-\ndepsu
=\underbrace{\big(\x[j]-\xu[j]\big)-\big(\X^{(-j)}-\Xu[-j]\big)^\top\gammaZero(u)}_{=:I_t}
\;-\;\underbrace{\X\mjtop\bm d_t}_{=:II_t}.
\]

\emph{Term $I_t$.}
By H\"older's inequality and Lemma~\ref{lem. asymp. nodewise model}(vi),
\[
|I_t|\le\big(1+\|\gammaZero(u)\|_1\big)\,\|\X-\Xu\|_\infty\le C\big(1+\sqrt{\sj}\big)\,\|\X-\Xu\|_\infty ,
\]
so that, by Assumption~\ref{as. local stationarity}(i),
\[
\E\big[|I_t|^\rho\big]^{1/\rho}\le C\big(1+\sqrt{\sj}\big)\,\rate\,\Delta_{t,n}.
\]

\emph{Term $II_t$.}
For $k\notin S_{j,\local}(u)$, $\gamma^0_{j,k}$ vanishes on $[u-\local,u+\local]$ by definition of $S_{j,\local}(u)$. Since $|t/n-u|\le\local$, the vector $\bm d_t$ is therefore supported on $S_{j,\local}(u)$.
By Lemma~\ref{lem. asymp. nodewise model}(v) and the Cauchy--Schwarz inequality,
\[
\|\bm d_t\|_1\le\sqrt{\sj}\,\|\bm d_t\|_2\le C_\gamma L_\Theta\sqrt{\sj}\,|t/n-u| .
\]
By Minkowski's inequality,
\[
\E\big[|II_t|^\rho\big]^{1/\rho}\le\sum_{k\neq j}|d_{t,k}|\,\E\big[|\x[k]|^\rho\big]^{1/\rho}\le\|\bm d_t\|_1\max_{k}\E\big[|\x[k]|^\rho\big]^{1/\rho}.
\]
Since $\rho\le2\bar\order$, Assumption~\ref{as. NED}(i) and Lyapunov's inequality give $\E[|\xu[k]|^\rho]^{1/\rho}\le C$, and Assumption~\ref{as. local stationarity}(i) gives
\[
\E\big[|\x[k]|^\rho\big]^{1/\rho}\le\E\big[|\xu[k]|^\rho\big]^{1/\rho}+\E\big[\|\X-\Xu\|_\infty^\rho\big]^{1/\rho}\le C\big(1+\rate\,\Delta_{t,n}\big).
\]
Hence, using $|t/n-u|\le\Delta_{t,n}$ and $|t/n-u|\le1$,
\[
\E\big[|II_t|^\rho\big]^{1/\rho}\le C\sqrt{\sj}\,L_\Theta\,|t/n-u|\big(1+\rate\,\Delta_{t,n}\big)\le C\sqrt{\sj}\,(L_\Theta+\rate)\,\Delta_{t,n}.
\]

\emph{Conclusion.}
By Minkowski's inequality, and since $1+\sqrt{\sj}\le2\sqrt{\sbar}$ (as $\sbar\ge1$) and $L_\Theta+\rate\le2(\rate\vee L_\Theta)$,
\[
\E\big[|\ndeps-\ndepsu|^\rho\big]^{1/\rho}\le C\sqrt{\sbar}\,(\rate\vee L_\Theta)\,\Delta_{t,n},
\]
which is the claim, with $\ndeps$ inheriting the lag order $\lagorder$ of $\X$.

\subsection{Proof of Theorem~\ref{th. linear representation}}

Fix $j\in\mathcal J$, and let
$\mathcal B_{n,j}(u):=\sqrt{nh}\,\bhat\Theta_j(u)\bm\omega(u)$.
Since $\lambda_j(u)\asymp\lambda(u)$, the ratios $\lambda_j(u)/\lambda(u)$ and $\lambda(u)/\lambda_j(u)$ are bounded and absorbed into the constants. Recall that $\smax\ge\sbar\ge1$.
By Assumption~\ref{as. precision matrix}(i), Assumption~\ref{as. compatibility condition} holds with $\localComp^2=1/C$, both for the \llasso\ and for the nodewise regressions.

Let $\mathcal E_n$ be the event defined in the notation of this appendix, with $\nu_n$ chosen in Step~2.
Since $\Ssetbar\subseteq\Sset$, we have $\mathcal E_n\subseteq\Fset\cap\Sset\cap\mathcal A_\local(u)$, so that \eqref{eq. oracle final} in the proof of Theorem~\ref{th. error bound} holds on $\mathcal E_n$.
Moreover, for $n$ large enough, Lemma~\ref{lem. asymp. nodewise lasso}(iii) applies on $\mathcal E_n$ (see Step~2) and gives $\hat\tau_j^2(u)\ge\tau_j^2(u)/2\ge1/(2C)$.
Hence $\bhat\Theta_j(u)\X=\hat v_{j,t}/\hat\tau_j^2(u)$ on $\mathcal E_n$, and since $\bm\Theta_j(u)\X=\check v_{j,t}/\tau_j^2(u)$, \eqref{eq. S5 decomposition} gives
\[
\sqrt{nh}\,(\hat b_j(u)-\beta^0_j(u))-\frac{1}{\sqrt{nh}}\sumt\kernel\bm\Theta_j(u)\X\varepsilon\tn-\mathcal B_{n,j}(u)
=R_{1,j}+R_{2,j}-\Delta_{n,j}(u),
\]
where
\begin{align*}
R_{1,j}&:=\frac{1}{\hat\tau_j^2(u)\sqrt{nh}}\sumt\kernel(\hat v_{j,t}-\check v_{j,t})\varepsilon\tn,\\
R_{2,j}&:=\Big(\frac{1}{\hat\tau_j^2(u)}-\frac{1}{\tau_j^2(u)}\Big)\frac{1}{\sqrt{nh}}\sumt\kernel\check v_{j,t}\varepsilon\tn .
\end{align*}
It suffices to show that, on $\mathcal E_n$, $R_{1,j}$, $R_{2,j}$ and $\Delta_{n,j}(u)$ are bounded by deterministic sequences tending to zero and $|\mathcal B_{n,j}(u)|\le C\sqrt{nh^3\localEig\s}$ (Step~1), and that $\p(\mathcal E_n)\to1$ (Step~2).

\paragraph{Step 1: deterministic part.}
Throughout this step, we work on $\mathcal E_n$, with $n$ large enough.

\emph{Remainder.}
By H\"older's inequality, Lemma~\ref{lem. asymp. nodewise lasso}(i) and \eqref{eq. oracle final},
\begin{align*}
|\Delta_{n,j}(u)|
&\le\sqrt{nh}\,\big\|\bm e_j^\top-\bhat\Theta_j(u)\hat{\bm\Sigma}(u)\big\|_\infty\|\BetahatU-\BetazeroU\|_1\\
&\le C\sqrt{nh}\,\frac{\lambda_j(u)}{\hat\tau_j^2(u)}\Big(\lambda(u)\s+\frac{\localEig\,\s\,h^2}{\lambda(u)}\Big) \\
&\le C\sqrt{nh}\,\big(\lambda(u)^2+\localEig h^2\big)\s ,
\end{align*}
which vanishes by (c).

\emph{Estimation error of the projection.}
Since $\hat v_{j,t}-\check v_{j,t}=\X\mjtop(\gammaZero(u)-\hat{\bm\gamma}_j(u))$, H\"older's inequality on $\Fset$ and Lemma~\ref{lem. asymp. nodewise lasso}(ii) give
\begin{align*}
|R_{1,j}|
&=\frac{\sqrt{nh}}{\hat\tau_j^2(u)}\Big|(\gammaZero(u)-\hat{\bm\gamma}_j(u))^\top\frac{\bm X\mjtop\kernelmat\bm\varepsilon}{nh}\Big|
\le C\sqrt{nh}\,\lambda(u)\,\|\hat{\bm\gamma}_j(u)-\gammaZero(u)\|_1\\
&\le C\sqrt{nh}\Big(\lambda(u)\lambda_j(u)\sj+\frac{\lambda(u)}{\lambda_j(u)}\Lambda_{j,\local}(u)\sj h^2\Big)
\le C\sqrt{nh}\,\big(\lambda(u)^2+\bar\Lambda_\local(u)h^2\big)\sbar ,
\end{align*}
which vanishes by (c).

\emph{Estimation error of the nodewise variance.}
Since $\check v_{j,t}=\tau_j^2(u)\,\bm\Theta_j(u)\X$,
\[
\frac{1}{\sqrt{nh}}\sumt\kernel\check v_{j,t}\varepsilon\tn
=\tau_j^2(u)\sqrt{nh}\;\bm\Theta_j(u)\,\frac{\bm X^\top\kernelmat\bm\varepsilon}{nh}.
\]
On $\Fset$, H\"older's inequality, Lemma~\ref{lem. asymp. nodewise model}(vi) and $\sbar\ge1$ give
\[
\Big|\frac{1}{\sqrt{nh}}\sumt\kernel\check v_{j,t}\varepsilon\tn\Big|
\le\tau_j^2(u)\sqrt{nh}\,\|\bm\Theta_j(u)\|_1\Big\|\frac{\bm X^\top\kernelmat\bm\varepsilon}{nh}\Big\|_\infty
\le C\sqrt{nh}\,\sqrt{\sbar}\,\lambda(u).
\]
By Lemma~\ref{lem. asymp. nodewise lasso}(iii) and $\hat\tau_j^2(u)\tau_j^2(u)\ge1/(2C^2)$,
\[
\Big|\frac{1}{\hat\tau_j^2(u)}-\frac{1}{\tau_j^2(u)}\Big|
=\frac{|\hat\tau_j^2(u)-\tau_j^2(u)|}{\hat\tau_j^2(u)\,\tau_j^2(u)}
\le C\big(\lambda(u)\sqrt{\sbar}+\sqrt{\sbar\,\bar\Lambda_\local(u)}\,h+\nu_n\big).
\]
Combining the two displays,
\[
|R_{2,j}|
\le C\sqrt{nh}\,\big(\lambda(u)^2\sbar+\lambda(u)\,\sbar\sqrt{\bar\Lambda_\local(u)}\,h+\lambda(u)\sqrt{\sbar}\,\nu_n\big).
\]
The first term vanishes by (c), and so does the second, since
$\sqrt{nh}\,\lambda(u)\,\sbar\sqrt{\bar\Lambda_\local(u)}\,h=\big(\sqrt{nh}\,\lambda(u)^2\sbar\big)^{1/2}\big(\sqrt{nh}\,\bar\Lambda_\local(u)\sbar h^2\big)^{1/2}$.
The third vanishes by the choice of $\nu_n$ in Step~2.

\emph{Bias.}
By the Cauchy--Schwarz inequality,
\begin{align*}
|\mathcal B_{n,j}(u)|
&=\sqrt{nh}\,\Big|\meant\kernel\big(\bhat\Theta_j(u)\X\big)\,\X^\top\big(\Beta^0(t/n)-\BetazeroU\big)\Big|\\
&\le\sqrt{nh}\,\Big(\meant\kernel\big(\bhat\Theta_j(u)\X\big)^2\Big)^{1/2}\\
&\qquad\times\Big(\meant\kernel\big(\X^\top(\Beta^0(t/n)-\BetazeroU)\big)^2\Big)^{1/2}
\le\frac{\sqrt{nh}\,B_n}{\sqrt{\hat\tau_j^2(u)}},
\end{align*}
where the last inequality uses Lemma~\ref{lem. asymp. nodewise lasso}(i) and the definition \eqref{eq. approximation bias definition} of $B_n$.
By Step~1 of the proof of Theorem~\ref{th. error bound}, $B_n^2\le C\localEig\,\s\,h^2$ on $\mathcal A_\local(u)$. Since $\hat\tau_j^2(u)\ge1/(2C)$,
\[
|\mathcal B_{n,j}(u)|\le C\sqrt{nh^3\,\localEig\,\s}.
\]

\paragraph{Step 2: probabilistic part.}
We show that $\p(\mathcal E_n)\to1$. Since $|\mathcal J|\le C$, it suffices to show that each event in $\mathcal E_n$ has probability tending to one.
Assumption~\ref{as. bound on s} gives $nh\to\infty$. Together with $\smax\ge1$, conditions (a)--(c) then give
\begin{equation}\label{eq. th3 consequences}
h\rate\to0\ \text{by (a)},\qquad
\frac{h(\rateErr\vee\rate^v)}{\lambda(u)}\to0\ \text{by (b)},\qquad
\lambda(u)^2\le\frac{\sqrt{nh}\,\lambda(u)^2\smax}{\sqrt{nh}}\to0\ \text{by (c)}.
\end{equation}
Since $u\in(0,1)$, $h\to0$ and $\smax/(nh)\to0$ under Assumption~\ref{as. bound on s}, the conditions $h\le2(u\wedge(1-u))$, $nh\ge C_0\s/\kappa$ and $nh\ge C_0\smax/\bar\kappa$ of Lemmas~\ref{lem. error. event S}, \ref{lem. asymp. event S} and~\ref{lem. asymp. event V} hold for $n$ large enough.
As in Step~2 of the proof of Theorem~\ref{th. error bound}, we may assume that Assumption~\ref{as. bound on s} holds with a sufficiently small constant, at the cost of replacing $\vanishing$ by $M\vanishing\to0$.

\emph{Empirical process events.}
By Lemmas~\ref{lem. error. event F} and~\ref{lem. asymp. event F},
\[
\p\big(\Fset^c\big)+\max_{j\in\mathcal J}\p\big(\Fsetj^c\big)
\le C\Big(\frac{h\big(\rate+(\rateErr\vee\rate^v)\,p^{1/(2\bar\order)}\big)}{\lambda(u)}
+h\rate\,\frac{h(\rateErr\vee\rate^v)}{\lambda(u)}
+\Big(\frac{p^{1/\order}}{\lambda(u)\sqrt{nh}}\Big)^{\order}\Big).
\]
The first and third terms vanish by (b), and the second by \eqref{eq. th3 consequences}.

\emph{Covariance event.}
By Lemma~\ref{lem. asymp. event S},
$\p(\Ssetbar^c)\le C\big(\smax h\rate\,p^{1/(2\bar\order)}+(\smax h\rate)(h\rate)+\vanishing\big)\to0$ by (a).

\emph{Bias events.}
By Lemmas~\ref{lem. error. event A} and~\ref{lem. asymp. event A}, since $\s,\sj\le\smax\le p$,
\[
\p\big(\mathcal A_\local(u)^c\big)+\max_{j\in\mathcal J}\p\big(\mathcal A_{j,\local}(u)^c\big)
\le C\big(\smax h\rate\,p^{1/(2\bar\order)}+(\smax h\rate)(h\rate)+\vanishing\big)\to0
\]
by (a).

\emph{Nodewise variance event.}
Let
\[
a_n:=(nh)^{-1/2}+h\rate^v+\lambda(u)\sqrt{\sbar}+\sqrt{nh}\,\lambda(u)\sqrt{\sbar}\,h\rate^v .
\]
Each term tends to zero.
The first does since $nh\to\infty$, and the second by \eqref{eq. th3 consequences}, since $h\rate^v=\lambda(u)\cdot h\rate^v/\lambda(u)$.
The third does since $\lambda(u)^2\sbar\le(nh)^{-1/2}\big(\sqrt{nh}\,\lambda(u)^2\smax\big)\to0$ by (c).
The fourth equals $\big(\sqrt{nh}\,\lambda(u)^2\sqrt{\sbar}\big)\big(h\rate^v/\lambda(u)\big)$ and vanishes by (b) and (c), since $\sqrt{\sbar}\le\smax$.
Let $\nu_n:=a_n^{-1/2}\big((nh)^{-1/2}+h\rate^v\big)$. Then
$\nu_n\le a_n^{1/2}\to0$ and
\[
\sqrt{nh}\,\lambda(u)\sqrt{\sbar}\,\nu_n=a_n^{-1/2}\big(\lambda(u)\sqrt{\sbar}+\sqrt{nh}\,\lambda(u)\sqrt{\sbar}\,h\rate^v\big)\le a_n^{1/2}\to0,
\]
as required in Step~1.
Moreover, $\nu_n\ge a_n^{-1/2}(nh)^{-1/2}\ge C_0/(nh)$ for $n$ large enough, and Lemma~\ref{lem. asymp. event V} gives
\[
\p\big(\mathcal V_{\nu_n}(u)^c\big)\le C\Big(\frac{h\rate^v(1+h\rate^v)}{\nu_n}+\frac{1}{\nu_n^{\order}(nh)^{\order/2}}\Big)
\le C\big(a_n^{1/2}(1+h\rate^v)+a_n^{\order/2}\big)\to0 .
\]

\emph{Condition of Lemma~\ref{lem. asymp. nodewise lasso}(iii).}
By (c), $\lambda_j(u)^2\sj+\Lambda_{j,\local}(u)\sj h^2\le C(nh)^{-1/2}\sqrt{nh}\big(\lambda(u)^2+\bar\Lambda_\local(u)h^2\big)\smax\to0$, and $\nu_n\to0$. Hence the condition of Lemma~\ref{lem. asymp. nodewise lasso}(iii) holds for $n$ large enough.

Combining Steps~1 and~2 proves the theorem.

\subsection{Proof of Corollary~\ref{cor. asymptotic normality}}

By Theorem~\ref{th. linear representation}, $|\mathcal B_{n,j}(u)|\le C\sqrt{nh^3\localEig\,\s}$ with probability tending to one, and this bound vanishes by (e). Since $|\mathcal J|\le C$, it follows that
\[
\sqrt{nh}\,\big(\hat{\bm b}_{\mathcal J}(u)-\Beta^0_{\mathcal J}(u)\big)=\frac{1}{\sqrt{nh}}\sumt\kernel\bm\Theta_{\mathcal J}(u)\X\varepsilon\tn+o_\p(1).
\]

\emph{Replacement by the stationary approximation.}
We apply Lemma~\ref{lem. tech. transfer} with $\bm A_{t,n}=\bm\Theta_{\mathcal J}(u)\X$ ($p_A=|\mathcal J|$) and $B_{t,n}=\varepsilon\tn$ ($p_B=1$), with stationary approximations $\bm\Theta_{\mathcal J}(u)\Xu$ and $\epsu$, and verify its conditions.
\begin{itemize}
\item[(i)] By Lemma~\ref{lem. asymp. nodewise model}(vi) and $\sbar\ge1$, $|\bm\Theta_j(u)(\X-\Xu)|\le\|\bm\Theta_j(u)\|_1\|\X-\Xu\|_\infty\le C\sqrt{\sbar}\,\|\X-\Xu\|_\infty$. Hence, by Assumption~\ref{as. local stationarity}(i) and Lyapunov's inequality, condition (i) holds with rate $C\sqrt{\sbar}\,\rate$ and lag order $\lagorder$.
\item[(ii)] Condition (ii) holds with rate $\rateErr$, by Assumption~\ref{as. local stationarity}(ii) and Lyapunov's inequality.
\item[(iii)] Since $\bm\Theta_j(u)\Xu=\ndepsu/\tau_j^2(u)$, Assumption~\ref{as. projection errors} and Lemma~\ref{lem. asymp. nodewise model}(vi) give $\E|\bm\Theta_j(u)\Xu|^{2\bar\order}\le C$, and $\E|\epsu|^{2\bar\order}\le C$ by Assumption~\ref{as. NED}(i).
\end{itemize}
Since $h\le2\local$, $\lagorder\le Cnh$ and $nh\ge1$ for $n$ large, Lemma~\ref{lem. tech. transfer} gives
\begin{align*}
&\E\Big[\max_{j\in\mathcal J}\Big|\frac{1}{\sqrt{nh}}\sumt\kernel\bm\Theta_j(u)\big(\X\varepsilon\tn-\Xu\epsu\big)\Big|\Big]\\
&\qquad\le C\sqrt{nh^3}\,\big(\sqrt{\sbar}\,\rate+\rateErr\big)\big(1+h\rateErr\big),
\end{align*}
which vanishes by (d), since $h\rateErr\to0$ by (b) and (c).
By Markov's inequality, and since $\bm\Theta_{\mathcal J}(u)\Xu\epsu=\tilde{\bm w}\tn(u)$,
\[
\sqrt{nh}\,\big(\hat{\bm b}_{\mathcal J}(u)-\Beta^0_{\mathcal J}(u)\big)=\frac{1}{\sqrt{nh}}\sumt\kernel\tilde{\bm w}\tn(u)+o_\p(1).
\]

\emph{Joint statement.}
By Assumption~\ref{as. long-run variance}, $\|\bm V_n(u)^{-1/2}\|\le C$, so the $o_\p(1)$ term remains $o_\p(1)$ after multiplication by $\bm V_n(u)^{-1/2}$.
Lemma~\ref{lem. asymp. clt} and Slutsky's theorem then give
\[
\bm Z_n:=\sqrt{nh}\;\bm V_n(u)^{-1/2}\big(\hat{\bm b}_{\mathcal J}(u)-\Beta^0_{\mathcal J}(u)\big)\overset{d}{\longrightarrow}\mathcal N(\bm0,\bm I_{|\mathcal J|}).
\]

\emph{Marginal statement.}
Fix $j\in\mathcal J$. Since $V_{n,jj}(u)\ge1/C$ by Assumption~\ref{as. long-run variance}, the vector $\bm\alpha_n:=\bm V_n(u)^{1/2}\bm e_j/V_{n,jj}(u)^{1/2}$ is well defined, $\|\bm\alpha_n\|_2=1$, and
\[
\bm\alpha_n^\top\bm Z_n=\sqrt{nh}\,\big(\hat b_j(u)-\beta^0_j(u)\big)/V_{n,jj}(u)^{1/2}.
\]
By compactness of the unit sphere, every subsequence has a further subsequence along which $\bm\alpha_n\to\bm\alpha$, for some unit vector $\bm\alpha$.
Along this further subsequence, $\bm Z_n=O_\p(1)$ gives $\bm\alpha_n^\top\bm Z_n=\bm\alpha^\top\bm Z_n+o_\p(1)\overset{d}{\to}\mathcal N(0,1)$.
Hence the convergence holds along the whole sequence. Since $\Phi$ is continuous, P\'olya's theorem gives uniformity in $z$.

\section{Proofs of Technical Lemmas}
\label{app. proofs of lemmas}

\subsection{Proof of Lemma \ref{lem. tech. triplex inequality}}

Let $\bar K:=\sup_x|K(x)|$ and $N:=\{t\in\{1,\dots,n\}:|t/n-u|\le h/2\}$, which contains every $t$ with $K((t/n-u)/h)\neq0$.
Since $N$ is a set of consecutive integers, we write $N=\{t_1,\dots,t_{\tilde n}\}$ with $t_s=t_1+s-1$, and $\tilde n\le nh+1\le 2nh$.
For $s\in\Z$, let $\mathcal G_s:=\mathcal F_{t_1+s-1}$, and for $s=1,\dots,\tilde n$ let $\varrho_s:=K((t_s/n-u)/h)\,w_{t_s}\chi_{t_s}$, which is $\mathcal G_s$-measurable.
Then $\sumt\kernel w_t(\chi_t-\E\chi_t)=\sum_{s=1}^{\tilde n}(\varrho_s-\E\varrho_s)$.
Let $\xi':=nh\,\xi/\tilde n\ge\xi/2$.
Theorem~1 of \citet{jiang2009uniform} gives
\begin{align*}
\p\Big(\Big|\sum_{s=1}^{\tilde n}(\varrho_s-\E\varrho_s)\Big|>\tilde n\xi'\Big)
&\le 2q\exp\Big(-\frac{\tilde n\xi'^2}{288q^2\alpha_n^2}\Big)
+\frac{6}{\tilde n\xi'}\sum_{s=1}^{\tilde n}\E\big|\E(\varrho_s\mid\mathcal G_{s-q})-\E\varrho_s\big| \\
&\quad+\frac{15}{\tilde n\xi'}\sum_{s=1}^{\tilde n}\E\big[|\varrho_s|\mathbbm 1(|\varrho_s|>\alpha_n)\big].
\end{align*}
For the first term, $\tilde n\xi'^2=(nh)^2\xi^2/\tilde n\ge nh\,\xi^2/2$.
For the second term, $\tilde n\xi'=nh\,\xi$ and $\mathcal G_{s-q}=\mathcal F_{t_s-q}$.
Since the weights are deterministic, Lyapunov's inequality and the mixingale property give
\[
\E\big|\E(\varrho_s\mid\mathcal G_{s-q})-\E\varrho_s\big|
\le \bar K\,\big\|\E(\chi_{t_s}\mid\mathcal F_{t_s-q})-\E\chi_{t_s}\big\|_{\tilde r}\le Cq^{-\nedsize},
\]
so that the second term is at most $6\tilde n Cq^{-\nedsize}/(nh\,\xi)\le Cq^{-\nedsize}/\xi$.
For the third term, $|\varrho_s|\le|\varrho_s|^r\alpha_n^{1-r}$ on $\{|\varrho_s|>\alpha_n\}$, so that
$\E[|\varrho_s|\mathbbm 1(|\varrho_s|>\alpha_n)]\le\alpha_n^{1-r}\bar K^r\sup_t\E|\chi_t|^r\le C\alpha_n^{1-r}$.
The third term is therefore at most $C\alpha_n^{1-r}/\xi$.

\subsection{Proof of Lemma \ref{lem. tech. eps X mixingale}}

\begin{proof}
Write $Y:=\tilde W\tn\pj(u)$ and $Z:=\tilde W\tn^{(k)}(u)$.

(i) By the Cauchy--Schwarz inequality and Assumption~\ref{as. NED}(i), $\E|YZ|^{\bar\order}\le(\E|Y|^{2\bar\order}\,\E|Z|^{2\bar\order})^{1/2}\le C$.

(ii) Let $\mathcal G^\ell\tn:=\sigma\{\bm U_{t-\ell,n}(u),\dots,\bm U_{t+\ell,n}(u)\}$ and $\E^\ell[\cdot]:=\E[\cdot\mid\mathcal G^\ell\tn]$.
For every $\mathcal G^\ell\tn$-measurable $W$, the conditional Jensen inequality gives $\|YZ-\E^\ell[YZ]\|_{L^{\order}}\le2\|YZ-W\|_{L^{\order}}$.
Taking $W=\E^\ell[Y]\,\E^\ell[Z]$ and using H\"older's inequality,
\[
\big\|YZ-\E^\ell[YZ]\big\|_{L^{\order}}
\le2\Big(\|Y-\E^\ell Y\|_{L^{2\order}}\|Z\|_{L^{2\order}}+\|Y\|_{L^{2\order}}\|Z-\E^\ell Z\|_{L^{2\order}}\Big),
\]
where we used $\|\E^\ell Y\|_{L^{2\order}}\le\|Y\|_{L^{2\order}}$.
Since $\order<\bar\order$, $Y$ and $Z$ are $L_{2\order}$-bounded by Assumption~\ref{as. NED}(i), and they are $L_{2\order}$-NED of size $-\nedsize$ with bounded constants by Assumption~\ref{as. NED}(ii). Hence $YZ$ is $L_{\order}$-NED of size $-\nedsize$ with bounded constants.

(iii) By Assumption~\ref{as. NED}(ii), $\chi\tn$ is $\mathcal F^U\tn(u)$-measurable, and $\{\bm U\tn(u)\}$ is $\alpha$-mixing of size $-\nedsize/(1/\order-1/\bar\order)$.
By (i), (ii) and Theorem~17.5 of \citet{davidson2021stochastic},  edition
$\{\chi\tn-\E\chi\tn\}$ is an $L_{\order}$-mixingale with respect to $\{\mathcal F^U\tn(u)\}$ of size
$-\min\big\{\nedsize,\ \nedsize(1/\order-1/\bar\order)/(1/\order-1/\bar\order)\big\}=-\nedsize$, with constants bounded by $C$.
A size of $-\nedsize$ with $\nedsize\ge1$ means that the mixingale sequence is $O(\ell^{-1-\eta})$ for some $\eta>0$, which is summable.
\end{proof}

\subsection{Proof of Lemma \ref{lem. tech. transfer}}
\label{proof. lem. tech. transfer}

\begin{proof}
Let $\bar K:=\sup_x|K(x)|$ and $N_u:=\{t:\,|t/n-u|\le h/2\}$.
By Assumption~\ref{as. kernel}, $K((t/n-u)/h)=0$ for $t\notin N_u$, and $|N_u|\le nh+1\le 2nh$ since $nh\ge1$.
Since $h\le 2\local$, every $t\in N_u$ satisfies $|t/n-u|\le\local$, so (i) and (ii) hold for $t\in N_u$.
Moreover, since $\lagorder^a\vee\lagorder^b\le Cnh$,
\begin{equation}\label{eq. transfer window}
|t/n-u|+\lagorder^a/n\le Ch
\quad\text{and}\quad
|t/n-u|+\lagorder^b/n\le Ch,
\qquad t\in N_u .
\end{equation}

Write $\bm\Delta^A_t:=\bm A_{t,n}-\tilde{\bm A}_{t,n}(u)$ and $\bm\Delta^B_t:=\bm B_{t,n}-\tilde{\bm B}_{t,n}(u)$.
For all $j,k$,
\[
A^{(j)}_{t,n}B^{(k)}_{t,n}-\tilde A^{(j)}_{t,n}(u)\tilde B^{(k)}_{t,n}(u)
=\Delta^{A,(j)}_t\,\tilde B^{(k)}_{t,n}(u)+\tilde A^{(j)}_{t,n}(u)\,\Delta^{B,(k)}_t+\Delta^{A,(j)}_t\,\Delta^{B,(k)}_t ,
\]
so that
\[
\max_{j,k}\big|A^{(j)}_{t,n}B^{(k)}_{t,n}-\tilde A^{(j)}_{t,n}(u)\tilde B^{(k)}_{t,n}(u)\big|
\le \|\bm\Delta^A_t\|_\infty\|\tilde{\bm B}_{t,n}(u)\|_\infty
+\|\tilde{\bm A}_{t,n}(u)\|_\infty\|\bm\Delta^B_t\|_\infty
+\|\bm\Delta^A_t\|_\infty\|\bm\Delta^B_t\|_\infty
=: M_t .
\]
By the triangle inequality, the left-hand side of the lemma is bounded by
\begin{equation}\label{eq. transfer sum}
\E\Big[\frac{1}{nh}\sum_{t\in N_u}\Big|K\Big(\frac{t/n-u}{h}\Big)\Big|\,M_t\Big]
\le \frac{\bar K}{nh}\sum_{t\in N_u}\E[M_t]
\le 2\bar K\max_{t\in N_u}\E[M_t].
\end{equation}

We bound the three terms of $\E[M_t]$ for $t\in N_u$.
By (iii),
\[
\E\big[\|\tilde{\bm B}_{t,n}(u)\|_\infty^{2\bar\order}\big]\le\sum_{k=1}^{p_B}\E\big|\tilde B^{(k)}_{t,n}(u)\big|^{2\bar\order}\le C p_B ,
\]
and similarly $\E[\|\tilde{\bm A}_{t,n}(u)\|_\infty^{2\bar\order}]\le Cp_A$.
Let $r':=2\bar\order/(2\bar\order-1)$ be the conjugate exponent of $2\bar\order$. Since $\bar\order>2$, we have $r'<2$.
H\"older's inequality with exponents $r'$ and $2\bar\order$, followed by Lyapunov's inequality, gives
\begin{align*}
\E\big[\|\bm\Delta^A_t\|_\infty\|\tilde{\bm B}_{t,n}(u)\|_\infty\big]
&\le \E\big[\|\bm\Delta^A_t\|_\infty^{r'}\big]^{1/r'}\,\E\big[\|\tilde{\bm B}_{t,n}(u)\|_\infty^{2\bar\order}\big]^{1/(2\bar\order)}\\
&\le \E\big[\|\bm\Delta^A_t\|_\infty^{2}\big]^{1/2}\,(Cp_B)^{1/(2\bar\order)}
\le C\rate^a h\,p_B^{1/(2\bar\order)},
\end{align*}
where the last step uses (i) and \eqref{eq. transfer window}.
In the same way, (ii) gives $\E[\|\tilde{\bm A}_{t,n}(u)\|_\infty\|\bm\Delta^B_t\|_\infty]\le C\rate^b h\,p_A^{1/(2\bar\order)}$.
For the last term, the Cauchy--Schwarz inequality, (i), (ii) and \eqref{eq. transfer window} give
\[
\E\big[\|\bm\Delta^A_t\|_\infty\|\bm\Delta^B_t\|_\infty\big]
\le \E\big[\|\bm\Delta^A_t\|_\infty^2\big]^{1/2}\,\E\big[\|\bm\Delta^B_t\|_\infty^2\big]^{1/2}
\le C\rate^a\rate^b h^2 .
\]
Hence $\max_{t\in N_u}\E[M_t]\le Ch\big(\rate^a p_B^{1/(2\bar\order)}+\rate^b p_A^{1/(2\bar\order)}+h\rate^a\rate^b\big)$, and the result follows from \eqref{eq. transfer sum}.
\end{proof}

\subsection{Proof of Lemma \ref{lem. tech. score}}
\label{ProofA. Stochastic part}
Let $N_u:=\{t\in\{1,\dots,n\}:|t/n-u|\le h/2\}$.
By Assumption~\ref{as. kernel}, $\kernel=0$ for $t\notin N_u$, $|\kernel|\le\bar K$, and $|N_u|\le 2nh$.
Since $h\le2\local$, every $t\in N_u$ satisfies $|t/n-u|\le\local$, so the stationary approximations are defined for all summands.
For each $k$, we decompose
\begin{align*}
\meant\kernel A^{(k)}_{t,n}e_{t,n}
&=\underbrace{\meant\kernel\big(A^{(k)}_{t,n}e_{t,n}-\tilde A^{(k)}_{t,n}(u)\,\tilde e_{t,n}(u)\big)}_{=:R^{(k)}_1}  \\
&+ \underbrace{\meant\kernel\tilde A^{(k)}_{t,n}(u)\,\tilde e_{t,n}(u)}_{=:R^{(k)}_2},
\end{align*}
so that
\[
\p\Big(\max_{k}\Big|\meant\kernel A^{(k)}_{t,n}e_{t,n}\Big|>\xi\Big)
\le \p\Big(\max_k|R^{(k)}_1|>\xi/2\Big)+\p\Big(\max_k|R^{(k)}_2|>\xi/2\Big).
\]

\noindent\textbf{Transfer term.}
We apply Lemma~\ref{lem. tech. transfer} with $\bm B_{t,n}=e_{t,n}$, $p_B=1$, $\rate^a=\rate$, $\rate^b=\zeta_n$ and lag orders $\lagorder^a,\lagorder^e\le Cnh$.
Its conditions (i) and (ii) follow from (i) and (ii) by Lyapunov's inequality, since $\rho\ge2$, and its condition (iii) holds by Assumption~\ref{as. NED}(i).
Markov's inequality then gives
\[
\p\Big(\max_k|R^{(k)}_1|>\xi/2\Big)\le\frac{2}{\xi}\,\E\Big[\max_k|R^{(k)}_1|\Big]
\le\frac{Ch\big(\rate+\zeta_n\,p_A^{1/(2\bar m)}+h\,\rate\,\zeta_n\big)}{\xi}.
\]
\noindent\textbf{Stationary term.}
By Lemma~\ref{lem. tech. eps X mixingale}, applied to $(\tilde{\bm A}_{t,n}(u)^\top,\tilde e_{t,n}(u))^\top$, and the orthogonality condition, $\{\tilde A^{(k)}_{t,n}(u)\,\tilde e_{t,n}(u)\}_t$ is a centred $L_{\order}$-mixingale with constants $c_t\le C$ and a summable mixingale sequence.
Since the weights $\kernel$ are deterministic, bounded, and vanish for $|t/n-u|>h/2$, the weighted process is an $L_{\order}$-mixingale with constants $\kernel c_t$, and $\sum_{t=1}^n \kernel^2 c_t^2\le Cnh$.
By the union bound, Markov's inequality, and the maximal inequality of \citet[Lemma~2]{hansen1991strong},
\begin{align*}
\p\Big(\max_k|R^{(k)}_2|>\xi/2\Big)
\le\sum_{k=1}^{p_A}\frac{2^{\order}}{(nh\,\xi)^{\order}}\,
\E\Big|\sum_{t=1}^n\kernel\,\tilde A^{(k)}_{t,n}(u)\,\tilde e_{t,n}(u)\Big|^{\order} \\
\le\frac{C\,p_A}{(nh\,\xi)^{\order}}\Big(\sum_{t=1}^n \kernel^2c_t^2\Big)^{\order/2}
\le\frac{C\,p_A}{\xi^{\order}(nh)^{\order/2}}.
\end{align*}

Combining the two bounds proves the claim.

\subsection{Proof of Lemma \ref{lem. tech. covariance}}

Since $h\le 2\local$, the stationary approximations are defined for all $t$ with $\kernel\neq0$.
Because $\tilde{\bm A}_{t,n}(u)$ is stationary in $t$, $\E[\tilde A^{(j)}_{t,n}(u)\tilde A^{(k)}_{t,n}(u)]=\Sigma^A_{j,k}(u)$ for all $t$.
For each pair $(j,k)$, we decompose
\begin{align*}
&\meant\kernel w^{(j,k)}_t\big(A^{(j)}_{t,n}A^{(k)}_{t,n}-\Sigma^A_{j,k}(u)\big)\\
&\qquad=\underbrace{\meant\kernel w^{(j,k)}_t\big(A^{(j)}_{t,n}A^{(k)}_{t,n}-\tilde A^{(j)}_{t,n}(u)\tilde A^{(k)}_{t,n}(u)\big)}_{=:R^{(j,k)}_1}  \\
&\qquad\quad+\underbrace{\meant\kernel w^{(j,k)}_t\big(\tilde A^{(j)}_{t,n}(u)\tilde A^{(k)}_{t,n}(u)-\Sigma^A_{j,k}(u)\big)}_{=:R^{(j,k)}_2},
\end{align*}
and bound $\p(\max_{j,k}|R^{(j,k)}_1|>\xi/2)$ and $\p(\max_{j,k}|R^{(j,k)}_2|>\xi/2)$ separately.

\noindent\textbf{Transfer term.}
We apply Lemma~\ref{lem. tech. transfer} with $\bm B_{t,n}=\bm A_{t,n}$, $p_A=p_B=q$ and $\rate^a=\rate^b=\rate$.
Its conditions (i) and (ii) follow from the $(\rate,\local)$-stationarity of $\bm A_{t,n}$ by Lyapunov's inequality, and its condition (iii) holds by Assumption~\ref{as. NED}(i).
The proof of Lemma~\ref{lem. tech. transfer} bounds every summand in absolute value, so its conclusion still holds after inserting the weights $|w^{(j,k)}_t|\le1$.
Markov's inequality then gives
\[
\p\Big(\max_{j,k}|R^{(j,k)}_1|>\xi/2\Big)\le\frac{2}{\xi}\,\E\Big[\max_{j,k}|R^{(j,k)}_1|\Big]
\le\frac{Ch\big(\rate\,q^{1/(2\bar m)}+h\,\rate^2\big)}{\xi}.
\]

\noindent\textbf{Stationary term.}
The probability $\p(\max_{j,k}|R^{(j,k)}_2|>\xi/2)$ is nonincreasing in $\xi$, so it suffices to show that it is at most $C\vanishing$ at $\xi=\xi_n$.
By Lemma~\ref{lem. tech. eps X mixingale}, applied to $\tilde{\bm A}_{t,n}(u)$, $\chi_t:=\tilde A^{(j)}_{t,n}(u)\tilde A^{(k)}_{t,n}(u)$ is $L_{\order}$-bounded and $\chi_t-\E\chi_t$ is an $L_{\order}$-mixingale of size $-\nedsize$ with bounded constants, uniformly in $(j,k)$.
Lemma~\ref{lem. tech. triplex inequality}, with integer parameter $\ell\ge1$ and truncation level $\alpha>0$, combined with the union bound over the $q^2$ pairs, gives
\[
\p\Big(\max_{j,k}|R^{(j,k)}_2|>\xi_n/2\Big)
\le Cq^2\Big(\ell\,e^{-nh\xi_n^2/(C\ell^2\alpha^2)}+\frac{\ell^{-\nedsize}}{\xi_n}+\frac{\alpha^{1-\order}}{\xi_n}\Big).
\]
Let $x:=q^2/(\vanishing\,\xi_n)$, and choose $\ell:=\lceil x^{1/\nedsize}\rceil$ and $\alpha:=x^{1/(\order-1)}$.
Then the second and third terms are each at most $C\vanishing$.
The first term is at most $C\vanishing$ if
\begin{equation}\label{eq. cov first term}
\ell\,\alpha\,\sqrt{\log(q^2\ell/\vanishing)}\le c\,\sqrt{nh}\,\xi_n .
\end{equation}
Since $\xi_n\ge c(nh)^{-1/2}$ and $\vanishing\ge(nh)^{-1/2}$, the ratio $q^2\ell/\vanishing$ is polynomial in $q$ and $\sqrt{nh}$, so $\log(q^2\ell/\vanishing)\le C\log(q\vee\sqrt{nh})$.
If $x\ge1$, then $\ell\alpha\le 2x^{1/\nedsize+1/(\order-1)}$, and \eqref{eq. cov first term} holds if
\[
\Big(\frac{q^2}{\vanishing\,\xi_n}\Big)^{\frac1\nedsize+\frac{1}{\order-1}}\sqrt{\log(q\vee\sqrt{nh})}\le c\sqrt{nh}\,\xi_n
\iff
\xi_n^{\frac1\nedsize+\frac{\order}{\order-1}}\ge C\,\vanishing^{-\left(\frac1\nedsize+\frac1{\order-1}\right)}\,\frac{q^{\frac2\nedsize+\frac2{\order-1}}\sqrt{\log(q\vee\sqrt{nh})}}{\sqrt{nh}} ,
\]
which is the definition of $\xi_n$, because $\big(\frac1\nedsize+\frac1{\order-1}\big)\big/\big(\frac1\nedsize+\frac{\order}{\order-1}\big)=\frac{\nedsize+\order-1}{\nedsize\order+\order-1}$.
If $x<1$, then $\ell=1$, $\alpha<1$ and $\xi_n>q^2/\vanishing\ge1$, so \eqref{eq. cov first term} holds for $n$ large enough.
Hence $\p(\max_{j,k}|R^{(j,k)}_2|>\xi/2)\le C\vanishing$ for all $\xi\ge\xi_n$, and combining the two bounds proves the claim.

\subsection{Proof of Lemma \ref{lem. error. event F}}
\label{proof. lem. error. event F}

\begin{proof}
Apply Lemma~\ref{lem. tech. score} with $\bm A_{t,n}=\X$, $e_{t,n}=\varepsilon\tn$, $p_A=p$ and $\xi=\lambda(u)/4$.
Conditions (i) and (ii) of that lemma hold by Assumption~\ref{as. local stationarity}.
The orthogonality condition $\E[\Xu\epsu]=\bm0$ and the moment and dependence conditions hold by Assumption~\ref{as. NED}.
\end{proof}

\subsection{Proof of Lemma \ref{lem. error. event S}}
\label{proof. lem. error. event S}

\begin{proof}
Let $c_n:=\meant\kernel$. With unit weights, $\meant\kernel\big(\x[j]\x[k]-\Sigma_{j,k}(u)\big)=\hat\Sigma_{j,k}(u)-c_n\Sigma_{j,k}(u)$, so that, by \eqref{eq. riemann},
\[
\s\,\|\hat{\bm\Sigma}(u)-\bm\Sigma(u)\|_{\max}\le\s\,\|\hat{\bm\Sigma}(u)-c_n\bm\Sigma(u)\|_{\max}+\frac{C\s}{nh}\max_j\Sigma_{j,j}(u).
\]
The last term is at most $\kappa/2$ since $nh\ge C_0\s/\kappa$.
Assumption~\ref{as. bound on s} ensures that $\kappa/(2\s)\ge\xi_n$ with $q=p$, and Lemma~\ref{lem. tech. covariance} with $\bm A_{t,n}=\X$ and $\xi=\kappa/(2\s)$ gives the claim.
\end{proof}

\subsection{Proof of Lemma \ref{lem. error. event A}}
\label{proof. lem. error. event A}

\begin{proof}
By Assumption~\ref{as. Lipschitz}, the weights $w^{(j,k)}_t:=a_{t,j}a_{t,k}$ are deterministic and satisfy $|w^{(j,k)}_t|\le1$.
Since $\xi_n$ is nondecreasing in $q$ and $\s\le p$, Assumption~\ref{as. bound on s} gives $\kappa/\s\ge\xi_n$ with $q=\s$.
Lemma~\ref{lem. tech. covariance} with $\bm A_{t,n}=(\x[k])_{k\in S_\local(u)}$, these weights and $\xi=\kappa/\s$ gives the claim.
No Riemann step is needed, since $\mathcal A_\local(u)$ is centred at $\bm\Sigma(u)$ inside the weighted sum.
\end{proof}

\subsection{Proof of Lemma \ref{lem. asymp. nodewise model}}
\label{proof. lem. asymp. nodewise model}

\begin{proof}
Fix $j\in\mathcal J$, $u$ and $n$, and write $\tilde{\bm\Sigma}:=\bm\Sigma(u)$, $\bm\Theta:=\bm\Theta(u)$, $\tau_j^2:=\tau_j^2(u)$ and $\bm\gamma:=\gammaZero(u)$ when no confusion arises.

(i)
By Assumption~\ref{as. NED}(i), $\E\Xu=\bm0$, so that $\E\ndepsu=\E\xu[j]-\bm\gamma^\top\E\Xu[-j]=0$.
The function $\bm g\mapsto\E(\xu[j]-\Xu[-j]^\top\bm g)^2$ is convex, differentiable and finite, since second moments are bounded. Hence its minimizer $\bm\gamma$ satisfies the first-order condition $\E\big[\Xu[-j](\xu[j]-\Xu[-j]^\top\bm\gamma)\big]=\bm0$, that is, $\E[\ndepsu\xu[k]]=0$ for all $k\neq j$.

(ii)
By the Cauchy--Schwarz inequality, Assumption~\ref{as. projection errors} and Assumption~\ref{as. NED}(i),
$\E|\ndepsu\xu[k]|^{\bar\order}\le\big(\E|\ndepsu|^{2\bar\order}\,\E|\xu[k]|^{2\bar\order}\big)^{1/2}\le C$, and similarly with $\epsu$ in place of $\xu[k]$.

(iii) and (iv)
By Assumptions~\ref{as. NED} and~\ref{as. projection errors}, the vector $(\Xu^\top,\epsu,\ndepsu)^\top$ satisfies Assumption~\ref{as. NED}: it is centred by (i), it is $\mathcal F^U\tn(u)$-measurable since $\ndepsu$ is a linear function of $\Xu$, and its coordinates have bounded $2\bar\order$-th moments and are $L_{2\order}$-NED of size $-\nedsize$ with bounded constants.
Lemma~\ref{lem. tech. eps X mixingale} therefore applies to the products $\ndepsu\xu[k]$ and $\ndepsu\epsu$.
No centring is needed, since $\E[\ndepsu\xu[k]]=0$ for $k\neq j$ by (i), and $\E[\ndepsu\epsu]=\E[\xu[j]\epsu]-\bm\gamma^\top\E[\Xu[-j]\epsu]=0$ by Assumption~\ref{as. NED}(i).

(vi)
By Assumption~\ref{as. precision matrix}(i), $\|\bm\Theta\|\le C$.
By block inversion of $\tilde{\bm\Sigma}$, $\Theta_{j,j}=1/\tau_j^2$ and $\Theta_{j,k}=-\gamma_k/\tau_j^2$ for $k\neq j$, as stated in Section~\ref{sec. asymptotic analysis}.
Hence $\tau_j^2=1/\Theta_{j,j}\ge1/\|\bm\Theta\|\ge1/C$.
Taking $\bm g=\bm0$ in the definition of $\bm\gamma$ gives $\tau_j^2\le\E\xu[j]^2\le C$ by Assumption~\ref{as. NED}(i).
Moreover, $\|\bm\gamma\|_2=\tau_j^2\|\bm\Theta_{j,-j}\|_2\le\tau_j^2\|\bm\Theta\|\le C$.
For $k\notin S_{j,\local}(u)$, $\gamma^0_{j,k}$ vanishes on a neighbourhood of $u$ by definition of $S_{j,\local}(u)$, so $\bm\gamma$ is supported on $S_{j,\local}(u)$, and $\|\bm\gamma\|_1\le\sqrt{\sj}\,\|\bm\gamma\|_2\le C\sqrt{\sj}$.
Finally, $\|\bm\Theta_j\|_1=(1+\|\bm\gamma\|_1)/\tau_j^2\le C(1+\sqrt{\sj})\le C(1+\sqrt{\sbar})$.

(v)
By (vi), $\gammaZero(u)=-\tau_j^2(u)\,\bm\Theta_{j,-j}(u)$, so that
\[
\gammaZero(u)-\gammaZero(u')
=-\tau_j^2(u)\big(\bm\Theta_{j,-j}(u)-\bm\Theta_{j,-j}(u')\big)-\big(\tau_j^2(u)-\tau_j^2(u')\big)\bm\Theta_{j,-j}(u').
\]
Since $\tau_j^2=1/\Theta_{j,j}$,
\[
|\tau_j^2(u)-\tau_j^2(u')|=\tau_j^2(u)\,\tau_j^2(u')\,|\Theta_{j,j}(u')-\Theta_{j,j}(u)|\le C^2\|\bm\Theta_j(u)-\bm\Theta_j(u')\|_2 .
\]
Using $\tau_j^2(u)\le C$, $\|\bm\Theta_{j,-j}(u')\|_2\le\|\bm\Theta(u')\|\le C$ and Assumption~\ref{as. precision matrix}(ii),
\[
\|\gammaZero(u)-\gammaZero(u')\|_2\le C(1+C^2)\,\|\bm\Theta_j(u)-\bm\Theta_j(u')\|_2\le C_\gamma L_\Theta\,|u-u'| .
\]
\end{proof}

\subsection{Proof of Lemma \ref{lem. asymp. nodewise lasso}}
\label{proof. lemma nodewise lasso}

\begin{proof}
(i) The KKT conditions of the nodewise lasso give $\meant\kernel\x[k]\hat v_{j,t}=\lambda_j(u)\hat\kappa_k$ for $k\neq j$, with $\|\hat{\bm\kappa}\|_\infty\le1$ and $\hat{\bm\kappa}^\top\hat{\bm\gamma}_j(u)=\|\hat{\bm\gamma}_j(u)\|_1$.
Since $\bhat\Theta_j(u)\X=\hat v_{j,t}/\hat\tau_j^2(u)$, we have $(\bhat\Theta_j(u)\hat{\bm\Sigma}(u))_k=\lambda_j(u)\hat\kappa_k/\hat\tau_j^2(u)$ for $k\neq j$, and
\[
(\bhat\Theta_j(u)\hat{\bm\Sigma}(u))_j=\frac{1}{\hat\tau_j^2(u)}\Big(\meant\kernel\hat v_{j,t}^2+\lambda_j(u)\|\hat{\bm\gamma}_j(u)\|_1\Big)=1
\]
by the definition of $\hat\tau_j^2(u)$. This proves the first bound.
The second follows from $\meant\kernel\hat v_{j,t}^2\le\hat\tau_j^2(u)$.

(ii) The nodewise regression $\x[j]=\X\mjtop\gammaZero(t/n)+\ndeps$ has the form \eqref{eq. model}, with active set $S_{j,\local}(u)$ and Lipschitz constant $C_\gamma L_\Theta$ by Lemma~\ref{lem. asymp. nodewise model}(v).
Step~1 of the proof of Theorem~\ref{th. error bound} applies verbatim, with $\Fsetj$, $\Ssetbar$, $\mathcal A_{j,\local}(u)$ and $\bar\kappa$ in place of $\Fset$, $\Sset$, $\mathcal A_\local(u)$ and $\kappa$.
Assumption~\ref{as. precision matrix}(i) gives a compatibility constant of at least $1/C$, and $\Lambda_{j,\local}(u)$ replaces $\localEig$.

(iii) Write $\hat v_{j,t}=\ndeps+d_t$ with $d_t:=\X\mjtop(\gammaZero(t/n)-\hat{\bm\gamma}_j(u))$, and let $P_{n,j}:=\meant\kernel d_t^2$, which is bounded in (ii). Then
\begin{align*}
\hat\tau_j^2(u)-\tau_j^2(u)&=\Big(\meant\kernel\ndeps^2-\tau_j^2(u)\Big)+\frac{2}{nh}\sumt\kernel\ndeps d_t\\
&\quad+P_{n,j}+\lambda_j(u)\|\hat{\bm\gamma}_j(u)\|_1 .
\end{align*}
On $\mathcal V_\nu$, the first term is at most $\nu$, and by the Cauchy--Schwarz inequality the second is at most $2(\tau_j^2(u)+\nu)^{1/2}P_{n,j}^{1/2}$.
By (ii), $P_{n,j}^{1/2}\le C\big(\lambda_j(u)\sqrt{\sj}+\sqrt{\sj\Lambda_{j,\local}(u)}\,h\big)$.
By Lemma~\ref{lem. asymp. nodewise model}(vi) and (ii),
\begin{align*}
\lambda_j(u)\|\hat{\bm\gamma}_j(u)\|_1&\le\lambda_j(u)\|\gammaZero(u)\|_1+\lambda_j(u)\|\hat{\bm\gamma}_j(u)-\gammaZero(u)\|_1\\
&\le C\lambda_j(u)\sqrt{\sj}+C\big(\lambda_j(u)^2\sj+\Lambda_{j,\local}(u)\,\sj\, h^2\big).
\end{align*}
Since $\lambda_j(u)^2\sj+\Lambda_{j,\local}(u)\sj h^2\le c\le1$, we have $\lambda_j(u)^2\sj\le\lambda_j(u)\sqrt{\sj}$ and $\Lambda_{j,\local}(u)\sj h^2\le\sqrt{\sj\Lambda_{j,\local}(u)}\,h$. Combining these bounds with $\tau_j^2(u)+\nu\le C$ gives the first claim.
Since $\tau_j^2(u)\ge1/C$ by Lemma~\ref{lem. asymp. nodewise model}(vi), the right-hand side is at most $\tau_j^2(u)/2$ for $c$ small enough, which gives the lower bound on $\hat\tau_j^2(u)$.
\end{proof}

\subsection{Proof of Lemma \ref{lem. asymp. event F}}
\label{proof. lem. asymp. event F}

\begin{proof}
Apply Lemma~\ref{lem. tech. score} with $\bm A_{t,n}=\X\mj$, $e_{t,n}=\ndeps$, $p_A=p-1$ and $\xi=\lambda_j(u)/4$.
Its conditions (i) and (ii) hold by Assumption~\ref{as. local stationarity}(i) and Proposition~\ref{prop. projection errors}, with rates $\rate$ and $\rate^v$ and lag order $\lagorder$.
The orthogonality condition $\E[\Xu[-j]\ndepsu]=\bm0$ holds by Lemma~\ref{lem. asymp. nodewise model}(i).
Assumption~\ref{as. NED} holds for $(\Xu[-j]^\top,\ndepsu)^\top$ by Assumptions~\ref{as. NED} and~\ref{as. projection errors}, since $\ndepsu$ is a linear function of $\Xu$ and $\E\ndepsu=0$ by Lemma~\ref{lem. asymp. nodewise model}(i).
\end{proof}

\subsection{Proof of Lemma \ref{lem. asymp. event S}}
\label{proof. lem. event S}

\begin{proof}
The bound follows as in the proof of Lemma~\ref{lem. error. event S}, with $\smax$ and $\bar\kappa$ in place of $\s$ and $\kappa$. Assumption~\ref{as. bound on s} with $\smax$ in place of $\s$ ensures that $\bar\kappa/(2\smax)\ge\xi_n$ with $q=p$, and $nh\ge C_0\smax/\bar\kappa$ for $n$ large enough.
\end{proof}

\subsection{Proof of Lemma \ref{lem. asymp. event A}}
\label{proof. lem. event A}

\begin{proof}
Lemma~\ref{lem. asymp. nodewise model}(v) gives, for $|t/n-u|\le h/2$,
\[
|\gamma^0_{j,k}(t/n)-\gamma^0_{j,k}(u)|\le\|\gammaZero(t/n)-\gammaZero(u)\|_2\le C_\gamma L_\Theta\,|t/n-u|,
\]
so that the weights $g_{t,jk}$ are deterministic and satisfy $|g_{t,jk}|\le1$.
Since $\xi_n$ is nondecreasing in $q$ and $\sj\le\smax\le p$, we have $\bar\kappa/\sj\ge\xi_n$ with $q=\sj$.
The claim then follows as in the proof of Lemma~\ref{lem. error. event A}, with $\s$, $\kappa$, $S_\local(u)$ and $a_{t,j}$ replaced by $\sj$, $\bar\kappa$, $S_{j,\local}(u)$ and $g_{t,jk}$.
\end{proof}

\subsection{Proof of Lemma \ref{lem. asymp. event V}}
\label{proof. lem. asymp. event V}

\begin{proof}
Fix $j\in\mathcal J$. Since $\E\ndepsu^2=\tau_j^2(u)$ for all $t$,
\begin{align*}
&\meant\kernel\ndeps^2-\tau_j^2(u)
=\underbrace{\meant\kernel\big(\ndeps^2-\ndepsu^2\big)}_{=:I_j} \\
&+\underbrace{\meant\kernel\big(\ndepsu^2-\tau_j^2(u)\big)}_{=:II_j}
+\tau_j^2(u)\Big(\meant\kernel-1\Big).
\end{align*}
By \eqref{eq. riemann} and $\tau_j^2(u)\le C$ (Lemma~\ref{lem. asymp. nodewise model}(vi)), the last term is at most $C/(nh)$, which is at most $\nu/2$ for $C_0$ large enough. Hence
\[
\p\Big(\Big|\meant\kernel\ndeps^2-\tau_j^2(u)\Big|>\nu\Big)\le\p\big(|I_j|>\nu/4\big)+\p\big(|II_j|>\nu/4\big).
\]

\emph{Approximation part.}
By Proposition~\ref{prop. projection errors}, $\ndeps$ is $(\rate^v,\local)$-stationary with stationary approximation $\ndepsu$ and lag order $\lagorder\le Cnh$, and $\E|\ndepsu|^{2\bar\order}\le C$ by Assumption~\ref{as. projection errors}.
Lemma~\ref{lem. tech. transfer} with $\bm A_{t,n}=\bm B_{t,n}=\ndeps$ and $p_A=p_B=1$, whose conditions (i) and (ii) follow from Proposition~\ref{prop. projection errors} by Lyapunov's inequality, gives $\E|I_j|\le Ch\,\rate^v(1+h\,\rate^v)$, so that, by Markov's inequality,
\[
\p\big(|I_j|>\nu/4\big)\le\frac{Ch\,\rate^v(1+h\,\rate^v)}{\nu}.
\]

\emph{Stationary part.}
By Lemma~\ref{lem. tech. eps X mixingale}, applied to $(\Xu^\top,\epsu,\ndepsu)^\top$ as in the proof of Lemma~\ref{lem. asymp. nodewise model}(iii), $\{\ndepsu^2-\tau_j^2(u)\}$ is $L_{\bar\order}$-bounded and an $L_{\order}$-mixingale of size $-\nedsize$ with respect to $\mathcal F^U\tn(u)$, with mixingale constants bounded by $C$.
Since the weights $\kernel$ are deterministic and bounded, $\{\kernel(\ndepsu^2-\tau_j^2(u))\}$ is an $L_{\order}$-mixingale with constants at most $C\kernel$, and $\sumt\kernel^2\le Cnh$.
The moment inequality for mixingales of \citet[Lemma~2]{hansen1991strong}, which applies since $\order\ge2$ and $\nedsize\ge1$, gives
\[
\E\Big|\sumt\kernel\big(\ndepsu^2-\tau_j^2(u)\big)\Big|^{\order}\le C\Big(\sumt\kernel^2\Big)^{\order/2}\le C(nh)^{\order/2},
\]
so that, by Markov's inequality,
\[
\p\big(|II_j|>\nu/4\big)\le\frac{C(nh)^{\order/2}}{(nh\,\nu)^{\order}}=\frac{C}{\nu^{\order}(nh)^{\order/2}}.
\]

The claim follows from a union bound over $j\in\mathcal J$, since $|\mathcal J|\le C$.
\end{proof}

\subsection{Proof of Lemma \ref{lem. asymp. clt}}
\label{proof. lem. asymp. clt}

\begin{proof}
Let $\bm S_n:=(nh)^{-1/2}\sumt\kernel\tilde{\bm w}\tn(u)$.
By the Cram\'er--Wold device, it suffices to show that $\bm\alpha^\top\bm V_n(u)^{-1/2}\bm S_n\overset{d}{\to}\mathcal N(0,1)$ for every fixed unit vector $\bm\alpha\in\R^{|\mathcal J|}$.
Fix such an $\bm\alpha$, let $\bm b_n:=\bm V_n(u)^{-1/2}\bm\alpha$, and let $N_u:=\{t:\,|t/n-u|\le h/2\}$ with $k_n:=|N_u|$.
Since $K$ vanishes outside $[-1/2,1/2]$,
\[
\bm\alpha^\top\bm V_n(u)^{-1/2}\bm S_n=\sum_{t\in N_u}X_{n,t},
\qquad
X_{n,t}:=(nh)^{-1/2}\,\kernel\,\bm b_n^\top\tilde{\bm w}\tn(u).
\]
Since $nh\to\infty$, we have $k_n\ge nh-1\to\infty$ and, for $n$ large enough, $k_n\le nh+1\le2nh$.
We apply Theorem~24.6 and Corollary~24.7 of \citet{davidson2021stochastic} to the array $\{X_{n,t}:t\in N_u\}$, with constants $c_{n,t}:=(nh)^{-1/2}$, $r:=\bar\order$, and the $\sigma$-fields $\mathcal F^U\tn(u)$ of Assumption~\ref{as. NED}(ii). We verify their conditions.

\emph{(a) Measurability, mean and variance.}
Recall that $\tilde w_{j,t,n}(u)=\bm\Theta_j(u)\Xu\epsu=\ndepsu\epsu/\tau_j^2(u)$, where $\ndepsu=\xu[j]-\Xu[-j]^\top\gammaZero(u)$ is a fixed linear function of $\Xu$.
Hence $X_{n,t}$ is $\mathcal F^U\tn(u)$-measurable by Assumption~\ref{as. NED}(ii).
Moreover, $\E[\ndepsu\epsu]=\E[\xu[j]\epsu]-\gammaZero(u)^\top\E[\Xu[-j]\epsu]=0$ by Assumption~\ref{as. NED}(i), so that $\E X_{n,t}=0$.
By the definition of $\bm V_n(u)$,
\[
\E\Big[\Big(\sum_{t\in N_u}X_{n,t}\Big)^2\Big]=\bm b_n^\top\,\E\big[\bm S_n\bm S_n^\top\big]\,\bm b_n=\bm\alpha^\top\bm V_n(u)^{-1/2}\bm V_n(u)\bm V_n(u)^{-1/2}\bm\alpha=1 .
\]

\emph{(b) Uniform $L_r$-boundedness.}
By Assumption~\ref{as. long-run variance}, $\|\bm b_n\|_2\le\|\bm V_n(u)^{-1/2}\|\le C$.
By Lemma~\ref{lem. asymp. nodewise model}(vi), $\tau_j^2(u)\ge1/C$.
By the Cauchy--Schwarz inequality, Assumption~\ref{as. projection errors} and Assumption~\ref{as. NED}(i), $\E|\ndepsu\epsu|^{\bar\order}\le\big(\E|\ndepsu|^{2\bar\order}\,\E|\epsu|^{2\bar\order}\big)^{1/2}\le C$.
Since $|\mathcal J|\le C$ and $|K|\le\bar K$, Minkowski's inequality gives
\[
\sup_{n,t}\E\big|X_{n,t}/c_{n,t}\big|^{\bar\order}\le C .
\]

\emph{(c) Near-epoch dependence and mixing.}
By Lemma~\ref{lem. asymp. nodewise model}(iv), $\ndepsu\epsu$ is $L_{\order}$-NED of size $-\nedsize$ on $\{\bm U\tn(u)\}$ with bounded constants.
Since $X_{n,t}/c_{n,t}$ is a linear combination of these products, with at most $|\mathcal J|\le C$ terms and coefficients bounded by $C\bar K$, it is $L_{\order}$-NED of size $-\nedsize$. Since $\order>2$ and $\nedsize\ge1$, it is therefore $L_2$-NED of size $-1$.
Finally, $\{\bm U\tn(u)\}$ is $\alpha$-mixing of size $-\nedsize/(1/\order-1/\bar\order)$. Since $1/\order-1/\bar\order<1/2-1/\bar\order=(\bar\order-2)/(2\bar\order)$ and $\nedsize\ge1$, this size is smaller than $-2\bar\order/(\bar\order-2)$, and hence smaller than $-r/(r-2)$.

\emph{(d) Constants.}
With $M_n:=\max_{t\in N_u}c_{n,t}=(nh)^{-1/2}$, we have $\sup_n k_nM_n^2=\sup_n k_n/(nh)\le2$.

Hence $\sum_{t\in N_u}X_{n,t}\overset{d}{\to}\mathcal N(0,1)$, and the Cram\'er--Wold device gives the claim.
\end{proof}

\section{Implementation Details}
\label{appendix. simulations Competing methods}

This appendix describes the implementation of the \llasso, of the desparsified \llasso\ and of the competing methods, and gives further details on the simulations and the data analyses.
Throughout, $K$ is the Epanechnikov kernel $K(v)=\tfrac32(1-4v^2)\mathbbm 1(|v|\le1/2)$, so that $h$ is the full width of the window $\{t:|t/n-u|\le h/2\}$.
We write $K_t(u):=K((t/n-u)/h)$ and let $\tilde n(u):=\big(\sum_tK_t(u)\big)^2/\sum_tK_t(u)^2$ be the effective sample size of the window, which is close to $nh/1.2$ at interior points.

\subsection{Local lasso}
\label{app. F local lasso}

We solve \eqref{eq. min problem} by coordinate descent with the R package \texttt{glmnet}, with observation weights $K_t(u)$.
The penalty is selected in two steps.
First, at each point $u_g$ of a grid, $\hat\lambda_g$ minimizes the local BIC \eqref{eq. BIC criterion} over the default path of 100 penalties that \texttt{glmnet} computes for the unweighted lasso on the full sample.
Second, the pointwise choices are smoothed over $u$ by a kernel-weighted geometric mean,
\[
\lambda(u):=\exp\Big(\frac{\sum_g K\big((u_g-u)/h\big)\log\hat\lambda_g}{\sum_g K\big((u_g-u)/h\big)}\Big).
\]
Hence $\lambda(u)$ depends only on the grid points in the window of $u$.
For curves, the grid consists of $N=50$ equispaced points in $[1/(n+1),\,n/(n+1)]$.
When the estimator is needed at a few points only, such as $u=0.25$, $0.5$ and $0.75$ in Table~\ref{tab. inference}, the BIC is computed at 9 equispaced interior points of each window $[u-h/2,u+h/2]$, so that $\lambda(u)$ averages over the same neighbourhood as on a dense grid.

\subsection{Desparsified local lasso and long-run variance}
\label{sec. variance estimator}

The desparsified estimator \eqref{eq. desparsified lasso} uses the bandwidth $h=h^*n^{-1/15}$ for all its components, that is, the initial \llasso, the nodewise regressions and the kernel weights.
The penalties $\lambda(u)$ and $\lambda_j(u)$ are selected as in Section~\ref{app. F local lasso} at this bandwidth, with $\x[j]$ as response and $\X\mj$ as covariates for the nodewise regressions.
Only the rows $\bhat\Theta_j(u)$, $j\in\mathcal J$, are computed.

Let $\hat\varepsilon_t(u) := Y\tn - \X^\top\hat\Beta(u)$ denote the residuals of the \llasso\ and $\hat{\bm w}_t(u) := K_t(u)\,\bhat\Theta_{\mathcal{J}}(u)\,\X\,\hat\varepsilon_t(u)$, where $\bhat\Theta_j(u)\X=\hat v_{j,t}/\hat\tau_j^2(u)$ is the nodewise residual divided by $\hat\tau_j^2(u)$.
We estimate $\bm V_n(u)$ by the Bartlett estimator \citep{neweyWest1986}
\[
\bhat V(u) := \hat{\bm G}_0(u) + \sum_{\ell=1}^{L_n(u)}\Big(1 - \frac{\ell}{L_n(u)+1}\Big)\big(\hat{\bm G}_\ell(u) + \hat{\bm G}_\ell(u)^\top\big),
\qquad
\hat{\bm G}_\ell(u) := \frac{1}{nh}\sum_{t=\ell+1}^{n} \hat{\bm w}_t(u)\, \hat{\bm w}_{t-\ell}(u)^\top,
\]
with truncation lag $L_n(u):=\lfloor 4\,(\tilde n(u)/100)^{2/9}\rfloor$. This is the rule of thumb of \citet{newey1994automatic}, with the effective sample size of the window in place of the sample size. The Bartlett weights ensure that $\bhat V(u)$ is positive semidefinite.
Confidence intervals for $\beta_j^0(u)$ at level $1-\alpha$ are then given by
\[
\hat b_j(u) \pm \Phi^{-1}(1-\alpha/2)\,\sqrt{\hat V_{jj}(u)/(nh)}.
\]

\subsection{Group lasso and oracle}
\label{app. F group lasso}

We compare the \llasso\ with the local-constant version of the kernel group lasso of \citet{Lee2016}.
The coefficient curves are estimated jointly on the grid $u_1,\dots,u_N$ of Section~\ref{app. F local lasso} by minimizing
\begin{equation}\label{eq. grp min integral}
\frac1N\sum_{g=1}^N\frac{\sum_tK_t(u_g)\big(Y\tn-\X^\top\Beta_g\big)^2}{\sum_tK_t(u_g)}
+\lambda\sum_{j=1}^{p_n}\Big(\frac1N\sum_{g=1}^N\beta_{g,j}^2\Big)^{1/2}
\end{equation}
over $\Beta_1,\dots,\Beta_N\in\R^{p_n}$, with the same kernel and bandwidth $h^*$.
Criterion \eqref{eq. grp min integral} is a Riemann approximation of the criterion of \citet{Lee2016} without the derivative terms, which keeps the two estimators comparable.
Each group is a whole coefficient curve, so that a covariate is selected either on all of $[0,1]$ or nowhere.
We solve \eqref{eq. grp min integral} by block coordinate descent, with a sequential strong rule and warm starts along a decreasing geometric path of 30 penalties from $\lambda_{\max}$ to $0.01\,\lambda_{\max}$, followed by a check of the KKT conditions.
The penalty $\lambda$ is selected by five-fold cross-validation adapted to dependent data.

The oracle is the kernel-weighted least-squares estimator computed at each observation time $u_t = t/(n+1)$, with the same kernel and bandwidth $h^*$, on the covariates in the pointwise active set $\{j : \beta^0_j(u_t) \neq 0\}$, whose coefficients are nonzero at $u_t$ itself. Its other coefficients are set to zero.

\subsection{Simulation details}
\label{app. F simulations}

\textbf{Bandwidths.}
Table~\ref{tab. bandwidths} reports the bandwidths and the corresponding numbers of observations in a window.

\begin{table}[h]
\centering\small
\caption{Bandwidths $h^*=(p_n^{1/5}/n)^{1/3}$ of the \llasso\ and the group lasso, and $h=h^*n^{-1/15}$ of the desparsified \llasso, with the numbers of observations in a window, $nh^*$ and $nh$.}
\label{tab. bandwidths}
\begin{tabular}{rr cc cc}
\toprule
$n$ & $p_n$ & $h^*$ & $h$ & $nh^*$ & $nh$\\
\midrule
400  & 50   & 0.176 & 0.118 & 70  & 47\\
     & 200  & 0.193 & 0.130 & 77  & 52\\
     & 1000 & 0.215 & 0.144 & 86  & 58\\
\addlinespace
800  & 50   & 0.140 & 0.090 & 112 & 72\\
     & 200  & 0.153 & 0.098 & 123 & 79\\
     & 1000 & 0.171 & 0.109 & 137 & 87\\
\addlinespace
1600 & 50   & 0.111 & 0.068 & 178 & 109\\
     & 200  & 0.122 & 0.074 & 195 & 119\\
     & 1000 & 0.136 & 0.083 & 217 & 133\\
\bottomrule
\end{tabular}
\end{table}

\textbf{Data generation.}
The covariates start from their stationary distribution at $u=0$.
The tv-ARX process of Model~\ref{model time series} is started after a burn-in of 100 observations, during which the coefficients are kept at their values at $u=0$. 
Its autoregressive part is stable at every $u$, since the roots of $1-\alpha_1(u)z-\alpha_2(u)z^2$ have modulus at least $\sqrt2$. This bound is attained when $\alpha_1(u)=0$ and $\alpha_2(u)=-0.5$.

\textbf{Performance measures.}
The estimates on the grid are linearly interpolated to the time points $t/n$, $t=1,\dots,n$.
The $\ell_2$ and $\ell_1$ errors are $n^{-1}\sum_t\|\hat\Beta(t/n)-\Beta^0(t/n)\|_2^2$ and $n^{-1}\sum_t\|\hat\Beta(t/n)-\Beta^0(t/n)\|_1$.
At time $t/(n+1)$, a false negative is a coefficient with $\hat\beta_j(t/n)=0\neq\beta^0_j(t/n)$, and a false positive is a coefficient with $\hat\beta_j(t/n)\neq0=\beta^0_j(t/n)$. Their numbers are averaged over $t$.
Table~\ref{tab. estimation} reports medians over the 200 replications.
In Table~\ref{tab. inference}, the coverage is the share of the 200 replications in which the confidence interval at $u$ contains $\beta^0_j(u)$.

\subsection{Data analyses}
\label{app. F data}

Both analyses use the bandwidth rules of the simulations, $h^*=(p_n^{1/5}/n)^{1/3}$ for the \llasso\ and $h=h^*n^{-1/15}$ for the desparsified \llasso, together with the penalties of Section~\ref{app. F local lasso} and the variance estimator of Section~\ref{sec. variance estimator}.
The curves are computed on a grid of $N=200$ equispaced points.

\textbf{Phillips curve.}
The data are the September 2025 vintage of FRED-MD \citep{mccracken2016fred}.
The response is core PCE inflation twelve months ahead, $\pi_{t+12}$, where $\pi_t=1200\,\Delta\log\mathrm{PCEPILFE}_t$ is annualized monthly inflation.
The 118 predictors are the FRED-MD series with no missing values from January 1960 to July 2025.
They are transformed with the codes of \citet{mccracken2016fred}, except that no second differences are taken and a few rates, ratios and housing series are kept in levels or log levels; 29 series with extreme outliers are also compressed with an asinh transform.
The unemployment gap is the level of the unemployment rate minus the noncyclical rate of unemployment of the Congressional Budget Office, whose quarterly series is linearly interpolated to monthly frequency.
With $n=785$ and $p_n=249$, the bandwidths are $h^*\approx0.157$ and $h\approx0.100$, that is, windows of about 123 and 79 months.
The desparsified estimator is computed for the coefficients of the unemployment gap and of the first inflation lag.

\textbf{EEG data.}
The data are the stereo-EEG recordings of \citet{bernabei2023quantitative}, in which each contact is annotated as lying inside or outside the SOZ. They are available on OpenNeuro (ds004100, version 1.1.3).
We use the third ictal recording (run 3) of subject HUP162, from 20\,s before to 80\,s after clinical seizure onset.
The signals are not re-referenced. They are filtered causally, with a 1--40\,Hz band-pass, a 60\,Hz notch and an anti-aliasing low-pass filter, decimated from 1024\,Hz to 128\,Hz, and standardized with the mean and standard deviation of each contact over the baseline from 100\,s to 20\,s before onset.
Channels marked as bad in the dataset are excluded, which leaves 169 contacts with 12,800 samples each.
Each contact enters with lags 1 to 10, including the lags of the response LA3, so that $p_n=1690$ and $n=12{,}790$.
The bandwidths are $h^*\approx0.070$ and $h\approx0.037$, that is, windows of about 7.0\,s and 3.7\,s.

\begin{table}[h]
\centering\small
\caption{Bandwidths of the \llasso\ and the desparsified \llasso\ in the data analyses, with the number of observations in a window, $nh$, and the Kish effective sample size of the window centered at $u=1/2$.}
\label{tab. effective sample size}
\begin{tabular}{lrr lccc}
\toprule
Application & $n$ & $p_n$ & Estimator & Bandwidth & $nh$ & Effective sample size\\
\midrule
Phillips curve & 785    & 249  & \llasso\            & 0.157 & 123 & 103\\
               &        &      & desparsified \llasso\ & 0.100 & 79  & 66\\
\addlinespace
EEG            & 12,790 & 1690 & \llasso\            & 0.070 & 898 & 748\\
               &        &      & desparsified \llasso\ & 0.037 & 478 & 398\\
\bottomrule
\end{tabular}
\end{table}

\end{appendices}

\putbib[Lasso]        
\end{bibunit}

\end{document}